\documentclass{article}

\usepackage{arxiv}

\usepackage[utf8]{inputenc} 
\usepackage[T1]{fontenc}    
\usepackage{hyperref}       
\usepackage{url}            
\usepackage{booktabs}       
\usepackage{amsfonts}       
\usepackage{nicefrac}       
\usepackage{microtype}      
\usepackage{lipsum}		    
\usepackage{doi}
\usepackage{caption}
\usepackage{subcaption}
\usepackage{graphicx}
\usepackage{proba}
\usepackage{bm}
\usepackage{bbm}
\usepackage{enumitem}
\usepackage{xcolor}
\usepackage{afterpage}
\usepackage[authoryear]{natbib}
\usepackage[symbol]{footmisc}
\usepackage{amsmath}
\usepackage{cleveref}
\usepackage{lscape}
\usepackage{mwe}
\usepackage{adjustbox}
\usepackage{multirow}
\usepackage[flushleft]{threeparttable}
\usepackage{lineno}
\usepackage{soul}
\usepackage{array}

\newcolumntype{A}{ >{${}} r <{$} @{} >{${}} l <{$} } 

\usepackage{subfiles} 

\title{Data Mining of Telematics Data: \\Unveiling the Hidden Patterns in Driving Behaviour}

\author{
    Ian Weng Chan\thanks{Department of Statistical Sciences, University of Toronto. Ontario Power Building, 700 University Avenue, 9th Floor, Toronto, ON M5G 1Z5, Canada. Email addresses: \texttt{ianweng.chan@mail.utoronto.ca} (Ian Weng Chan), \texttt{spark.tseung@mail.utoronto.ca} (Spark C.~Tseung), \texttt{andrei.badescu@utoronto.ca} (Andrei L.~Badescu), \texttt{sheldon.lin@utoronto.ca} (X. Sheldon Lin).}\\ 
        \And
	{Spark C.~Tseung$^*$} \\
	\And
	{Andrei L.~Badescu$^*$} \\
    \And
	{X. Sheldon Lin$^*$} \\
}

\date{April 16, 2024}

\renewcommand{\headeright}{}
\renewcommand{\shorttitle}{Data Mining of Telematics Data: Unveiling the Hidden Patterns in Driving Behaviour}

\hypersetup{
pdftitle={Data Mining of Telematics Data},
pdfsubject={},
pdfauthor={},
pdfkeywords={},
}

\begin{document}
\maketitle

\begin{abstract}

With the advancement in technology, telematics data which capture vehicle movements information are becoming available to more insurers.  As these data capture the actual driving behaviour, they are expected to improve our understanding of driving risk and facilitate more accurate auto-insurance ratemaking.  In this paper, we analyze an auto-insurance dataset with telematics data collected from a major European insurer.  Through a detailed discussion of the telematics data structure and related data quality issues, we elaborate on practical challenges in processing and incorporating telematics information in loss modelling and ratemaking.  Then, with an exploratory data analysis, we demonstrate the existence of heterogeneity in individual driving behaviour, even within the groups of policyholders with and without claims, which supports the study of telematics data.  Our regression analysis reiterates the importance of telematics data in claims modelling; in particular, we propose a speed transition matrix that describes discretely recorded speed time series and produces statistically significant predictors for claim counts.  We conclude that large speed transitions, together with higher maximum speed attained, nighttime driving and increased harsh braking, are associated with increased claim counts.  Moreover, we empirically illustrate the learning effects in driving behaviour: we show that both severe harsh events detected at a high threshold and expected claim counts are not directly proportional with driving time or distance, but they increase at a decreasing rate.

\end{abstract}

\keywords{Usage-based Insurance, Vehicle Telematics, Data Mining, Feature Engineering, Principal Component Analysis}

\newpage
\section{Introduction}
\label{Introduction}

In classical auto-insurance ratemaking, insurers collect information on both the driver and the vehicle, such as driver’s age, years since driver’s licence obtained, vehicle brand and engine power, etc.  Actuaries then use these features (also known as covariates) for risk classification, prediction of claim counts and loss amounts, and determination of a premium that will be sufficient to cover future claims.  To facilitate subsequent discussion, these features will be described as `traditional'.  With the advancement in technology, vehicle movements can be monitored and recorded by on-board diagnostics installed in the vehicle or on the driver's smartphone.  Such information, including global positioning system (GPS) locations, driving speed and acceleration, etc., is referred to as (vehicle) telematics data and is becoming available to more insurers (\citet{eling_impact_2020}).  As these data capture the actual driving behaviour, they are expected to provide additional information on a driver's risk that has not been captured by traditional covariates.  Thus, the inclusion of telematics data should improve our understanding of driving risk and facilitate a more accurate auto-insurance ratemaking.  Usage-/telematics-based insurance is a rapidly growing field of research and there is an extensive publication. \citet{chauhan_bibliometric_2024} conduct a bibliometric review of telematics-based automobile insurance, while \citet{boylan_systematic_2024} provide a systematic review of the use of in-vehicle telematics in monitoring driving behaviours and risk, and we refer readers to the comprehensive lists of literature therein.

However, there are two major challenges.  The first challenge is what information from telematics data is useful and what features should be extracted.  Feature selection is a standard problem in classical insurance ratemaking as insurers have been trying to select traditional covariates which have high predictive power for the policyholders' risk levels, but this problem becomes more complicated as telematics data are recorded very frequently, ranging from minutes (as in this work) to even seconds (as in \citet{ma_use_2018} and \citet{gao_claims_2019}).  Data aggregation is often employed and it usually begins at the trip level.  While feature selection also depends on the specific data available to researchers, commonly considered variables include: numbers of harsh acceleration, harsh braking, cornering (with thresholds based on expert judgement), (fractions of) distances travelled in daytime/nighttime, on weekdays/weekends, on highways/ordinary roads, etc. over some duration (see \citet{paefgen_multivariate_2014}, \citet{ayuso_using_2016}, \citet{verbelen_unravelling_2018}, \citet{bian_good_2018}, \citet{jin_latent_2018}, \cite{denuit_multivariate_2019}, \citet{ayuso_improving_2019}, \citet{huang_automobile_2019}, \citet{sun_assessing_2020}, \citet{longhi_car_2020}).  Although most of these features have proven to be statistically significant predictors of accident occurrence, the use of these features can lead to loss of information.  For example, any changes and trends in telematics data over time will be neglected.  Recently, more researchers have started considering machine learning approaches on (raw) telematics time series, such as the use of speed-acceleration (\textit{v-a}) heatmaps (\citet{wuthrich_covariate_2017}, \citet{gao_claims_2019}, \citet{gao_boosting_2022}), pattern recognition (\citet{weidner_classification_2016}, \citet{weidner_telematic_2017}), modelling and prediction of time series using hidden Markov models (\citet{jiang_auto_2024}) and neural networks (\citet{fang_mocha_2021}), and automatic feature engineering using combined actuarial neural network (CANN) (\citet{duval_telematics_2024}).

The second challenge concerns how the telematics data should be incorporated in a modelling framework for insurance applications.  Although telematics data are abundant, claim occurrence remains rare, which leads to an inconsistency in data granularity.  In existing literature, telematics features are often aggregated at the trip level, and the yearly response (either the probability of incurring at least one claim, or the number of claims incurred) is regressed on these telematics features, leading to multiple observations for each policyholder in a year.  For example, if a policyholder reports no claims in the year, all his/her trips will have a response of 0 in the modelling of claim probability.  There are three major ways of incorporating telematics information into the model.  First, \citet{baecke_value_2017}, \citet{verbelen_unravelling_2018}, \citet{ma_use_2018}, \citet{jin_latent_2018}, \citet{ayuso_improving_2019}, \citet{huang_automobile_2019}, \citet{guillen_near-miss_2021} and \citet{duval_how_2022} treat telematics features as additional covariates and these features enter into the regression model in the same way as traditional covariates.  Second, \citet{ayuso_improving_2019} and \citet{gao_boosting_2022} use telematics features as a correction to the existing, classical pricing models by performing another regression with the original estimates as offsets.  Third, \citet{denuit_multivariate_2019} propose a multivariate credibility framework to model telematics features jointly with claim counts, which allows a posteriori correction on the expected claim counts using the information from observed telematics data.  However, a limitation often found in existing work is a mismatch between policy period and telematics observation period: researchers implicitly assume constant driving behaviour and make predictions for upcoming insurance periods based solely on observed driving history.  Recently, \citet{fang_mocha_2021} have empirically shown that such approach is sub-optimal: the use of predicted telematics metrics outperforms the use of historical averages, because the latter are biased when used to describe future driving patterns.

In this paper, we analyze an auto-insurance portfolio with vehicle telematics data collected in Romania, with the aim of better understanding telematics data, various patterns in individual's driving behaviour and its relationship with claim occurrence.  One noteworthy contribution of this paper, often overlooked in the telematics insurance literature, is the focus on data quality issues.  It is in general challenging to handle and process telematics data, due to the massive data volume and unfamiliarity of researchers and practitioners with such data.  Moreover, our dataset presents an additional challenge for analysis due to its complex, peculiar data structure and an above-average claim rate associated with highly heterogeneous driving behaviours.  We devote a section to elaborate on the data quality issues and provide sample data to illustrate the situations, in the hope that our work can serve as a preface on telematics data for researchers and practitioners who are interested in utilizing telematics raw data for insurance applications.

Moreover, we aim to tackle the challenge on what telematics features are useful.  By performing exploratory data analysis on both a portfolio level and an individual policyholder level, we also reveal why some covariates are less predictive of claim occurrence than expected.  On the one hand, self-reported (traditional) covariates can be a poor risk proxy when the policyholders' behaviour constantly deviate from what is declared.  In particular, we examine the different areas/regions driven by policyholders and unveil that the self-reported region of residence fails to proxy when the policyholder frequently drives outside of the region or even the country.  This discovery, made possible by telematics data, offers new insight to both actuarial researchers and practitioners.  On the other hand, drivers can exhibit very different driving habits (e.g. when trips are usually made) and/or behaviour (e.g. average and maximum speed of a trip) despite the same number of reported claims, not to mention an individual's driving habits and behaviour can also evolve over time.

Our other contribution lies in feature engineering.  Motivated by the aforementioned observations, we introduce a \textbf{speed transition matrix} to summarize speed evolution of each driver and extract risk-related information.  When driving speed is discretely recorded and acceleration values are unavailable, as in our dataset, existing feature engineering methods, such as the \textit{v-a} heatmap introduced by \citet{wuthrich_covariate_2017}, are inapplicable.  Notably, our speed transition matrix provides a solution to such a situation.  We illustrate how the transition matrix can be used to capture and reveal different driving patterns, while maintaining interpretability even after undergoing dimension reduction.  In claims modelling, speed transition matrix supplements average and maximum driving speeds by capturing the stability in speed transitions, e.g. how often drivers harsh accelerate from 20 km/h to 80 km/h.  While the matrix's predictive power is robust with respect to different construction set-ups (the different speed bins), we provide guidelines on how the matrix can be constructed to best suit the portfolio (e.g. in accordance to the local speed limit).

We consider a Negative Binomial regression for its interpretability and readiness for feature selection.  Since the focus of this paper is on telematics feature engineering and selection, we do not consider the multivariate model proposed in \citet{denuit_multivariate_2019} due to the complexity when a large number of telematics features in different formats (discrete, continuous, categorical, compositional) are included.  Yet, we emphasize on the importance of using claim history and telematics data observed over the same period.  This approach reduces bias and better relates the two without making prior characterization of drivers.  In contrast, using telematics data from a previous, non-overlapping period implicitly assumes driving behaviour remains unchanged and has already characterized drivers, regardless of whether their driving risk has improved or worsened.  The proposed speed transition matrix effectively captures risk information from driving speed time series.  Our result shows that large speed transitions, higher maximum speed attained, nighttime driving and increased harsh braking are associated with increased (expected) claim counts, which encourages further study and modelling of these features.

Through this work, we would like to once again draw attention to the non-directly proportional relationship between claim counts and total driving time or distance, in line with \citet{boucher_pay-as-you-drive_2013} and \citet{boucher_exposure_2017}.  While expected claim counts keep increasing as policyholders drive, we find that on average, every additional mile or hour travelled is less risky than the previous.  Consequently, for accurate ratemaking, insurance premiums should be scaled with (approximately) the square root of mileage or total driving time instead of a linear scale.  On a similar note, \citet{cheng_pay_2023} suggest a concavity relation of premium and driving distance, and illustrate that Pay-As-You-Drive (PAYD) insurance will be more efficient than fuel tax in incentivizing people to drive less from a utility maximization perspective.

Studying changes in driving behaviour is essential to understand this change in driving risk.  \citet{perez_young_2019} find evidence of reduction in speeding among young drivers that have been involved in an accident, while \citet{af_changes_2012} show that driver celeration (speed change) steadily declines despite accident involvement.  Another important contribution of this work is to discover the existence of a similar pattern between severe harsh events (i.e., those detected at a higher threshold) and total driving time or distance, where it takes increasingly longer for the next harsh event to arrive.  Perceptual learning - the long-lasting changes in perception that result from practice or experience - has been shown to improve collision detection in both older and younger adults (\citet{deloss_behavioral_2015} and \citet{lemon_training_2017}).   Our finding further completes the discussion on learning effect - the idea that as drivers spend more time on the road, they become more experienced and can react to road conditions better.  Our analysis provides empirical evidence on how learning effect is reflected in driving behaviour, and ultimately translating into reduced claim arrivals.  While we have not found a formal definition of driver learning effect in the literature, motivated by empirical observations and analysis from data, we define and quantify it as the decreasing rate of severe harsh event arrivals as cumulative driving time (or distance) increases.

The remainder of this paper is organized as follows.  Section \ref{Data-Description-and-Challenges} gives an overview of the raw telematics data, and describes some data quality issues encountered and the cleaning procedure therefore employed.  Section \ref{Exploratory-Data-Analysis-and-Feature-Engineering} provides a detailed exploratory data analysis from both a portfolio level and an individual policyholder level and describes our feature engineering process.  Then, Section \ref{NBGLM} identifies important predictors of auto-insurance claim counts via a regression analysis, reveals a non-directly proportional relationship between (expected) claim counts and total driving time or distance, and discovers a similar relationship in severe harsh event arrivals.  Finally, Section \ref{Conclusion} concludes with some future directions on the research of telematics related to actuarial science.

\section{Data Description and Challenges}
\label{Data-Description-and-Challenges}

This section provides a description of the raw data which we will work on throughout this paper. We will discuss the challenges faced and the cleaning procedure employed to arrive at a more complete data structure.  While existing works such as \citet{gao_telematics_2021}, \citet{meng_improving_2022} and \citet{gao_what_2022} focus on missing data and the reliability of acceleration values from different sources, our discussion will address more general issues such as non-chronological records, difficulties in trip detection and unreliable trips.

\subsection{Overview of Raw Data}
\label{Overview-of-Raw-Data}

Our auto-insurance dataset originates from a major European insurer with all policies written in Romania.  We have information collected from over 1,600 telematics tracking devices, with tracking period beginning as early as mid 2017 and ending as late as mid 2020.  For each device, we have the following information:

\begin{enumerate}
    \item Raw telematics records
    \begin{itemize}
        \item Recorded at each minute are the following information: date and time, GPS coordinates, GPS speed and device status.  If a harsh driving event occurs, then the aforementioned information at that particular moment will also be recorded.
        \item Harsh driving events include harsh acceleration, braking, left and right cornering which are detected at 0.5G (G-force), where 1G is 9.8 m/s$^2$.  In addition, stronger and more abrupt harsh events are detected at 1.5G and are recorded separately from the aforementioned events.
    \end{itemize}
    \item Trip lists
    \begin{itemize}
        \item For each trip, the following are recorded: the date and time when the trip starts and ends (referred to as `trip start time' and `trip end time' hereafter), distance travelled, duration, road types (in proportions), average speed and maximum speed.
    \end{itemize}
\end{enumerate}

There are two immediate challenges at this stage of data processing.  On the one hand, the data volume of raw telematics records is massive.  The exact trip start time and trip end time are unclear based solely on these records, as will be discussed in Section \ref{Data-Cleaning-Procedure-and-Challenges}.  Moreover, since information is recorded at each minute, acceleration, average speed, distance travelled, and road type (e.g. urban road/highway) are hard or impossible to derive.  On the other hand, harsh events detected during each trip are not included in the trip lists.  Hence it is insufficient to rely solely on either one of the two data sources and there is a need to combine them to get a more complete summary of each trip.

We also have policy data which include information on both the policyholders and the vehicles.  These policy records have coverage beginning as early as mid 2017, and ending as late as end of 2021.  All policies are initially written for a one-year coverage, but some may have shorter coverage due to policy cancellation.  Since a telematics tracking device may be reused due to policy renewal and/or policy cancellation, we have more policies than devices.  Detailed discussions on the variables are given in Section \ref{Exploratory-Data-Analysis-and-Feature-Engineering}.

\subsection{Data Cleaning Procedure and Challenges}
\label{Data-Cleaning-Procedure-and-Challenges}

In the existing actuarial literature, datasets studied related to telematics are often much smaller (up to several thousands drivers and/or observed for a shorter period of time (ranging from weeks to a year) (\citet{baecke_value_2017}, \citet{bian_good_2018}, \citet{jin_latent_2018}, \citet{ma_use_2018}, \citet{denuit_multivariate_2019}, \citet{gao_claims_2019}, \citet{sun_assessing_2020}, \citet{gao_boosting_2022}).  Despite the relatively small portfolio size, we spend a considerable amount of time on data cleaning, partially due to the massive volume of raw telematics data, and partially due to some unexpected issues in the data.  While some researchers have also reported the burden of data preprocessing (\citet{gao_telematics_2021}, \citet{meng_improving_2022} and \citet{gao_what_2022}), discussions on the challenges in handling telematics data are generally lacking.  Hence we would like to devote this section to discuss the data cleaning procedure employed, with emphasis on issues encountered.  We hope that this can serve as both a preview and an alert for interested researchers.

As an illustration of potential issues in raw telematics data, sample trips are shown from Tables \ref{table: sample-no-issue} to \ref{table: sample-no-end}. 
 \texttt{DeviceId} is anonymized and GPS locations are removed for privacy concerns.

\begin{itemize}
    \item \textbf{Good sample:} Table \ref{table: sample-no-issue} is a sample without any recording issues: the previous trip ends with a `key off' event; the current trip starts with a `key on' event, no interruption in between, and ends with a `key off' event; and the next trip also begins with a `key on' event.
    \item \textbf{Non-chronological records:} Although telematics observations should be made sequentially, raw telematics data may not be recorded in chronological order in reality.  As shown in Table \ref{table: sample-time-flipped}, the observations between 18:58 to 19:03 enter the system at a later time.  Re-ordering by date and time should be straight-forward in any programming languages, but the massive data volume can produce a computational burden.
    \item \textbf{Invalid GPS records leading to device calibration:} Occasionally there are invalid GPS records where GPS positions are off, such as those recorded as $(0,0)$.  There are various reasons behind this, including weather conditions (e.g. heavy precipitation), external obstructions (e.g. mountains, tall buildings and bridges), and obstruction within the vehicle (e.g. metallic tinting).  Although the telematics tracking device will calibrate itself, it still complicates the systematic identification of a trip, especially when GPS invalidity occurs at trip start.  An example is shown in Table \ref{table: sample-device-calibrate}, where the `key on' event occurs at 13:42:40 but the device calibrates shortly after.
    \item \textbf{Uncertain start and end of a trip:} While it is natural to assume a trip begins with a `key on' event and ends with a `key off' event, it is not necessarily true in practice.  As demonstrated in Table \ref{table: sample-device-calibrate}, a `key on' event does not accurately mark the beginning of a trip when the telematics tracking device calibrates its GPS positions before the vehicle starts moving.  Similarly, a trip may end with prolonged idling instead of a `key off' event, as displayed in Table \ref{table: sample-no-end}.
\end{itemize}

\begin{table}[!ht]
\centering
\caption{Sample of a Trip Without Recording Issues}
\begin{tabular}{l|llrrl} \hline
 & DeviceId & TimeStamp & \multicolumn{1}{l}{GPSDirection} & \multicolumn{1}{l}{GPSSpeed} & EventDescription \\ \hline\hline
last trip & 2022123 & 07/16/2018 11:51:48 & 0 & 0 & KEY OFF \\ \hline
current trip & 2022123 & 07/16/2018 11:55:28 & 0 & 2 & KEY ON \\
 & 2022123 & 07/16/2018 11:56:28 & 0 & 2 & POSITION IN TIME \\
 & 2022123 & 07/16/2018 11:57:28 & 132 & 50 & POSITION IN TIME \\
 & 2022123 & 07/16/2018 11:58:28 & 144 & 60 & POSITION IN TIME \\
 & 2022123 & 07/16/2018 11:59:28 & 134 & 16 & POSITION IN TIME \\
 & 2022123 & 07/16/2018 12:00:28 & 32 & 21 & POSITION IN TIME \\
 & 2022123 & 07/16/2018 12:01:28 & 28 & 39 & POSITION IN TIME \\
 & 2022123 & 07/16/2018 12:02:28 & 38 & 35 & POSITION IN TIME \\
 & 2022123 & 07/16/2018 12:03:28 & 42 & 29 & POSITION IN TIME \\
 & 2022123 & 07/16/2018 12:04:28 & 38 & 11 & POSITION IN TIME \\
 & 2022123 & 07/16/2018 12:05:28 & 56 & 14 & POSITION IN TIME \\
 & 2022123 & 07/16/2018 12:06:28 & 0 & 4 & POSITION IN TIME \\
 & 2022123 & 07/16/2018 12:07:28 & 0 & 3 & POSITION IN TIME \\
 & 2022123 & 07/16/2018 12:08:28 & 0 & 1 & POSITION IN TIME \\
 & 2022123 & 07/16/2018 12:09:28 & 0 & 2 & POSITION IN TIME \\
 & 2022123 & 07/16/2018 12:10:28 & 0 & 2 & POSITION IN TIME \\
 & 2022123 & 07/16/2018 12:10:45 & 0 & 1 & KEY OFF \\ \hline
next trip & 2022123 & 07/17/2018 13:27:16 & 0 & 1 & KEY ON
\end{tabular}
\label{table: sample-no-issue}
\end{table}

\begin{table}[!ht]
\centering
\caption{Sample of a Trip Not in Chronological Order}
\begin{tabular}{l|llrrl} \hline
 & DeviceId & TimeStamp & \multicolumn{1}{l}{GPSDirection} & \multicolumn{1}{l}{GPSSpeed} & EventDescription \\ \hline\hline
 & 2022123 & 07/19/2018 18:56:21 & 296 & 38 & POSITION IN TIME \\
 & 2022123 & 07/19/2018 18:57:21 & 276 & 30 & POSITION IN TIME \\
 & 2022123 & 07/19/2018 19:04:21 & 254 & 85 & POSITION IN TIME \\
 & 2022123 & 07/19/2018 19:05:21 & 258 & 79 & POSITION IN TIME \\
 & 2022123 & 07/19/2018 19:06:21 & 258 & 79 & POSITION IN TIME \\
 & 2022123 & 07/19/2018 19:07:21 & 254 & 58 & POSITION IN TIME \\
 & 2022123 & 07/19/2018 19:08:21 & 236 & 64 & POSITION IN TIME \\ \hline
time ordering & 2022123 & \textbf{\textit{07/19/2018 18:58:21}} & 270 & 38 & POSITION IN TIME \\
 & 2022123 & 07/19/2018 18:59:21 & 246 & 42 & POSITION IN TIME \\
 & 2022123 & 07/19/2018 19:00:21 & 246 & 61 & POSITION IN TIME \\
 & 2022123 & 07/19/2018 19:01:21 & 246 & 79 & POSITION IN TIME \\
 & 2022123 & 07/19/2018 19:02:21 & 236 & 49 & POSITION IN TIME \\
 & 2022123 & 07/19/2018 19:03:21 & 252 & 81 & POSITION IN TIME \\ \hline
 & 2022123 & 07/19/2018 19:09:21 & 272 & 49 & POSITION IN TIME
\end{tabular}
\label{table: sample-time-flipped}
\end{table}

\begin{table}[!ht]
\centering
\caption{Sample of a Trip With Device Calibration}
\begin{tabular}{l|llrrl} \hline
 & DeviceId & TimeStamp & \multicolumn{1}{l}{GPSDirection} & \multicolumn{1}{l}{GPSSpeed} & EventDescription \\ \hline\hline
 & 2022123 & 04/02/2019 18:09:33 & 0   & 0   & KEY ON           \\
device calibration & 2022123 & 04/02/2019 18:09:37 & 0   & 0   & \textbf{\textit{FIX GPS OK}} \\
 & 2022123 & 04/02/2019 18:10:34 & 266 & 73  & POSITION IN TIME \\
 & 2022123 & 04/02/2019 18:11:34 & 264 & 85  & POSITION IN TIME \\
 & 2022123 & 04/02/2019 18:12:34 & 326 & 95  & POSITION IN TIME \\
 & 2022123 & 04/02/2019 18:13:34 & 326 & 94  & POSITION IN TIME \\
 & 2022123 & 04/02/2019 18:14:34 & 328 & 109 & POSITION IN TIME \\
 & 2022123 & 04/02/2019 18:15:34 & 308 & 98  & POSITION IN TIME \\
 & 2022123 & 04/02/2019 18:16:34 & 358 & 88  & POSITION IN TIME \\
 & 2022123 & 04/02/2019 18:17:34 & 8   & 92  & POSITION IN TIME \\
 & 2022123 & 04/02/2019 18:18:34 & 342 & 75  & POSITION IN TIME \\
 & 2022123 & 04/02/2019 18:19:34 & 0   & 1   & POSITION IN TIME \\
 & 2022123 & 04/02/2019 18:20:19 & 0   & 0   & KEY OFF \\ \hline
\end{tabular}
\label{table: sample-device-calibrate}
\end{table}

\begin{table}[!ht]
\centering
\caption{Sample of a Trip With No `Key Off' Event}
\begin{tabular}{l|llrrl} \hline
 & DeviceId & TimeStamp & \multicolumn{1}{l}{GPSDirection} & \multicolumn{1}{l}{GPSSpeed} & EventDescription \\ \hline\hline
last trip & 2022123 & 07/26/2018 09:36:32 & 170 & 22 & \textbf{\textit{POSITION IN TIME}} \\ \hline
current trip & 2022123 & 07/26/2018 09:53:31 & 0 & 1 & KEY ON \\
 & 2022123 & 07/26/2018 09:54:30 & 26 & 20 & POSITION IN TIME \\
 & 2022123 & 07/26/2018 09:55:30 & 32 & 34 & POSITION IN TIME \\
 & 2022123 & 07/26/2018 09:56:30 & 122 & 13 & POSITION IN TIME \\
 & 2022123 & 07/26/2018 09:57:30 & 0 & 1 & POSITION IN TIME \\
 & 2022123 & 07/26/2018 09:58:30 & 0 & 1 & POSITION IN TIME \\
 & 2022123 & 07/26/2018 09:59:30 & 0 & 1 & POSITION IN TIME \\
 & 2022123 & 07/26/2018 09:59:56 & 0 & 1 & KEY OFF \\ \hline
\end{tabular}
\label{table: sample-no-end}
\end{table}

Detected trips do not necessarily match the trip lists in Section \ref{Overview-of-Raw-Data} due to the last issue above, hence we begin with the trip list of each device instead.  For each device and for each trip on its list, we find the indices corresponding to the start and end of each trip in the raw telematics data, and summarize the numbers of detected harsh events and the maximum speed in between.  Ideally, we should be able to find the exact time-points of both the start and end; however, this can occasionally fail due to unforeseen reasons such as system error and delay, and the time-points from both sources can differ from a second to several days.  Hence, we search for the timestamp with the minimal absolute time difference.

\textit{Remark: We do recognize the fact that the trip lists which contain duration, distance, etc. do not come without effort - either the insurer which provided us the data have additional information from the telematics tracking device and/or the drivers, or they have put in effort to produce the lists.  In fact, how to reasonably identify a trip, such that it is neither terminated if the vehicle just stops for a while (e.g. to buy gas), nor it goes on forever even if the vehicle has already parked but a `key off' event is not registered, is a practical problem faced by the telematics industry.  Possible solutions include filtering for invalid GPS points and manually defining a threshold for idling time, say three minutes.}

For each device, we now have a list of trips with information including start and end date and time, distance travelled, duration, average and maximum speed, road types travelled (in proportions), and the numbers of detected harsh events.  However there are still some problematic observations, e.g. trips lasting only 3 seconds, with average speed of 1,560 km/h, and maximum speed of 0 km/h, etc.  To improve data quality for reasonable analysis, we filter the trips using the following criteria: duration of at least 3 minutes, average speed between 5 and 150 km/h, maximum speed of at least 10 km/h, and time difference between two sources within a minute.  We expect the filtered trips to be able to show at least some driving behaviour.

For each policy, we then extract the trips that are within the insurance coverage period, with attention paid on whether the policy is cancelled early.  We summarize the following on a policy level: start date of the first and last recorded trips, total number of trips, driving time and distance, numbers of detected harsh events, mean of the average speed of each trip, weighted by trip duration, and maximum speed attained.  Some literatures have suggested that three months of telematics data are sufficient to attain stable characterization of driving habits and behaviour.  \citet{baecke_value_2017} use telematics information from the most recent three months to model claim probability, hence they implicitly assume that driving habits and behaviour will remain unchanged for the next period.  \citet{duval_how_2022} model (annualized) claim probability using information from first three months of a year, but the use of average and fractions can undermine trends and variations.  In contrast, we filter policies such that the total deviation between telematics observation and insurance coverage periods is no more than three months, as to ensure not to miss a considerable amount of characterization information.  This leaves us with 1,458 policies and a total of 1.83 million trips.

\section{Exploratory Data Analysis and Feature Engineering}
\label{Exploratory-Data-Analysis-and-Feature-Engineering}

This section provides an exploratory data analysis and outlines the feature engineering process employed.  We also demonstrate the heterogeneity in policyholders' driving behaviour through an analysis of telematics data, from both a portfolio level and an individual policy level.

\subsection{Description of Cleaned Data}
\label{Description-of-Cleaned-Data}

The cleaned dataset has 1,458 policies from 1,073 unique policyholders.   Among these policyholders, 717 (66.8\%), 327 (30.5\%) and 29 (2.7\%) are observed for 1, 2 and 3 insurance periods (mostly one year, but can be shorter due to policy cancellation).  44 (3\%), 395 (27.1\%), 969 (66.5\%) and 50 (3.4\%) policies begin in years 2017, 2018, 2019 and 2020 respectively.  Table \ref{table: covariates-traditional} shows the claim counts and a list of traditional covariates, which are information on the policy, the policyholder and the vehicle; whilst Table \ref{table: covariates-telematics} shows a list of covariates derived from telematics data.  Table \ref{table: summary-stats} provides the summary statistics of numerical variables in the two tables.

There are limitations in some of the covariates, imposing the need of excluding them.  First, we see that \texttt{age} has 37\% non-numerical/missing values.  The reason is that these policies are written under a company entity and hence the policyholder's age is unreported.  Based solely on policyholders whose age data are available, we do not see a significant difference in policyholder's age with and without claims: the medians are 40.00 and 42.00, while the means are 42.44 and 43.86 respectively.  The mean claim rate in this group is 0.4651, which is close to the portfolio mean of 0.4465 (the claim counts are motor third-party liability and motor damage combined; more details are given in Section \ref{Motivation-for-Using-the-Negative-Binomial}).  It is worth mentioning that while some telematics datasets studied in actuarial science have focused on younger policyholders and hence an arguably more homogeneous group (\citet{boucher_exposure_2017}, \citet{ayuso_improving_2019}, \citet{denuit_multivariate_2019}, \citet{henckaerts_added_2022}), there is not such indication in our dataset.  Second, \texttt{engine\_cc}, \texttt{engine\_power} and \texttt{mileage} have unreliable minimum values.  While a zero engine capacity is clearly questionable, the usual range of engine power is above 100 horsepower, which is equivalent to 74 kilowatts.  Moreover, a zero \texttt{mileage} is also untrustable which suggests the need of mileage records from telematics instead of self-reported.  Finally, categorical variables such as \texttt{county} and \texttt{make} have many levels, while the latter may also have duplicates due to an unstandardized list of vehicle brands, hence we will exclude them and replace with other variables.

From a modelling perspective, we will further exclude the following: first, \texttt{avg\_distance} and \texttt{avg\_time} as to not average out trip characteristics beforehand; second, \texttt{num\_trips} due to the correlation between \texttt{total\_distance} or \texttt{total\_time} and \texttt{avg\_speed}; and finally, \texttt{prop\_other\_road} since it is compositional with \texttt{prop\_urban}, \texttt{prop\_extra\_urban} and \texttt{prop\_highway} (i.e., they sum to 100), and values of \texttt{prop\_other\_road} are relatively small (as reflected by the interquartile range at lower values), hence we will consider only the other three roadtypes.  For convenience, these proportions will be considered together and referred to as \texttt{prop\_roadtype} hereafter.

While most of the telematics covariates are extracted and/or summarized directly from the cleaned data in Section \ref{Data-Description-and-Challenges}, we introduce in Section \ref{Feature-Engineering-on-Telematics-Data} two additional sets of variables: principal components (PCs) of speed transition matrix and proportions of driving in different times of the day.

\begin{landscape}
\begin{table}[p]
\caption{Data Fields in the Cleaned Dataset - Traditional}
\centering
\begin{tabular}{l l c c}
\hline
\textbf{Name} & \textbf{Description} & \textbf{Range} & \textbf{Notes or Issues (if any)} \\ \hline\hline
\texttt{map\_id} & Unique identifier of policy & --- & 1458 policies\\
\texttt{unique\_id} & Unique identifier of policyholder & --- & 1073 policyholders \\
\hline
\texttt{claims} & Number of claims filed by the policy & [0, 6] & \begin{tabular}[c]{@{}c@{}} Motor third-party liability and motor damage combined; \\ 72\% observations with no claims \end{tabular} \\
\hline
\texttt{age} & Age of the policyholder & --- & 37\% observations have non-numerical values  \\
\texttt{county} & County of residence of the policyholder & Factor with 42 levels & \begin{tabular}[c]{@{}c@{}} A geographic region used for administrative purposes; \\ Dimension reduction is necessary \end{tabular} \\
\texttt{region1} & Region of residence of the policyholder & Factor with 9 levels & Alternative to \texttt{county} \\
\texttt{region2} & Region of residence of the policyholder & factor with 5 levels & Alternative to \texttt{county} \\
\hline
\texttt{car\_value} & Value of a new vehicle in insurer's domestic currency & [39172, 1368041] &  \\
\texttt{engine\_cc} & Vehicle's engine capacity in cubic centimeters & [0, 6592] & Minimum value of 0 is unreliable \\
\texttt{make} & Brand of the vehicle & Factor with 41 levels & Dimension reduction is necessary, duplicates may exist \\
\texttt{max\_weight} & Maximum weight of the vehicle in kilograms & [760, 3500] &  \\
\texttt{num\_seats} & Number of seats in the vehicle & [2, 9] &  \\
\texttt{engine\_power} & Engine power of the vehicle in kilowatts & [11, 467] & Minimum value of 11 is unreliable \\
\hline
\texttt{year} & Year in which insurance coverage begins & 2017, 2018, 2019, 2020 & \\
\texttt{insurance\_start} & Insurance coverage begin date & --- &  \\
\texttt{insurance\_end} & Insurance coverage end date (if not cancelled) & --- &  \\
\texttt{cancel\_date} & Cancellation date of the policy (if cancelled) & --- & \\
\texttt{policy\_period} & Duration of insurance coverage in years & [0.1236, 1.0027] & Corrected for policy cancellation \\
\texttt{mileage} & Self-reported mileage of the policy & [0, 526367] & Minimum of 0 is unreliable \\
\texttt{main\_cover} & Main coverage of the policy & Factor with 4 levels & Groups are highly imbalanced \\
\texttt{renewal} & Whether the policy is a new business or renewal & Factor with 2 levels & 74\% observations are new businesses \\
\texttt{usage} & Whether the vehicle is for personal or other uses & Factor with 2 levels & 69\% observations are for personal use \\ 
\texttt{vehicle\_age} & Age of the vehicle & [0, 11] & \\
\hline
\texttt{first\_trip} & Date of the first observed trip & --- &  \\
\texttt{last\_trip} & Date of the last observed trip & --- &  \\
\texttt{start\_diff} & \begin{tabular}[c]{@{}l@{}} Difference between start dates of telematics observation \\ and insurance coverage in days\end{tabular}
 & [0, 90] &  \\
\texttt{end\_diff} & \begin{tabular}[c]{@{}l@{}} Difference between end dates of telematics observation \\ and insurance coverage in days\end{tabular} & [0, 90] &  \\
\hline 
\end{tabular}
\label{table: covariates-traditional} 
\end{table}
\end{landscape}

\begin{landscape}
\begin{table}[p]
\caption{Data Fields in the Cleaned Dataset - Telematics}
\centering
\begin{tabular}{l l c c}
\hline
\textbf{Name} & \textbf{Description} & \textbf{Range} & \textbf{Notes or Issues (if any)} \\ 
\hline\hline
\texttt{avg\_distance} & Average distance travelled in each trip in kilometers & [1.64, 118.26] &  \\
\texttt{avg\_time} & Average duration of each trip in minutes & [6.46, 93.26] &  \\
\texttt{total\_distance} & Total distance travelled in all trips in kilometers & [237.8, 159997.7] &  \\
\texttt{total\_time} & Total driving time in all trips in minutes & [527.5, 145810.9] &  \\
\texttt{prop\_extra\_urban} & Percentage driven in extra urban areas & [0.03, 66.68] &  \\
\texttt{prop\_highway} & Percentage driven on highways & [0, 42.49] &  \\
\texttt{prop\_other\_road} & Percentage driven on other roads & [0, 16.53] &  \\
\texttt{prop\_urban} & Percentage driven in urban areas & [26.23, 99.97] &  \\
\hline
\texttt{avg\_speed} & Trip-duration-weighted mean of average speed of each trip in kilometers per hour & [26.23, 99.97] &  \\
\texttt{max\_speed} & Maximum speed attained throughout the policy in kilometers per hour & [57, 255] &  \\
\texttt{num\_acc} & Total number of harsh acceleration & [0, 4234] &  \\
\texttt{num\_brake} & Total number of harsh braking & [0, 1018] &  \\
\texttt{num\_left} & Total number of harsh left cornering & [0, 2756] &  \\
\texttt{num\_right} & Total number of harsh right cornering & [0, 2039] &  \\
\texttt{num\_severe} & Total number of severe harsh events & [0, 348] & Detected at 1.5 G-force \\
\texttt{num\_trips} & Total number of trips driven & [42, 11610] &  \\
\hline 
\texttt{pc1} & First principle component of the speed transition matrix & [-168.09, -41.71] & \\
\texttt{pc2} & Second principle component of the speed transition matrix & [11.52, 49.01] & \\
\hline
\texttt{prop\_0\_4} & Percentage driven between midnight to 3:59 a.m. & [0, 30.91] & \\
\texttt{prop\_4\_8} & Percentage driven between 4 to 7:59 a.m. & [0, 64.40] & \\
\texttt{prop\_8\_12} & Percentage driven between 8 to 11:59 a.m. & [1.29, 78.59] & \\
\texttt{prop\_12\_16} & Percentage driven between noon to 3:59 p.m. & [3.13, 73.82] & \\
\texttt{prop\_16\_20} & Percentage driven between 4 to 7:59 p.m. & [0.43, 49.38] & \\
\texttt{prop\_20\_24} & Percentage driven between 8 to 11:59 p.m. & [0, 39.67] & \\\hline
\end{tabular}
\label{table: covariates-telematics} 
\end{table}
\end{landscape}
\clearpage

\begin{table}[h]
\caption{Summary Statistics of Numerical Data Fields}
\centering
\begin{tabular}{l r r r r r r} 
\hline
\textbf{Name} & \textbf{Min} & \textbf{1st Qu.} & \textbf{Median} & \textbf{Mean} & \textbf{3rd Qu.} & \textbf{Max} \\ \hline\hline
\texttt{claims} & 0 & 0 & 0 & 0.4465 & 1 & 6 \\
\hline
\texttt{car\_value} & 39172 & 97558 & 134984 & 160472 & 190362 & 1368041 \\
\texttt{engine\_cc} & 0 & 1498 & 1968 & 1844 & 1997 & 6592 \\
\texttt{max\_weight} & 760 & 1840 & 2030 & 2158 & 2340 & 3500 \\
\texttt{num\_seats} & 2 & 5 & 5 & 4.95 & 5 & 9 \\
\texttt{engine\_power} & 11 & 88 & 108 & 115.7 & 133 & 467 \\ \hline
\texttt{policy\_period} & 0.1236 & 1.0000 & 1.0027 & 0.9350 & 1.0027 & 1.0027 \\
\texttt{mileage} & 0 & 900.5 & 28008 & 62148.4 & 107424.8 & 526367 \\
\texttt{vehicle\_age} & 0 & 0 & 1 & 2.29 & 4 & 11 \\ \hline
\texttt{start\_diff} & 0 & 0 & 3 & 5.93 & 7 & 90 \\
\texttt{end\_diff} & 0 & 0 & 1 & 12.36 & 11 & 90 \\ \hline
\texttt{avg\_distance} & 1.640 & 7.386 & 10.286 & 13.416 & 14.845 & 118.264 \\
\texttt{avg\_time} & 6.463 & 14.723 & 17.892 & 20.072 & 22.229 & 93.257 \\
\texttt{total\_distance} & 237.8 & 7407.2 & 11327.9 & 14687.3 & 17226.2 & 159997.7 \\
\texttt{total\_time} & 527.5 & 13112.8 & 19937.2 & 23237.0 & 28965.9 & 145810.9 \\
\texttt{prop\_extra\_urban} & 0.117 & 19.450 & 27.833 & 28.151 & 36.115 & 75.272 \\
\texttt{prop\_highway} & 0 & 1.565 & 8.507 & 13.857 & 21.411 & 75.877 \\
\texttt{prop\_other\_road} & 0 & 0.006 & 0.081 & 0.356 & 0.341 & 15.049 \\
\texttt{prop\_urban} & 7.214 & 46.35 & 57.928 & 57.635 & 68.994 & 99.883 \\ \hline
\texttt{avg\_speed} & 11.10 & 27.97 & 35.13 & 36.25 & 42.98 & 82.68 \\
\texttt{max\_speed} & 57 & 143 & 159 & 160.5 & 176 & 255 \\
\texttt{num\_acc} & 0 & 2 & 12 & 67.72 & 50 & 4234 \\
\texttt{num\_brake} & 0 & 15 & 31 & 49.57 & 59 & 1018 \\
\texttt{num\_left} & 0 & 1.25 & 8 & 36.26 & 27 & 2756 \\
\texttt{num\_right} & 0 & 1 & 6 & 23.94 & 21 & 2039 \\
\texttt{num\_severe} & 0 & 0 & 1 & 3.02 & 2 & 348 \\
\texttt{num\_trips} & 42 & 709 & 1150 & 1254 & 1595 & 11610 \\
\hline
\texttt{pc1} & -12.50 & -3.77 & -0.86 & 0 & 2.95 & 32.36 \\
\texttt{pc2} & -11.02 & -3.06 & -0.01 & 0 & 2.89 & 13.09 \\ \hline
\texttt{prop\_0\_4} & 0 & 0.084 & 0.477 & 1.303 & 1.384 & 30.906 \\
\texttt{prop\_4\_8} & 0 & 4.026 & 7.939 & 9.864 & 13.411 & 64.405 \\
\texttt{prop\_8\_12} & 1.289 & 20.029 & 25.166 & 26.046 & 30.987 & 78.592 \\
\texttt{prop\_12\_16} & 3.129 & 24.347 & 29.757 & 29.647 & 34.495 & 73.824 \\
\texttt{prop\_16\_20} & 0.425 & 21.318 & 26.329 & 26.182 & 31.708 & 49.380 \\
\texttt{prop\_20\_24} & 0 & 2.539 & 5.323 & 6.958 & 9.939 & 39.673 \\\hline
\end{tabular}
\label{table: summary-stats} 
\end{table}

\subsection{Feature Engineering on Telematics Data and the Speed Transition Matrix}
\label{Feature-Engineering-on-Telematics-Data}

First, we aim to extract more information from GPS speed, especially the time series structure in speed, as to supplement \texttt{avg\_speed} and \texttt{max\_speed}.  Our data structure does not permit the use of the \textit{v-a} heatmap proposed by \citet{wuthrich_covariate_2017}: we neither have acceleration data nor able to compute acceleration values because speed is recorded per minute in the raw GPS data, while acceleration is usually measured as the change in speed per second, hence it is impossible for us to produce the same \textit{v-a} heatmap.  Although the unavailability of per-second speed data may seem to be a limitation that exists only in our dataset, this is actually common in practice to reduce the burden of data transmission and storage.  Hence it is necessary to build general frameworks to process, analyze and model telematics data, such as speed, within the constraints of data availability (e.g. only minute-level data). 

We produce instead a speed transition matrix for each policy, where the rows indicate the current speed bin, and the columns specify the speed bin at the next minute, and each element in the matrix is the empirical probability of changing from one speed bin to another.  As one has to end up in some speed bin at the next minute (before the trip terminates), regardless of the current speed bin, each row of the matrix must sum to 1.  The bins are: [0, 0.5), [0.5, 10), [10, 20), [20, 30), ..., [120, 130), [130, Inf).  In particular, the first bin aims to capture the stationary state, the last bin is chosen according to the speed limit in the country where policies are written, and the intermediate bins are chosen to strike a balance between information granularity and dimensionality.  While we have chosen a constant bin width of 10 km/h based on expert judgement, we discuss in Section \ref{Robustness-of-Speed-Transition-Matrix-wrt-h} what bin width is suitable for the data, and show that the predictive power of the proposed matrix is in fact fairly robust to different bin widths.  \citet{pyrkov_extracting_2018} implement a similar transition matrix approach to describe human's locomotor activity.

For each policy, we consider the entire driving period (i.e., all trips).  For example, to calculate the (empirical) probability of changing from [0, 0.5) to [10, 20), we count the number of times the vehicle begins stationary and ends up between 10 to 20 km/h at the next minute, and then divide that number by the total number of times the vehicle is at rest (GPS speed recorded as 0).  A limitation here is that we only have and hence can only rely on information at the start and end of a minute.  For example, if the speeds at the start and end of a minute are 10 and 12, respectively, it is possible that the speed increased (say to 20) then decreased to 12 within that minute.  However, one can rest assured that any speed changes within a minute are mild and smooth, as harsh events would otherwise be detected and recorded.  Recall that, in addition to the observations made at minute intervals, any harsh events that occur will be promptly recorded at that specific moment.  To account for this, we assume harsh events are uniformly distributed over the minute interval and scale the number of transitions by the inverse of the time since the last observation.  Without loss of generality, consider the case that after the first observation at time $t_0 = 0$, there is a harsh event at time $t_1 = 20$ seconds, and the next observation is at the usual minute interval $t_2 = 60$ seconds.  We will scale the transition from $t_0$ to $t_1$ as $60/20 = 3$ instead of 1, and scale the transition from $t_1$ to $t_2$ as $60/40 = 1.5$.  Since the matrix is constructed on a minute basis, the number of transitions between two consecutive observations recorded at time $t_n$ and $t_{n+1}$ is scaled by $60/(t_{n+1} - t_{n})$.

\begin{figure*}[h]
\caption{First Two Principal Components (PCs) of Speed Transition Matrix with Fewer Speed Bins.  Patterns are consistent with the matrix with more speed bins and used for the regression analysis.}
    \centering
    \begin{subfigure}[b]{0.47\textwidth}
        \centering
        \includegraphics[width=\textwidth]{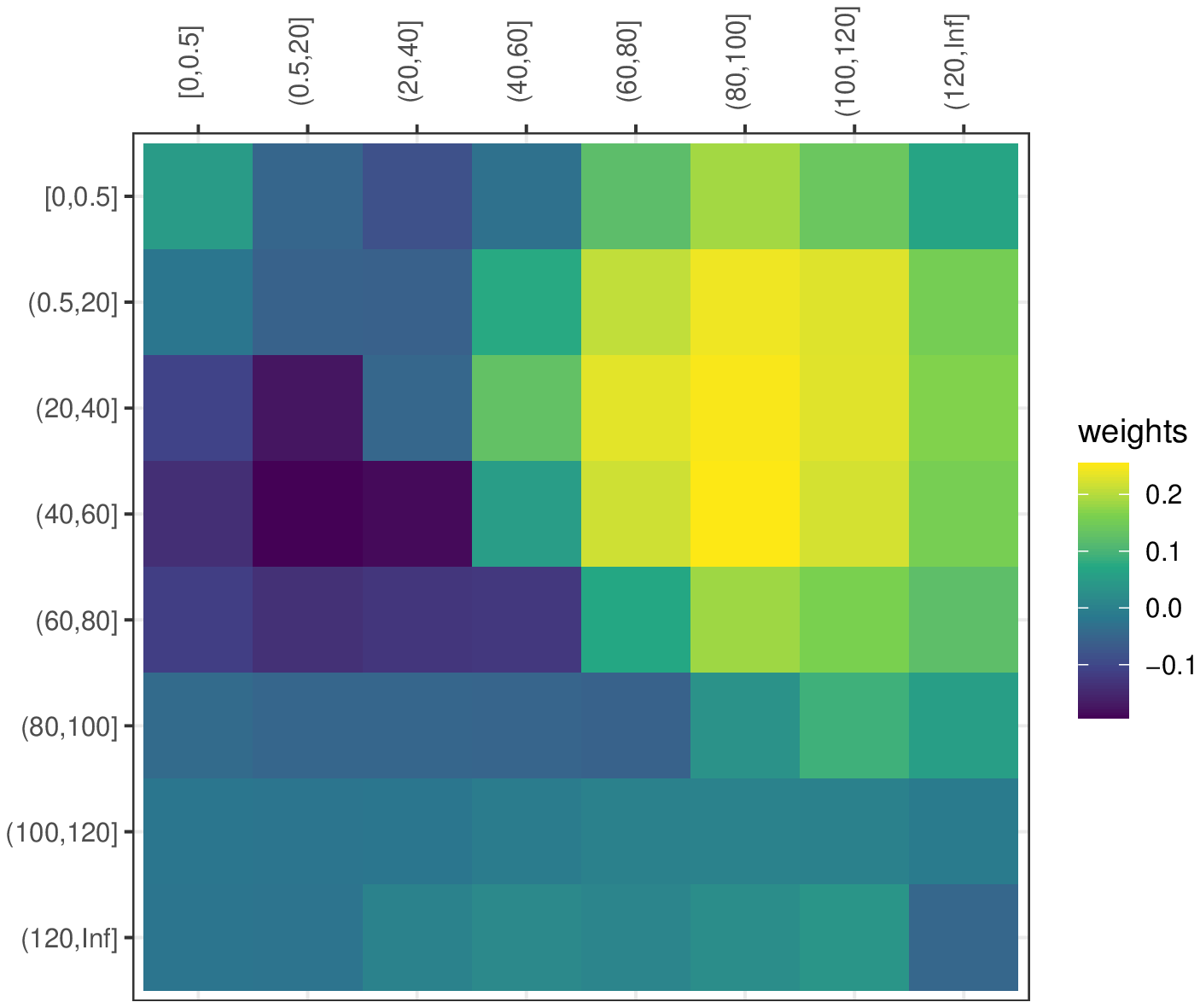}
        \caption{Loadings of PC1}
    \end{subfigure}
    \hfill
    \begin{subfigure}[b]{0.47\textwidth}  
        \centering 
        \includegraphics[width=\textwidth]{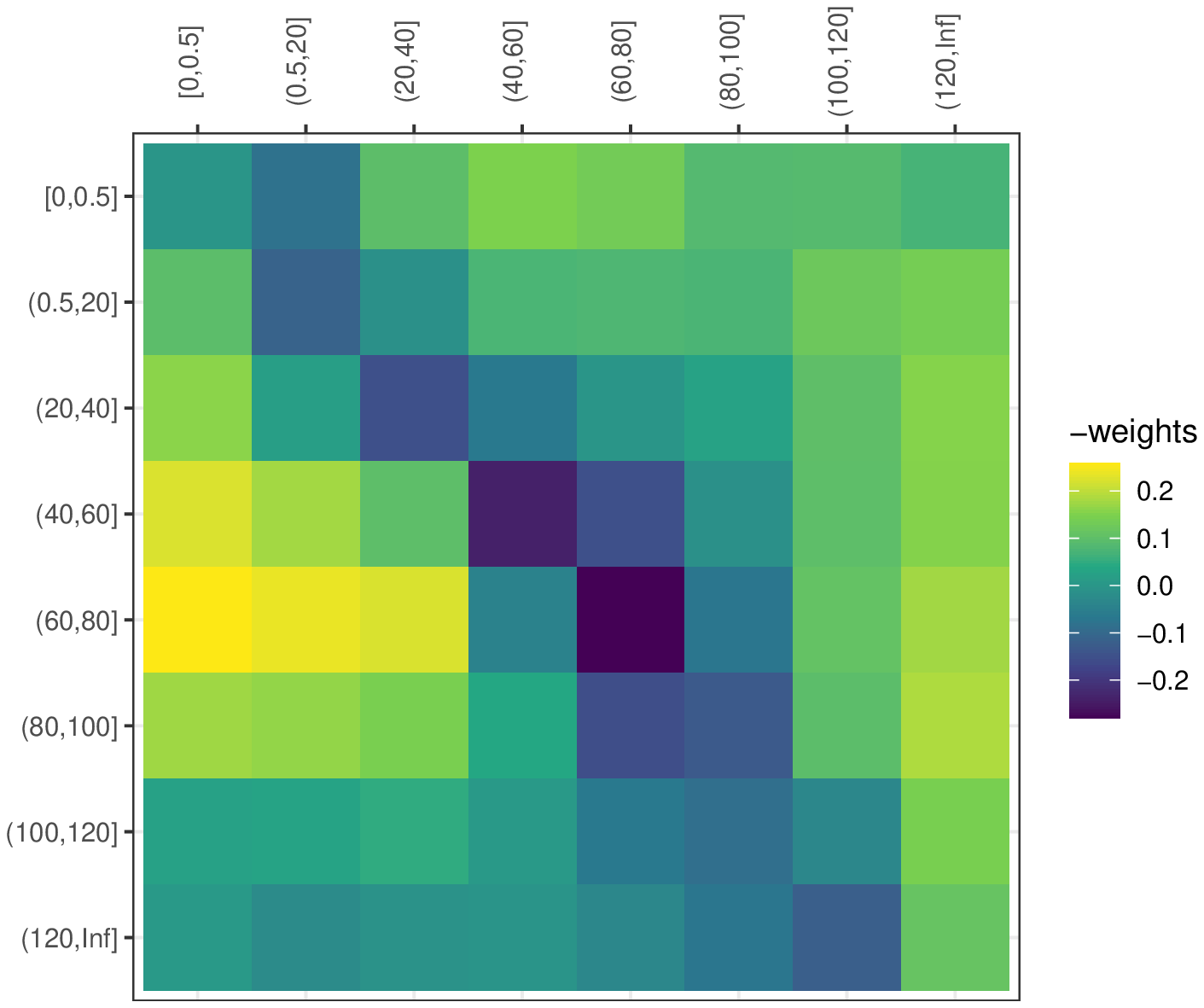}
        \caption{Loadings of PC2}
    \end{subfigure}
    \label{fig: pc-weights-fewer-bins}
\end{figure*}

This construction can also be a difference between our speed transition matrix and the \textit{v-a} heatmap, as our matrix considers all speeds while the latter is often truncated, e.g. \cite{gao_claims_2019} only consider [5, 20] km/h and \cite{gao_boosting_2022} consider (0, 80] km/h.  As will be shown in Section \ref{Individual-Heterogeneity}, the proposed transition matrix is able to capture a variety of patterns in individual driving behaviour.  In order to include this 15-by-15 matrix in regression models, we employ principal component analysis (PCA) to produce PCs that can be used as covariates.  To limit the number of covariates, as well as to see whether the PCs will actually supplement or substitute the commonly considered telematics features \texttt{avg\_speed} and \texttt{max\_speed}, we include only the first two PCs in our subsequent analysis.  While the use of PCA will unavoidably reduce the interpretability of the original features, we attempt to understand the general patterns captured by the two PCs.  Figure \ref{fig: pc-weights} visualizes the loadings of \texttt{pc1} and \texttt{pc2} via a heatmap, with yellow representing positive weights and dark blue representing negative weights.  As expected, the two PCs focus on different areas of the speed transition matrix: \texttt{pc1} emphasizes on accelerations, giving positive weights to the upper triangle, which are transitions to the higher speed bins; whilst \texttt{pc2} prioritizes decelerations and stationarity, giving positive weights to the few transitions from moderate speed bins to lower speed bins, and negative weights along the diagonal, which are the few transitions within the moderate speed bins.  We also provide the first two PC's loadings from a matrix with fewer speed bins (doubled the bin width) in Figure \ref{fig: pc-weights-fewer-bins}, which demonstrate that the patterns captured are consistent.

Moreover, we consider information on when the policyholder usually drives.  On the one hand, this is motivated by previous works that include this information and reveal its relevance to riskiness, such as driving during rush hours and/or at nighttime is more dangerous (\citet{jin_latent_2018}, \citet{ayuso_improving_2019}, \citet{duval_how_2022}).  On the other hand, we do observe from our data some differences in driving behaviour at different time of the day, as shown in Table \ref{table: driving-behaviour-timeslot}.  After allocating driving time of each trip to the different hours, we aggregate all trips for each policy and arrive at proportions of driving in each of the 24 hours.  To reduce the number of dimensions, we consider timeslots consisting of four hours each: midnight to 3:59 a.m., 4 to 7:59 a.m., 8 to 11:59 a.m., noon to 3:59 p.m., 4 to 7:59 p.m., and 8 to 11:59 p.m.  Since the proportions of driving in different timeslots are again compositional with one another, we exclude the last group \texttt{prop\_20\_24} without loss of generality.  For convenience, all these proportions will be considered together and referred to as \texttt{prop\_time} in subsequent discussion.

\subsection{Portfolio Level Distribution} 
\label{Portfolio-Level-Distribution} 

We first examine the GPS locations to understand the areas travelled by the drivers.  While all policies are written in Romania (referred to as the `home country' hereafter), trips are made all over Europe, thanks to the ease of travel granted by the European Union.  We report in Tables \ref{table: sort-travel-freq-with-claim} and \ref{table: sort-travel-freq-no-claim} the numbers of GPS observations and harsh events in the six most frequently travelled countries (anonymized for privacy reasons), for policies with and without claims respectively.  We see that 6.6\% and 9.4\% of the GPS records from the two groups (hence 8.5\% from the whole portfolio) are in fact made outside of the home country.

\begin{figure*}[h]
\caption{Grouped Boxplots of Proportion Driven in the Self-reported and Most-driven (indicated as `Max') Regions of Residence.  Red: With Claims, Blue: Without Claims.}
    \centering
    \includegraphics[width=\textwidth]{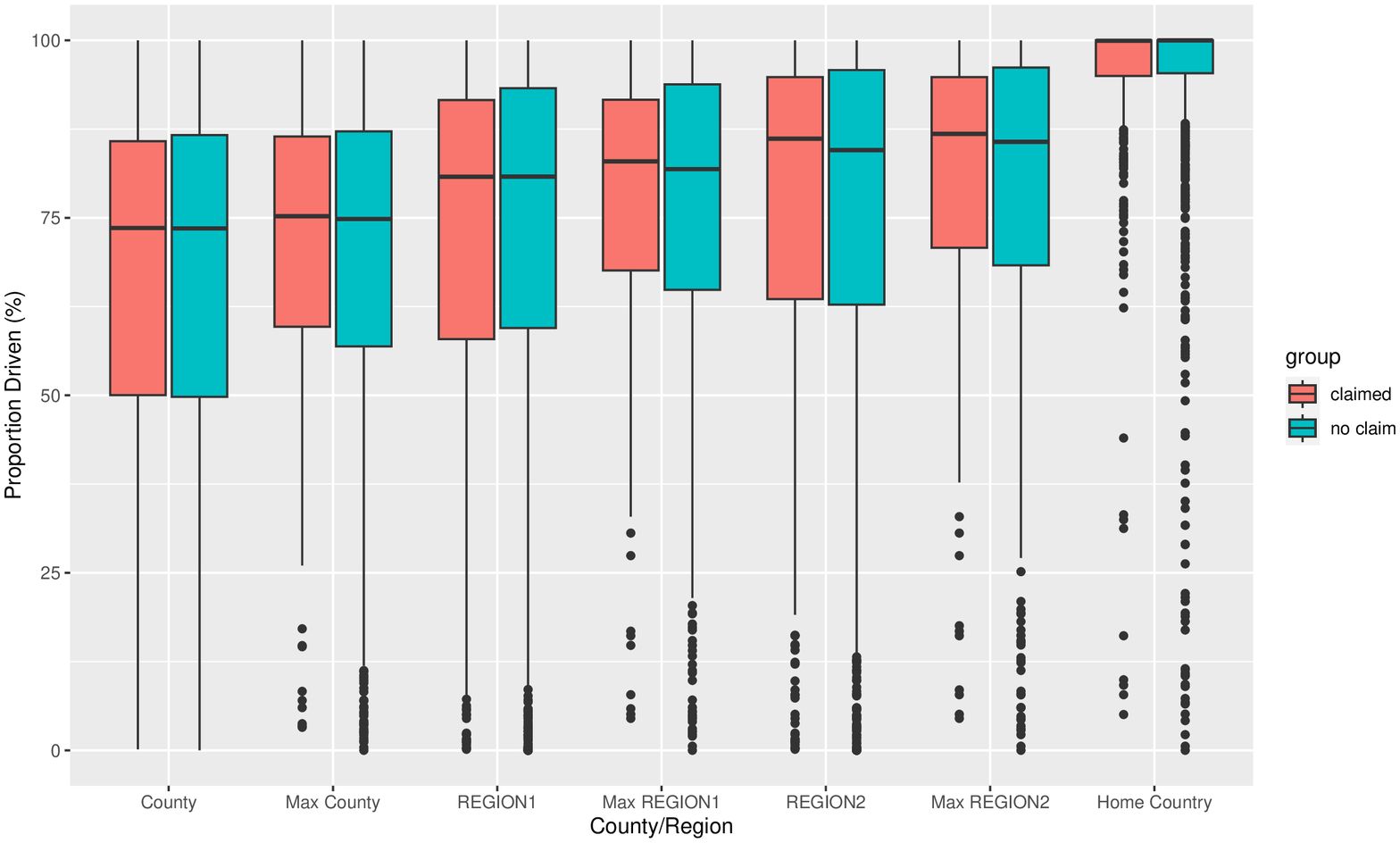}
    \label{fig: reported-vs-true-boxplots}
\end{figure*}

Furthermore, we explore how representative the policyholders' self-reported county and region of residence (within the home country) are for claims modelling, by computing the proportions of their driving time in these areas.  \texttt{County} is a geographic region in the home country for administrative or other purposes, while \texttt{region1} and \texttt{region2} are aggregates of county levels based on the NUTS classification (Nomenclature of territorial units for statistics).  In Figure \ref{fig: reported-vs-true-boxplots} we present the boxplots illustrating the driving time proportions in the self-reported areas and the most-driven areas (indicated as `Max'), and several observations can be made.  First, as the number of regional levels reduces, the proportion of driving time in self-reported areas increases, which aligns with expectations.  Second, the differences between the proportions of self-reported and most-driven areas reflect the fact that drivers do not necessarily drive the most in their self-reported regions of residence.  Third, the long tails of the boxplots, especially those for the home country, reveal that drivers allocate varying amounts of time in the home country, and in some cases, this proportion can be as low as close to 0\%.  This variation of driving patterns may be one of the reasons why the region of residence reported by the policyholders has a relatively low predictive power of claim counts/riskiness, as will be shown in Section \ref{NBGLM}.

\begin{table}[!h]
\caption{Countries with the Most GPS Observations - Policies with Claims} 
\centering
\begin{tabular}{llrrrr} \hline
\textbf{} & \textbf{Country} & \textbf{Total GPS Observations} & \textbf{Harsh Events} & \textbf{Travel Frequency (\%)} & \textbf{Harsh Event Rate (\%)} \\\hline\hline
1 & Home country & 10,837,451 & 101,601 & 93.43      & 0.94 \\
2 & Country A & 131,162    & 113     & 1.13          & 0.09 \\
3 & Country B & 121,394    & 137     & 1.05          & 0.11 \\
4 & Country C & 102,041    & 241     & 0.88          & 0.24 \\
5 & Country D & 99,243     & 170     & 0.86          & 0.17 \\
6 & Country E & 97,821     & 200     & 0.84          & 0.20 \\\hline
& ...     &            &         & \textit{0.39} &     
\end{tabular}
\label{table: sort-travel-freq-with-claim}
\end{table}

\begin{table}[!h]
\caption{Countries with the Most GPS Observations - Policies without Claims} 
\centering
\begin{tabular}{llrrrr} \hline
\textbf{} & \textbf{Country} & \textbf{Total GPS Observations} & \textbf{Harsh Events} & \textbf{Travel Frequency (\%)} & \textbf{Harsh Event Rate (\%)} \\\hline\hline
1 & Home country & 23,427,201 & 155,597 & 90.60      & 0.66 \\
2 & Country A & 756,033    & 248     & 2.92          & 0.03 \\
3 & Country B & 330,149    & 318     & 1.28          & 0.10 \\
4 & Country C & 254,300    & 846     & 0.98          & 0.33 \\
5 & Country D & 213,322    & 620     & 0.82          & 0.29 \\
6 & Country E & 181,073    & 175     & 0.70          & 0.10 \\\hline
  & ...     &            &         & \textit{0.47} &     
\end{tabular}
\label{table: sort-travel-freq-no-claim}
\end{table}

\begin{table}[h!]
\centering
\begin{threeparttable}
\caption{Two-Sample Means and Medians of Selected Covariates. Last column reports the p-value for testing the null hypothesis that the two means are equal.}
\begin{tabular}{l|rr|rr|rl} \hline
\textbf{} & \multicolumn{2}{c|}{\textbf{claimed}} & \multicolumn{2}{c|}{\textbf{no claim}} & \multicolumn{2}{c}{\textbf{p-value}} \\ 
\textbf{} & \multicolumn{1}{c}{\textbf{mean}} & \multicolumn{1}{c|}{\textbf{median}} & \multicolumn{1}{c}{\textbf{mean}} & \multicolumn{1}{c|}{\textbf{median}} & \multicolumn{2}{c}{{}{}} \\ \hline\hline
\cline{1-6}
\texttt{car\_value} & 169423.4 & 138355.8 & 157029.5 & 134202.3 & 0.064 & . \\
\texttt{vehicle\_age} & 2.2 & 1.0 & 2.3 & 1.0 & 0.402 &  \\
\texttt{max\_weight} & 2149.9 & 2040.0 & 2161.2 & 2030.0 & 0.668 &  \\
\texttt{policy\_period} & 1.0 & 1.0 & 0.9 & 1.0 & 0.000 & *** \\
\texttt{total\_distance} & 16370.2 & 13122.8 & 14040.0 & 10554.2 & 0.002 & * \\
\texttt{total\_time} & 26552.8 & 23091.6 & 21961.7 & 18850.8 & 0.000 & *** \\
\texttt{avg\_speed} & 36.1 & 34.8 & 36.3 & 35.2 & 0.747 &  \\
\texttt{max\_speed} & 168.0 & 166.0 & 157.7 & 155.0 & 0.000 & *** \\
\texttt{num\_acc} & 101.3 & 24.0 & 54.8 & 9.0 & 0.001 & ** \\
\texttt{num\_brake} & 67.2 & 43.0 & 42.8 & 28.0 & 0.000 & *** \\
\texttt{num\_left} & 55.6 & 11.0 & 28.8 & 7.0 & 0.003 & ** \\
\texttt{num\_right} & 34.4 & 8.0 & 19.9 & 5.0 & 0.015 & * \\
\texttt{num\_severe} & 3.4 & 1.0 & 2.9 & 1.0 & 0.398 &  \\
\texttt{num\_trips} & 1489.4 & 1330.0 & 1163.6 & 1076.0 & 0.000 & *** \\
\texttt{pc1} & 0.16 & -0.65 & -0.06 & -0.91 & 0.483 &  \\
\texttt{pc2} & 0.71 & 0.65 & -0.27 & -0.31 & 0.000 & *** \\\hline\hline
\end{tabular}
\begin{tablenotes}
\small
\item Significance codes: 0 `***', 0.001 `**', 0.01 `*', 0.05 `.', 0.1 ` ', 1
\end{tablenotes}
\label{table: two-sample-mean-median}
\end{threeparttable}
\end{table}

\begin{figure*}[h]
\caption{First Two Principal Components (PCs) of Speed Transition Matrix.  First Row: Loadings, Second Row: Grouped Histogram of Resulted PCs - Red: With Claims, Blue: Without Claim.}
    \centering
    \begin{subfigure}[b]{0.47\textwidth}
        \centering
        \includegraphics[width=\textwidth]{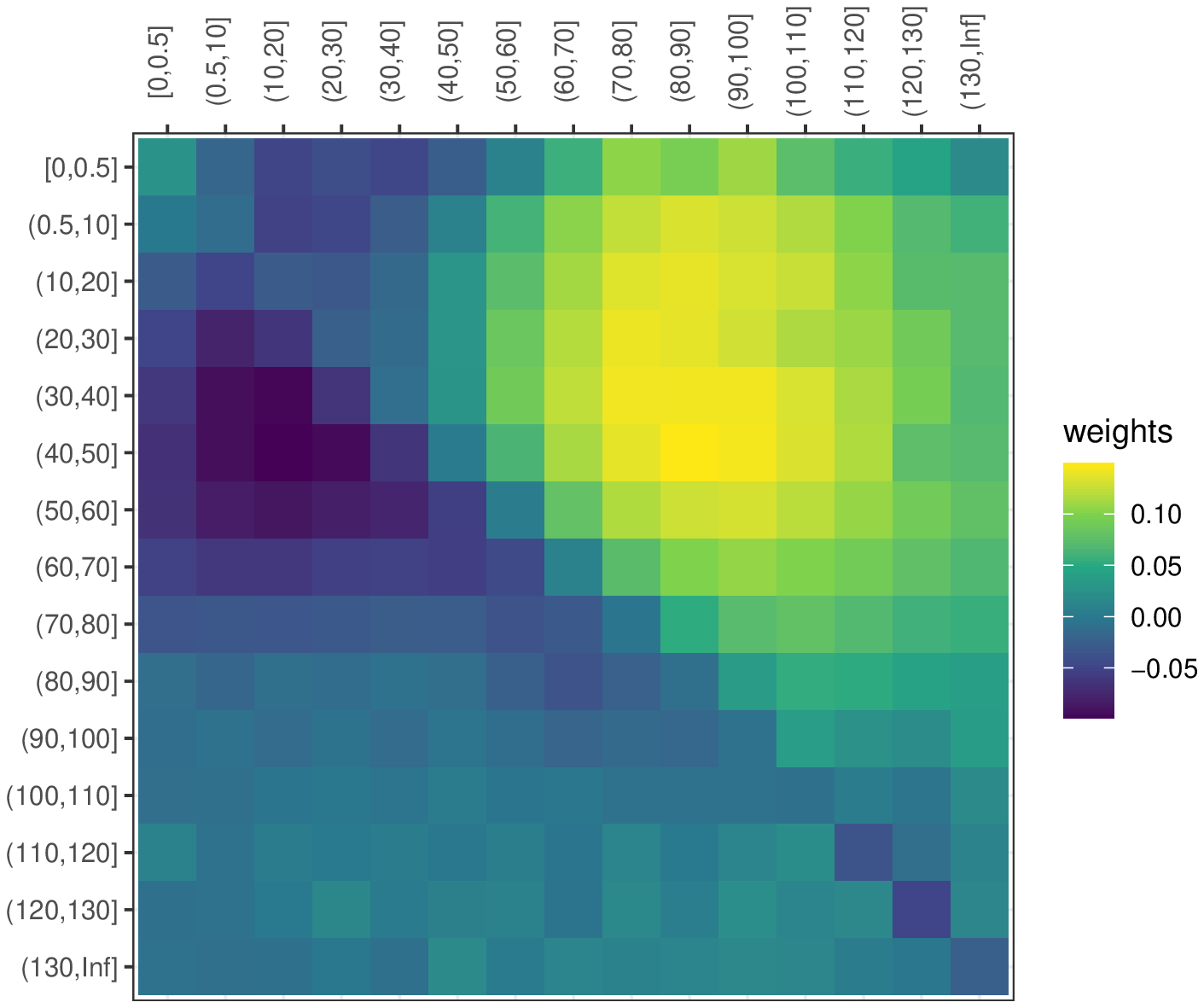}
        \caption{Loadings of PC1}
    \end{subfigure}
    \hfill
    \begin{subfigure}[b]{0.47\textwidth}  
        \centering 
        \includegraphics[width=\textwidth]{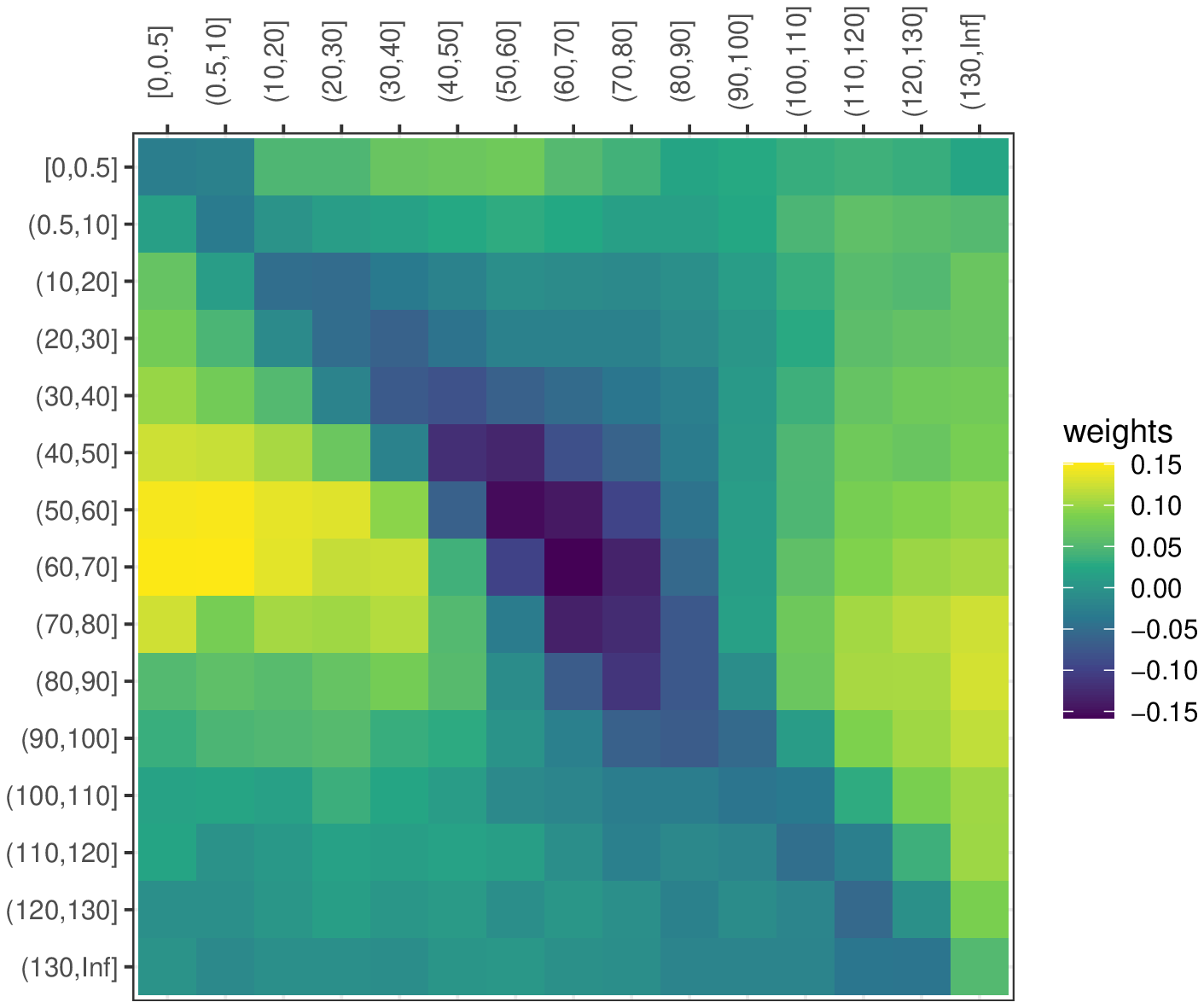}
        \caption{Loadings of PC2}
    \end{subfigure}
    \vskip\baselineskip
    \begin{subfigure}[b]{0.47\textwidth}   
        \centering 
        \includegraphics[width=\textwidth]{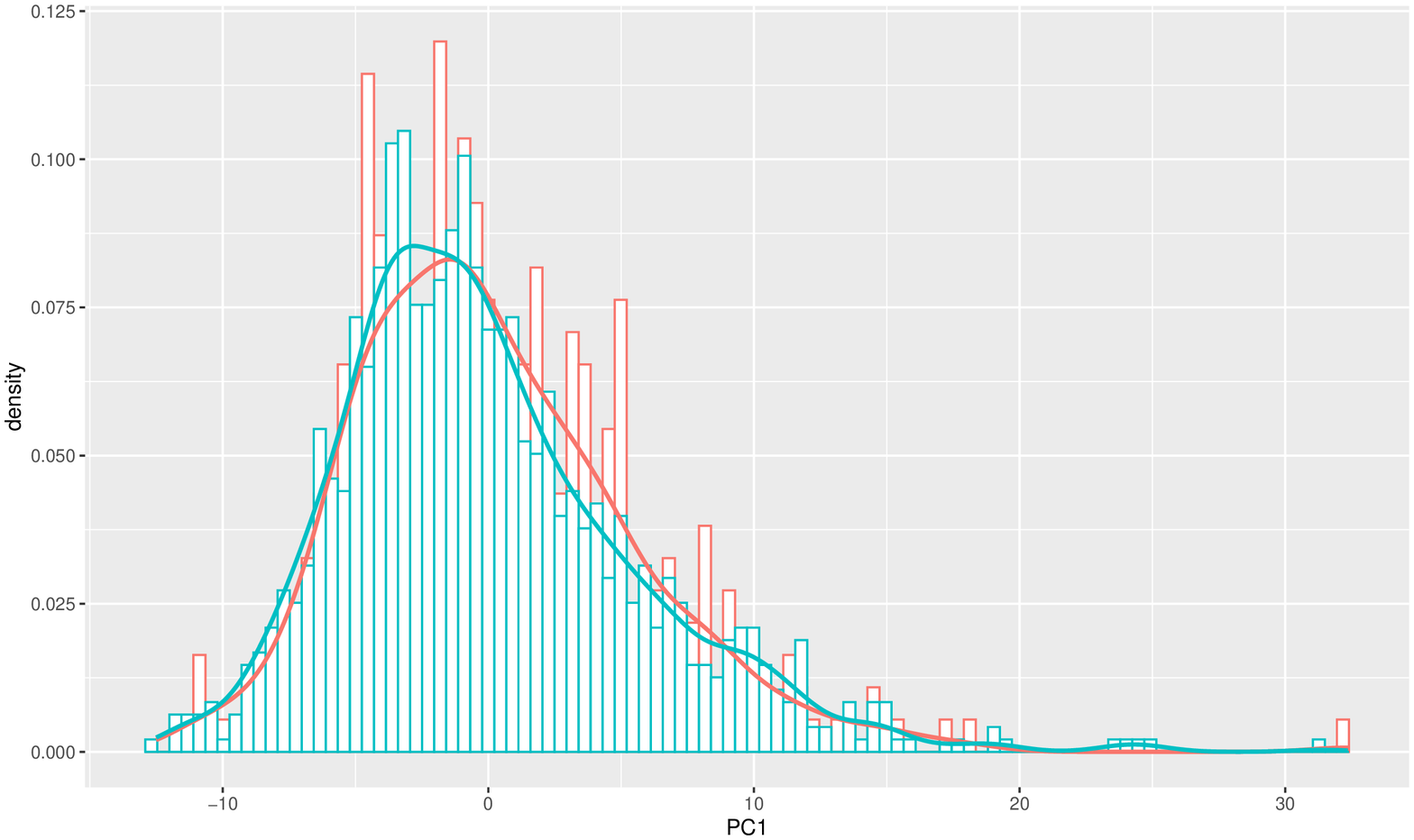}
        \caption{PC1 of Speed Transition Matrix}
    \end{subfigure}
    \hfill
    \begin{subfigure}[b]{0.47\textwidth}   
        \centering 
        \includegraphics[width=\textwidth]{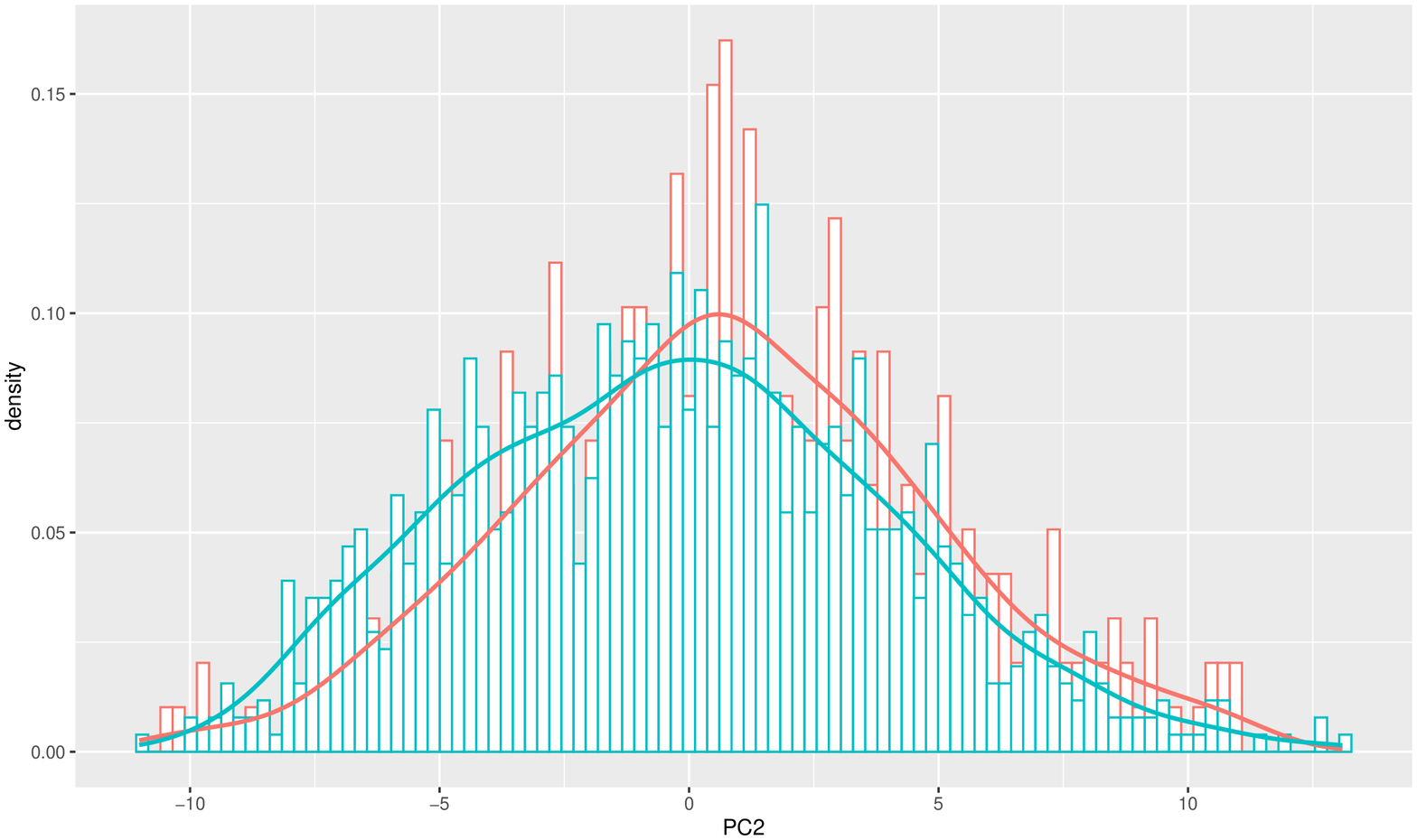}
        \caption{PC2 of Speed Transition Matrix}
    \end{subfigure}
    \label{fig: pc-weights}
\end{figure*}

\begin{figure*}[h]
\caption{Grouped Histograms of Selected Telematics Features.  Red: With Claims, Blue: Without Claim.}
    \centering
    \begin{subfigure}[b]{0.47\textwidth}
        \centering
        \includegraphics[width=\textwidth]{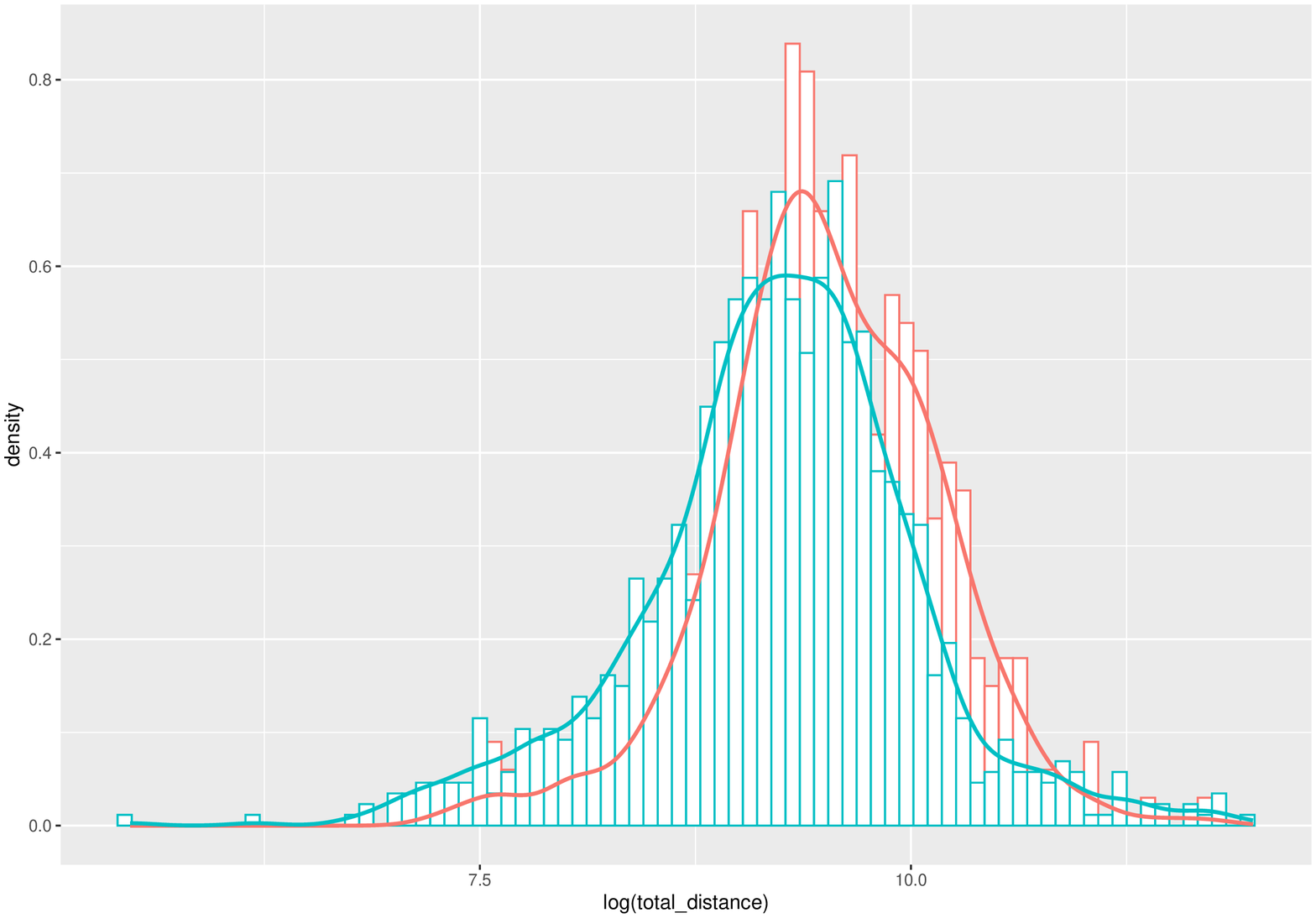}
        \caption{Logged Total Distance}
    \end{subfigure}
    \hfill
    \begin{subfigure}[b]{0.47\textwidth}  
        \centering 
\includegraphics[width=\textwidth]{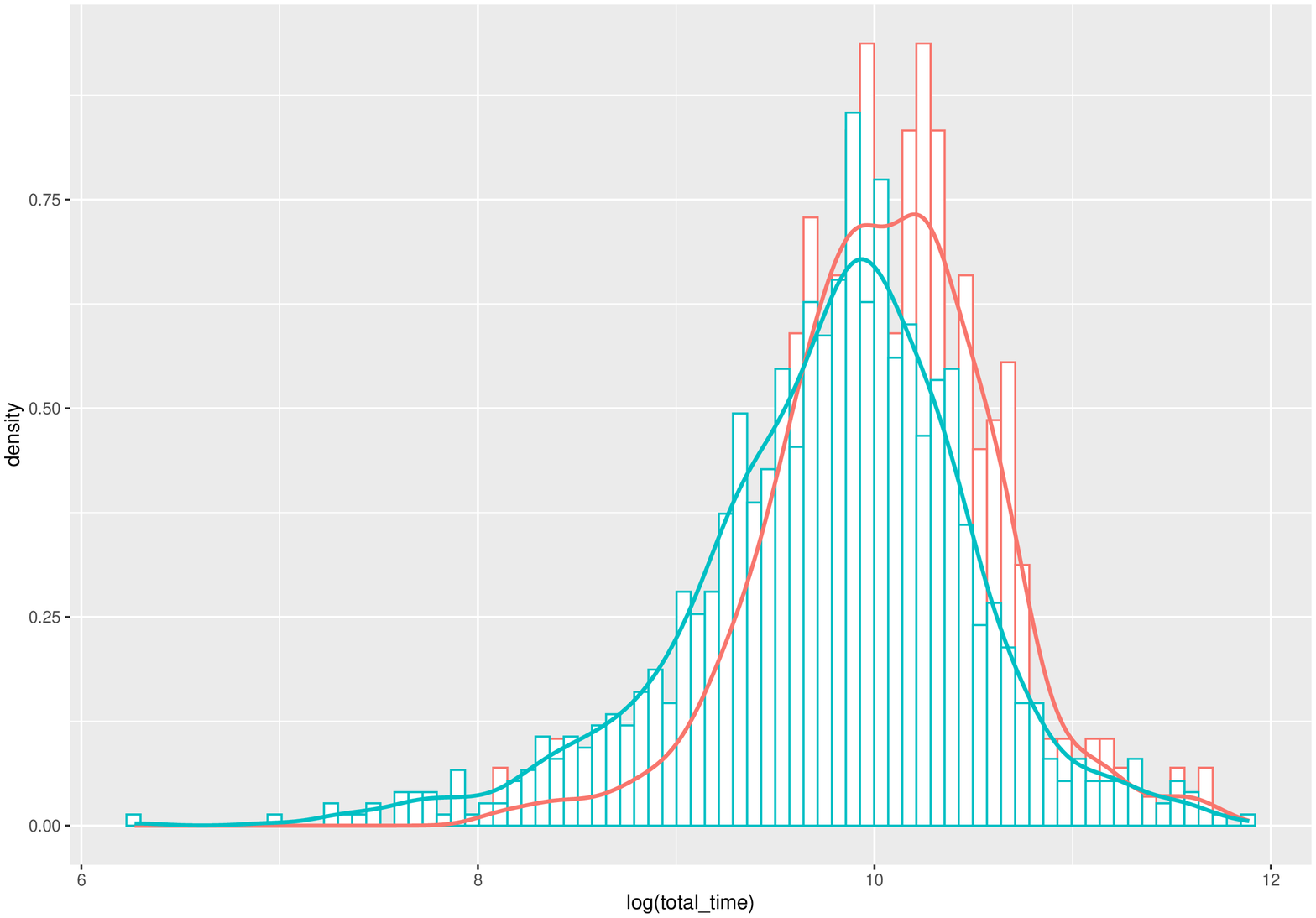}
        \caption{Logged Total Time}
    \end{subfigure}
    \vskip\baselineskip
    \begin{subfigure}[b]{0.47\textwidth}   
        \centering 
        \includegraphics[width=\textwidth]{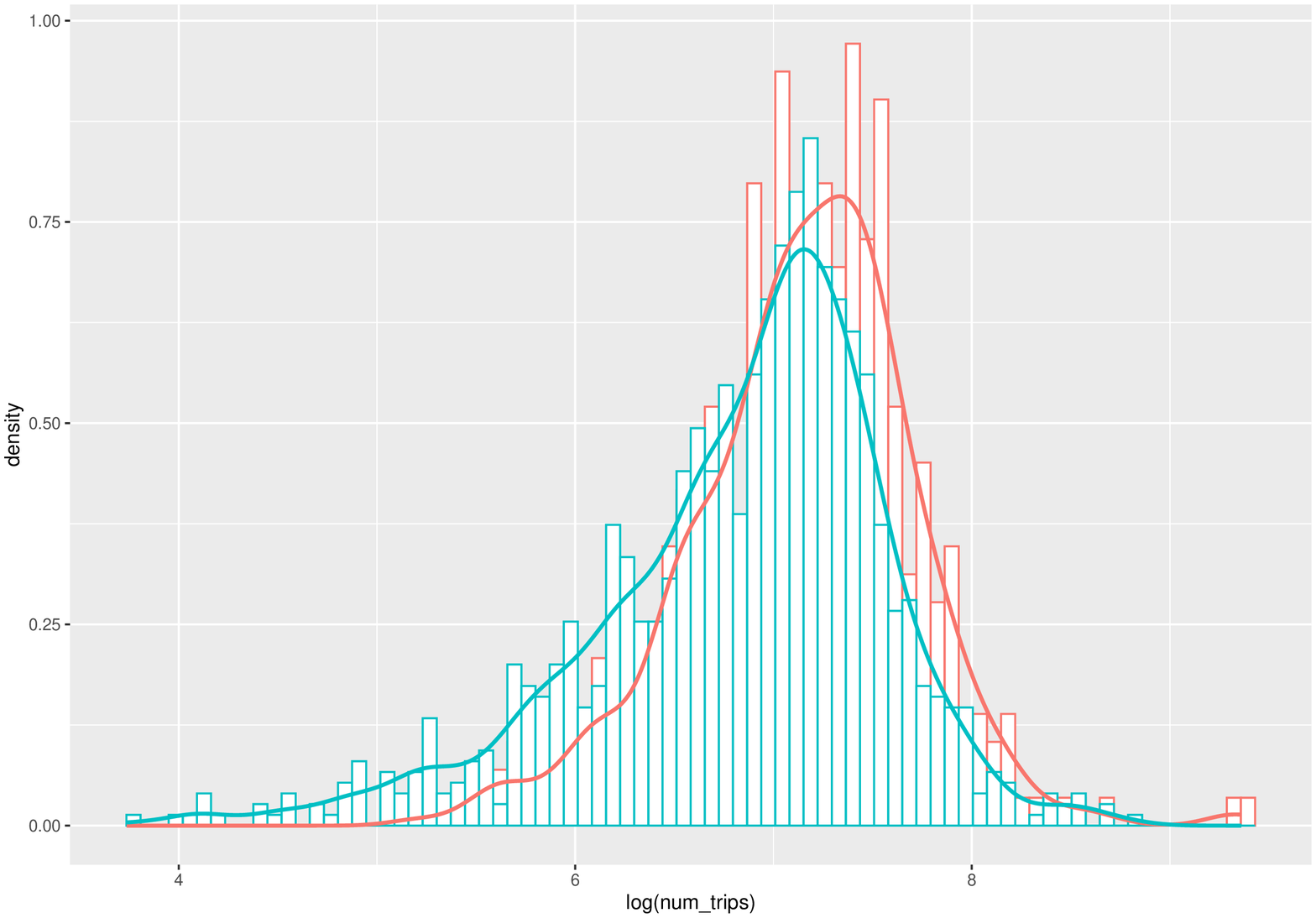}
        \caption{Logged Number of Trips}
    \end{subfigure}
    \hfill
    \begin{subfigure}[b]{0.47\textwidth}   
        \centering 
        \includegraphics[width=\textwidth]{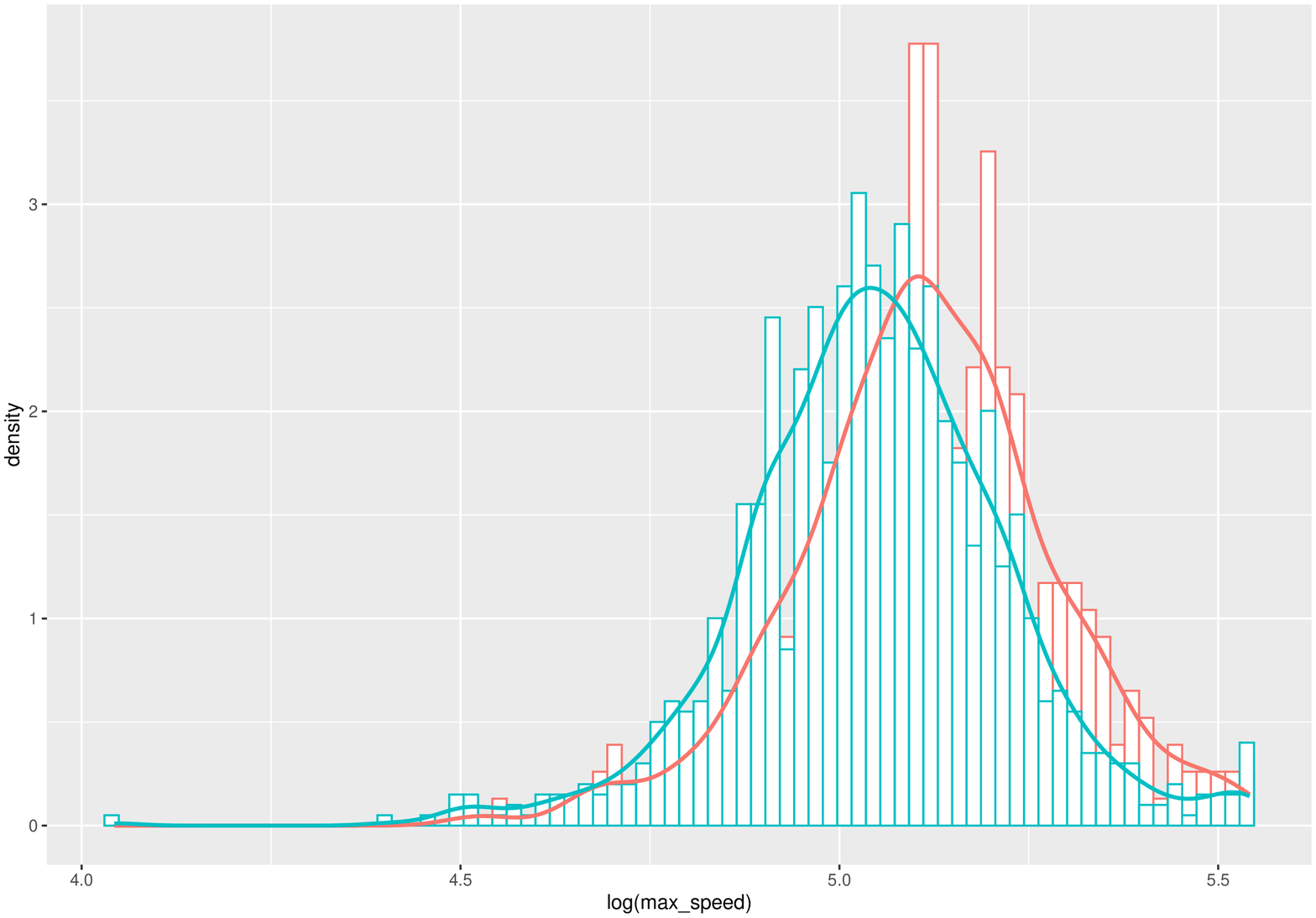}
        \caption{Logged Max Speed}
    \end{subfigure}
    \vskip\baselineskip
    \begin{subfigure}[b]{0.47\textwidth}   
        \centering 
        \includegraphics[width=\textwidth]{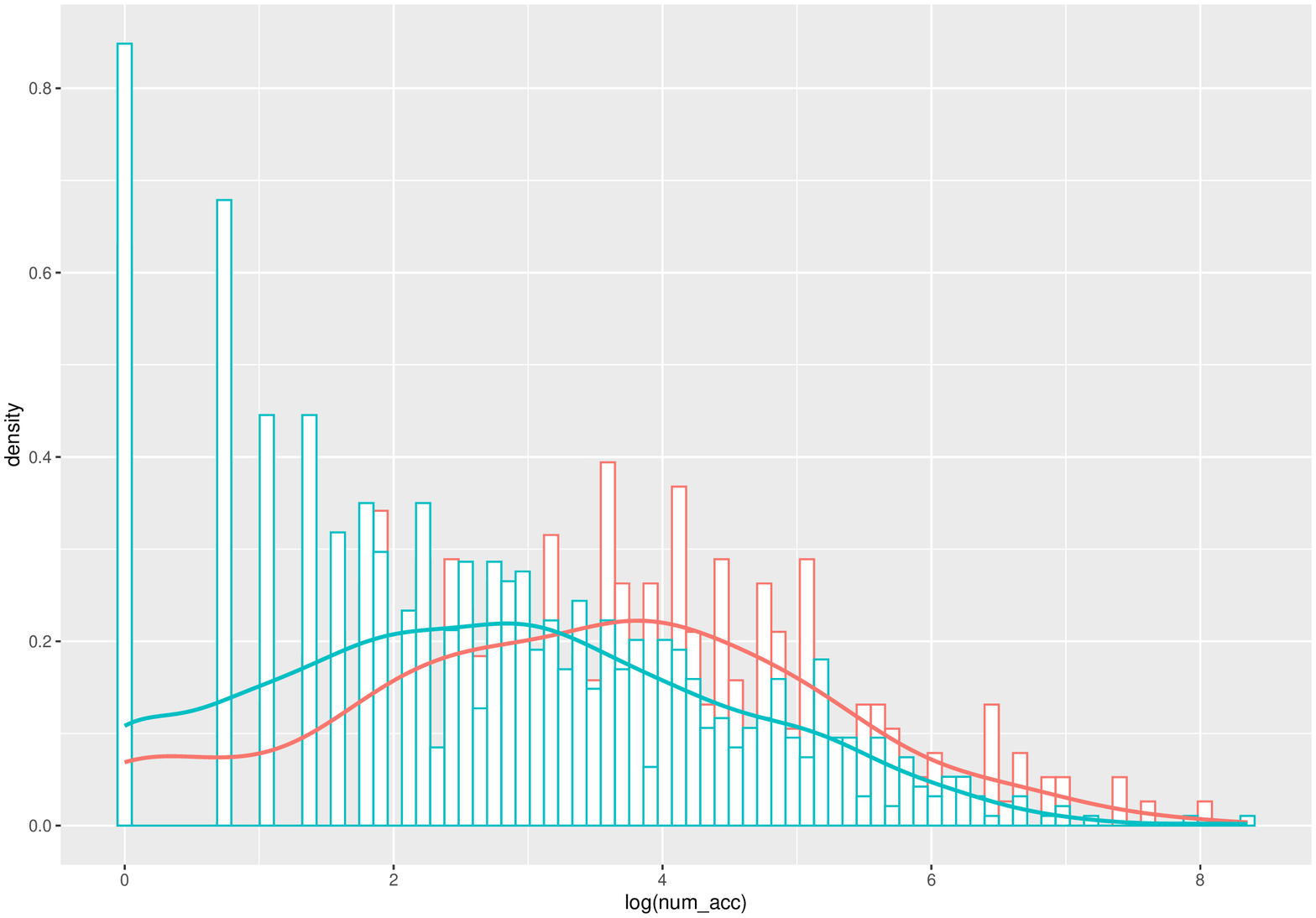}
        \caption{Logged Number of Harsh Accelerations}
    \end{subfigure}
    \hfill
    \begin{subfigure}[b]{0.47\textwidth}   
        \centering 
        \includegraphics[width=\textwidth]{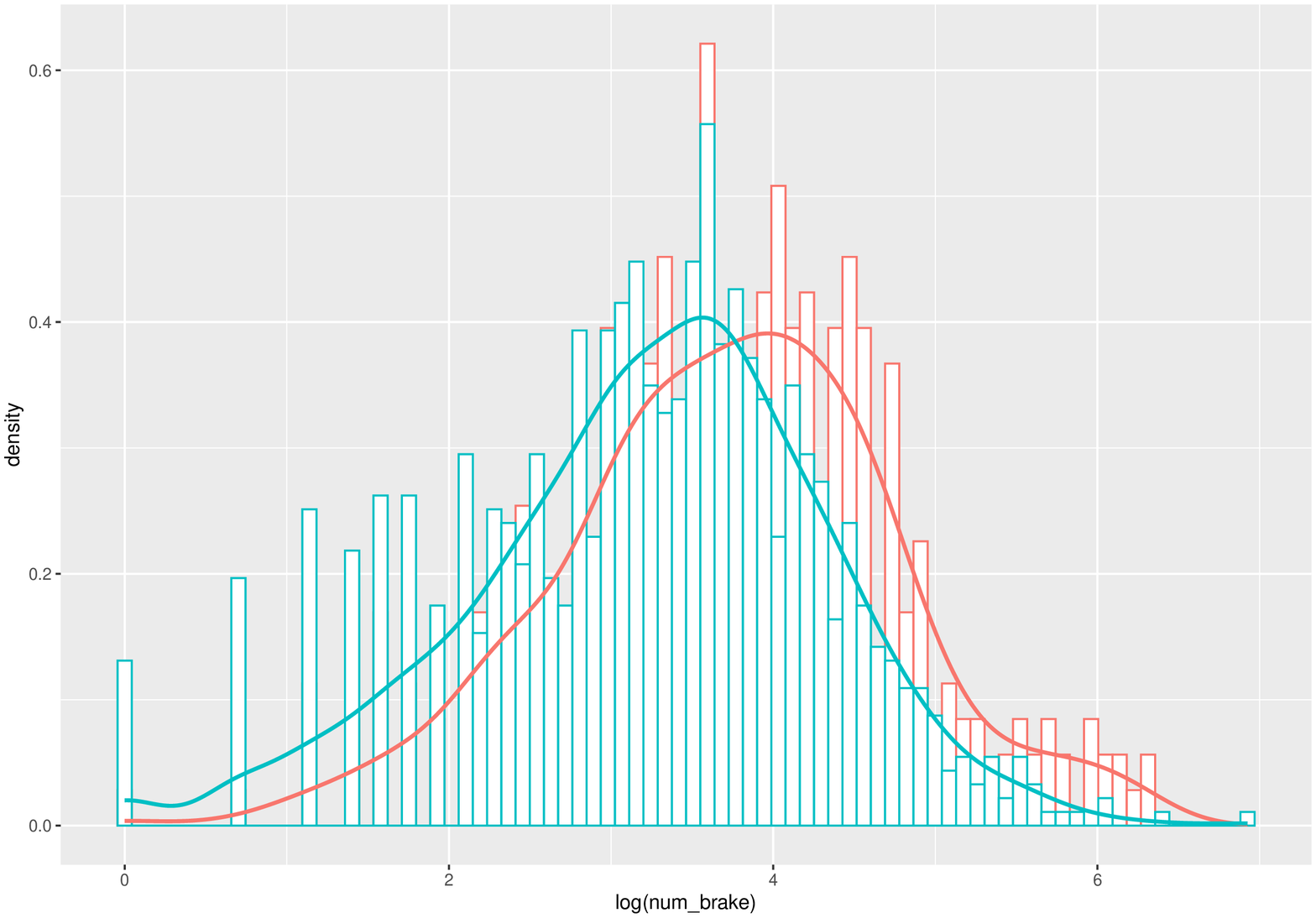}
        \caption{Logged Number of Harsh Braking}
    \end{subfigure}
    \label{fig: grouped-histograms-telematics}
\end{figure*}

\begin{figure*}[h]
\caption{Grouped Boxplots of Compositional Covariates.  Red: With Claims, Blue: Without Claims.}
    \centering
    \begin{subfigure}[b]{\textwidth}
        \centering
        \includegraphics[width=\textwidth]{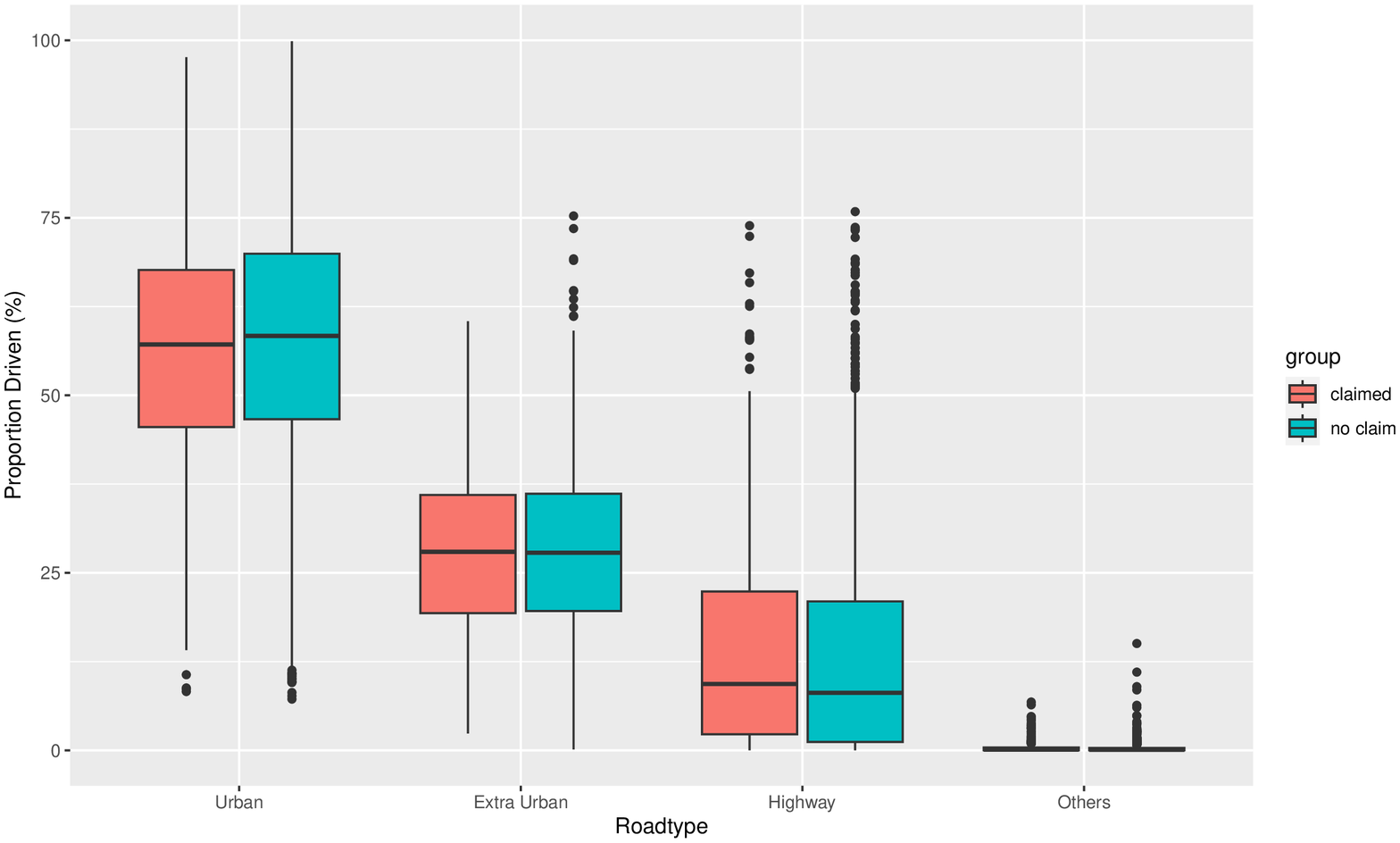}
        \caption{Proportion Driven on Different Roadtypes}
    \end{subfigure}
    \vskip\baselineskip
    \begin{subfigure}[b]{\textwidth}   
        \centering 
        \includegraphics[width=\textwidth]{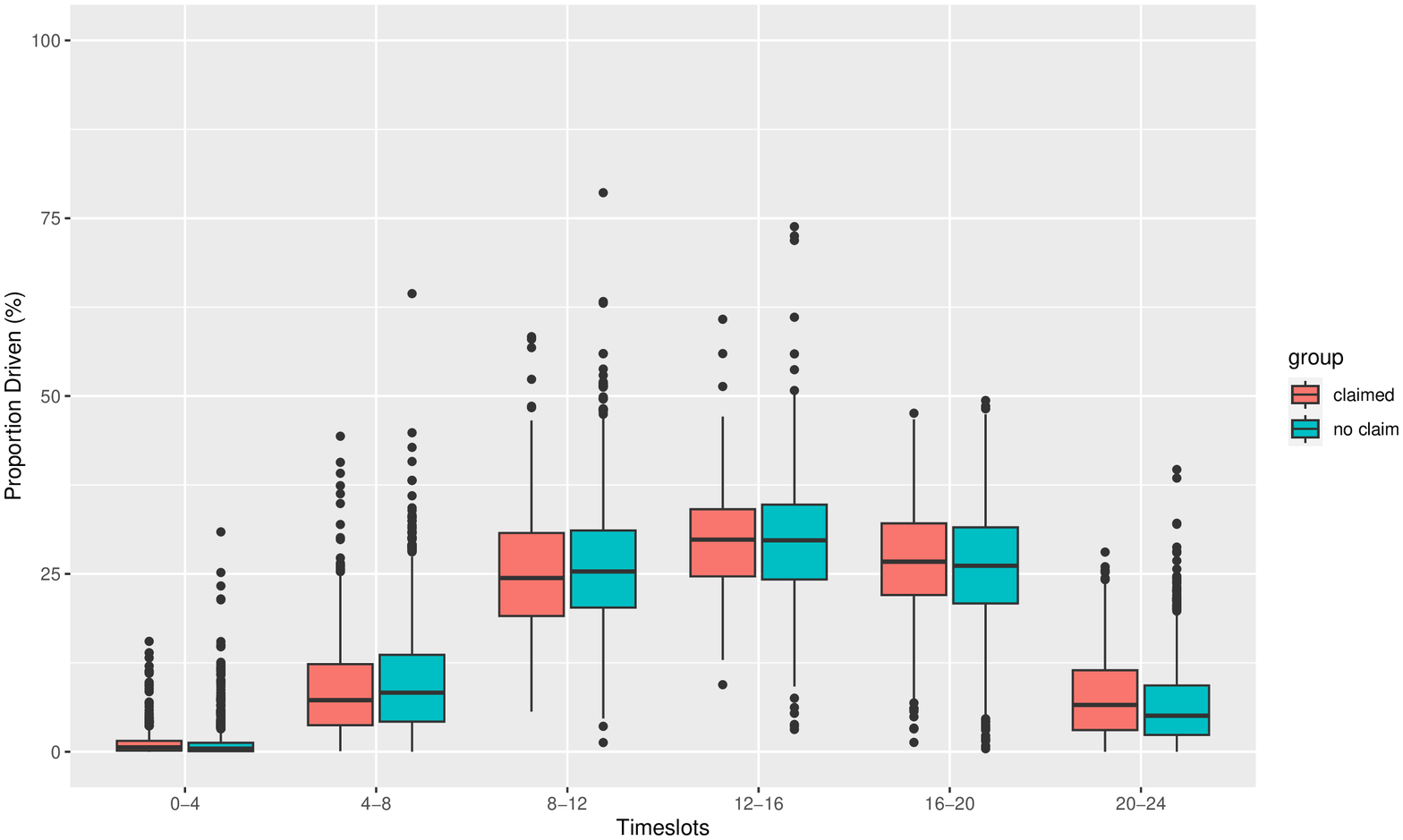}
        \caption{Proportion Driven in Different Timeslots}
    \end{subfigure}
    \label{fig: grouped-boxplots}
\end{figure*}

Classifying policies into `claimed' and `no claim' groups, in Table \ref{table: two-sample-mean-median} we report the two-sample means and medians of selected covariates, where the last column indicates whether the null hypothesis that the two means are equal can be rejected at a significance level of 5\%.  Then in Figures \ref{fig: pc-weights} and \ref{fig: grouped-histograms-telematics} we show grouped histograms of the telematics covariates having significantly different means, while in Figure \ref{fig: grouped-boxplots} we show grouped boxplots of compositional covariates.  As mentioned, \texttt{pc1} gives positive weights to the transitions towards higher speed bins, whilst \texttt{pc2} gives positive weights to the few transitions from moderate speed bins to lower speed bins, and negative weights along the diagonal, which are the few transitions within the moderate speed bins.  Hence large speed transitions will be represented by larger, positive values of \texttt{pc1} and \texttt{pc2}.  As displayed in the grouped histograms in Figure \ref{fig: pc-weights}, the group with claims has higher PC1 and PC2 on average, when compared to the group without claim, and the difference is more obvious in PC2.  This suggests that the two groups differ more in deceleration patterns than in acceleration patterns.  In Figure \ref{fig: grouped-histograms-telematics} we observe that the group with claims is driving longer on average, in terms of both distance and time, and is making more harsh accelerations and braking, with the difference in the latter being more distinct; this aligns with the interpretations from the PC's.  The maximum speed attained is also generally higher.  The proportions driven in different roadtypes and timeslots are generally close between the two groups.  Overall, the group with claims is driving slightly less in urban zones but more on highways, as well as more between 8 p.m. to midnight.  

The time when drivers begin their trips dominates the different timeslots when they are driving, and both are related to differences in driving behaviour.  In Table \ref{table: driving-behaviour-timeslot} we report the two-sample means of maximum speed and harsh event rate (i.e., per minute) in different timeslots.  For simplicity, we categorize trips mainly according to their start time, e.g. a trip is categorized in the first group if it begins between midnight to 2 a.m., or begins between 3 a.m. to 4 a.m. and ends before 4:15 a.m.  We observe that maximum speed and harsh event rate are generally higher during nighttime, and they are highest between midnight to 4 a.m., followed by 8 p.m. to midnight.  Yet, we cannot conclude that driving between midnight to 4 a.m. is the most dangerous since the number of trips begun in this timeslot is the lowest.

\begin{table}[h]
\caption{Two-Sample Means of Maximum Speed and Harsh Event Per Minute for Trips Beginning in Different Timeslots. Proportion indicates the number of trips with respect to all observed trips.}
\centering
\begin{tabular}{l|rrr|rrr} \hline
 & \multicolumn{3}{c|}{\textbf{claimed}} & \multicolumn{3}{c}{\textbf{no claim}} \\
\textbf{timeslot} & \multicolumn{1}{l}{\textbf{max speed}} & \multicolumn{1}{l}{\textbf{harsh event rate}} & \multicolumn{1}{l|}{\textbf{proportion}} & \multicolumn{1}{l}{\textbf{max speed}} & \multicolumn{1}{l}{\textbf{harsh event rate}} & \multicolumn{1}{l}{\textbf{proportion}} \\ \hline\hline
midnight - 4 a.m. & 71.6 & 0.023 & 0.3 & 77.5 & 0.013 & 0.5 \\
4 - 8 a.m. & 60.9 & 0.009 & 3.0 & 60.4 & 0.006 & 7.2 \\
8 a.m. - noon & 57.8 & 0.010 & 8.8 & 57.7 & 0.008 & 18.8 \\
noon - 4 p.m. & 58.4 & 0.011 & 9.9 & 59.0 & 0.008 & 20.0 \\
4 - 8 p.m. & 58.2 & 0.010 & 8.4 & 58.3 & 0.008 & 16.7 \\
8 p.m. - midnight & 62.9 & 0.019 & 2.5 & 63.5 & 0.011 & 3.8 \\\hline\hline
\end{tabular}
\label{table: driving-behaviour-timeslot}
\end{table}

\subsection{Individual Heterogeneity}
\label{Individual-Heterogeneity}

In this section we explore briefly the driving behaviour of individual drivers, through their driving speed and start time of each trip.

\begin{figure*}[h]
\caption{Trip-based Mean Speed for Policies With and Without Claims; Simple Clustering Based on Whether the Mode is Within (Blue) or Beyond (Red) 95\% Confidence Interval of the Group.}
    \centering
    \begin{subfigure}[b]{\textwidth}
        \centering
        \includegraphics[width=\textwidth]{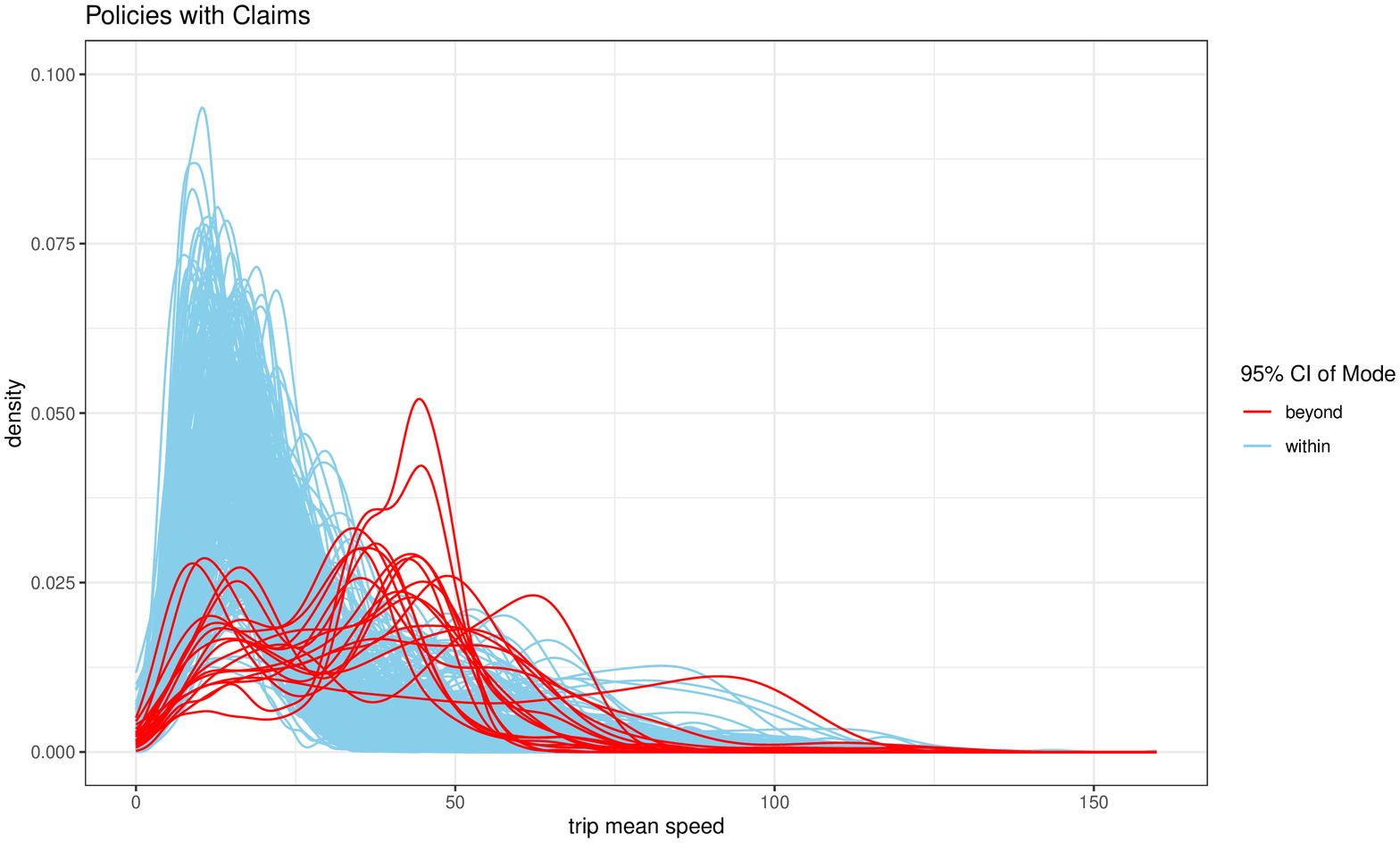}
        \caption{}
    \end{subfigure}
    \vskip\baselineskip
    \begin{subfigure}[b]{\textwidth}   
        \centering 
        \includegraphics[width=\textwidth]{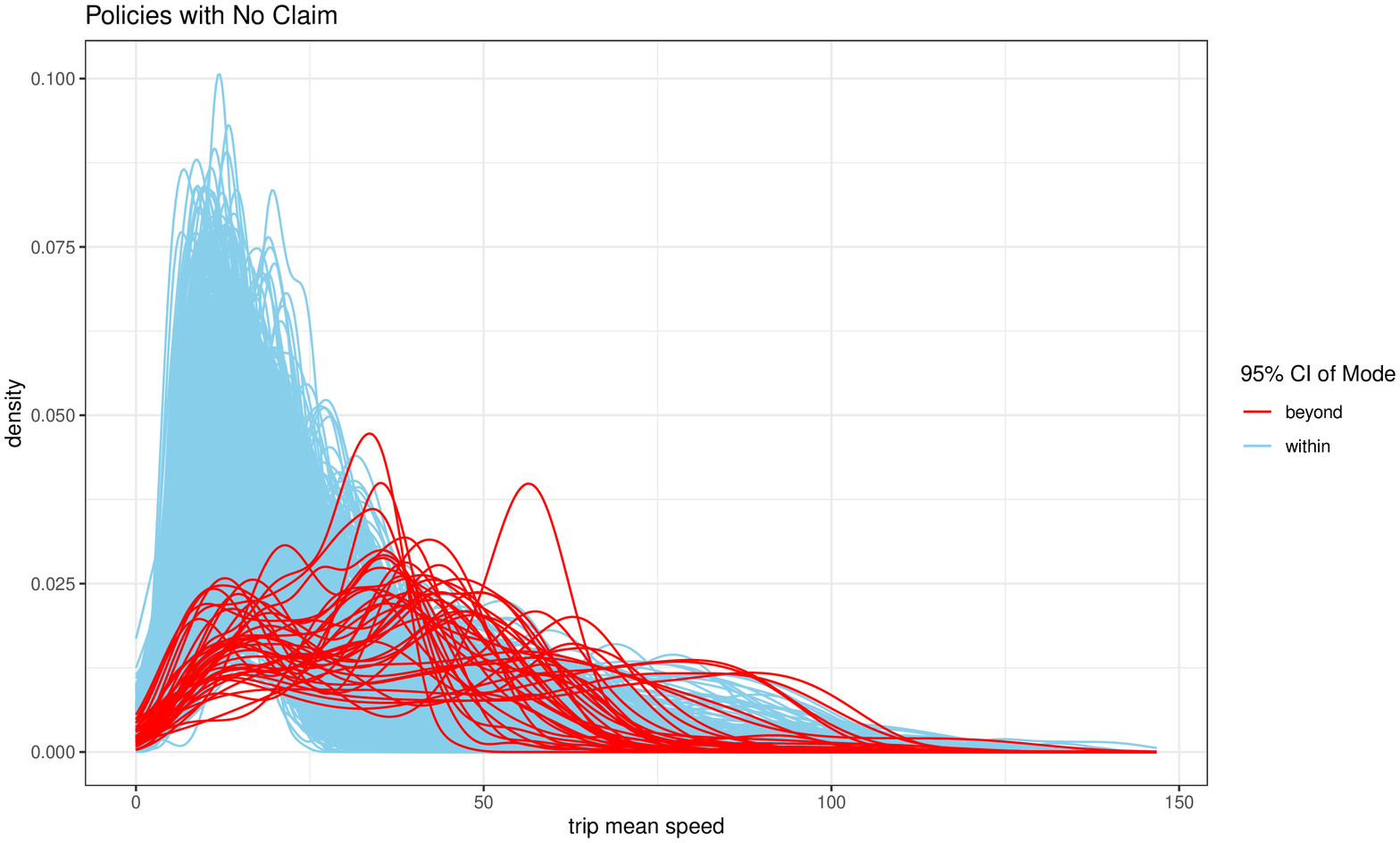}
        \caption{}
    \end{subfigure}
    \label{fig: trip-mean-speed}
\end{figure*}

\begin{figure*}[h]
\caption{Trip-based Maximum Speed for Policies With and Without Claims; Simple Clustering Based on Whether the Mode is Within (Blue) or Beyond (Red) 95\% Confidence Interval of the Group.}
    \centering
    \begin{subfigure}[b]{\textwidth}
        \centering
        \includegraphics[width=\textwidth]{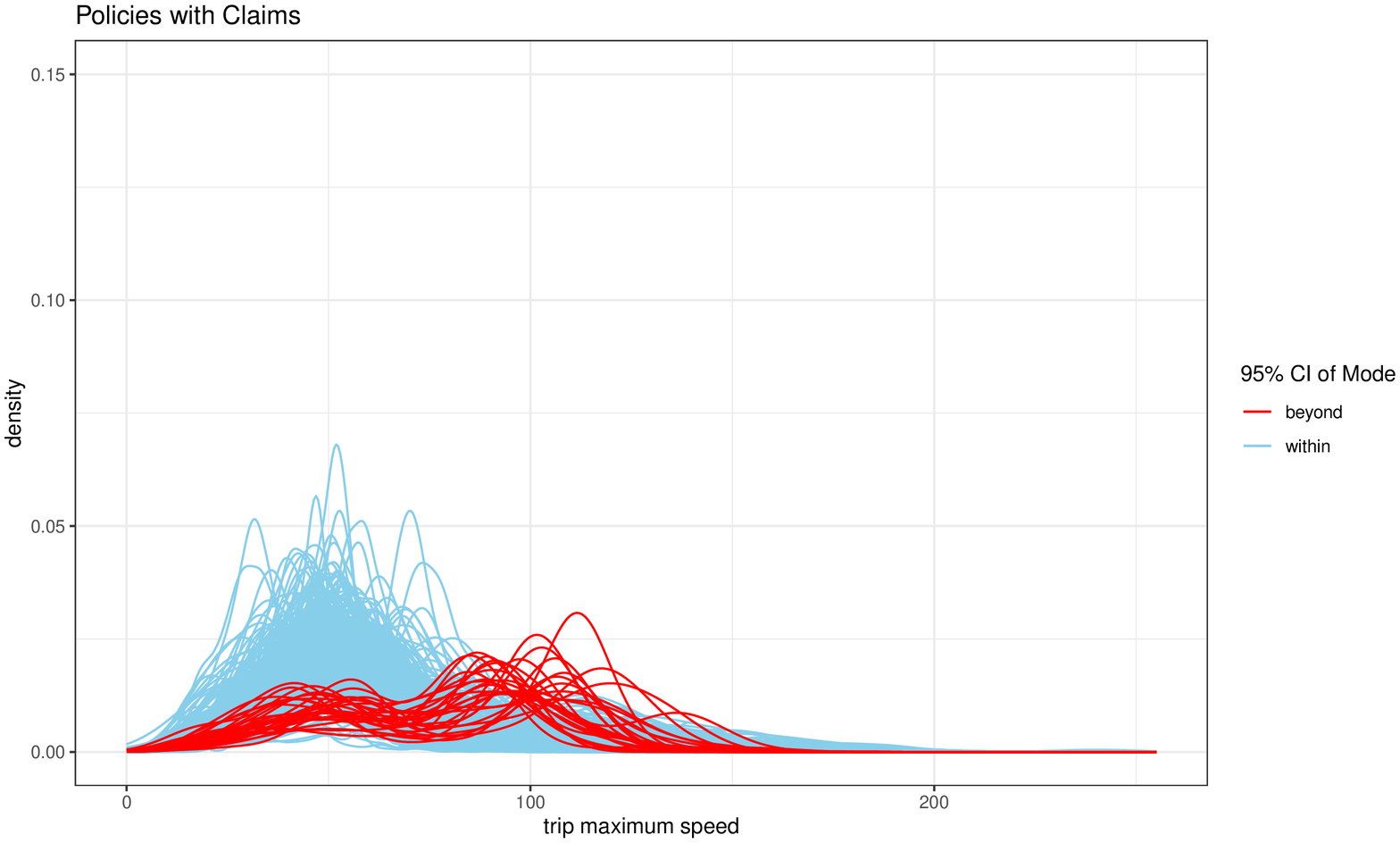}
        \caption{}
    \end{subfigure}
    \vskip\baselineskip
    \begin{subfigure}[b]{\textwidth}   
        \centering 
        \includegraphics[width=\textwidth]{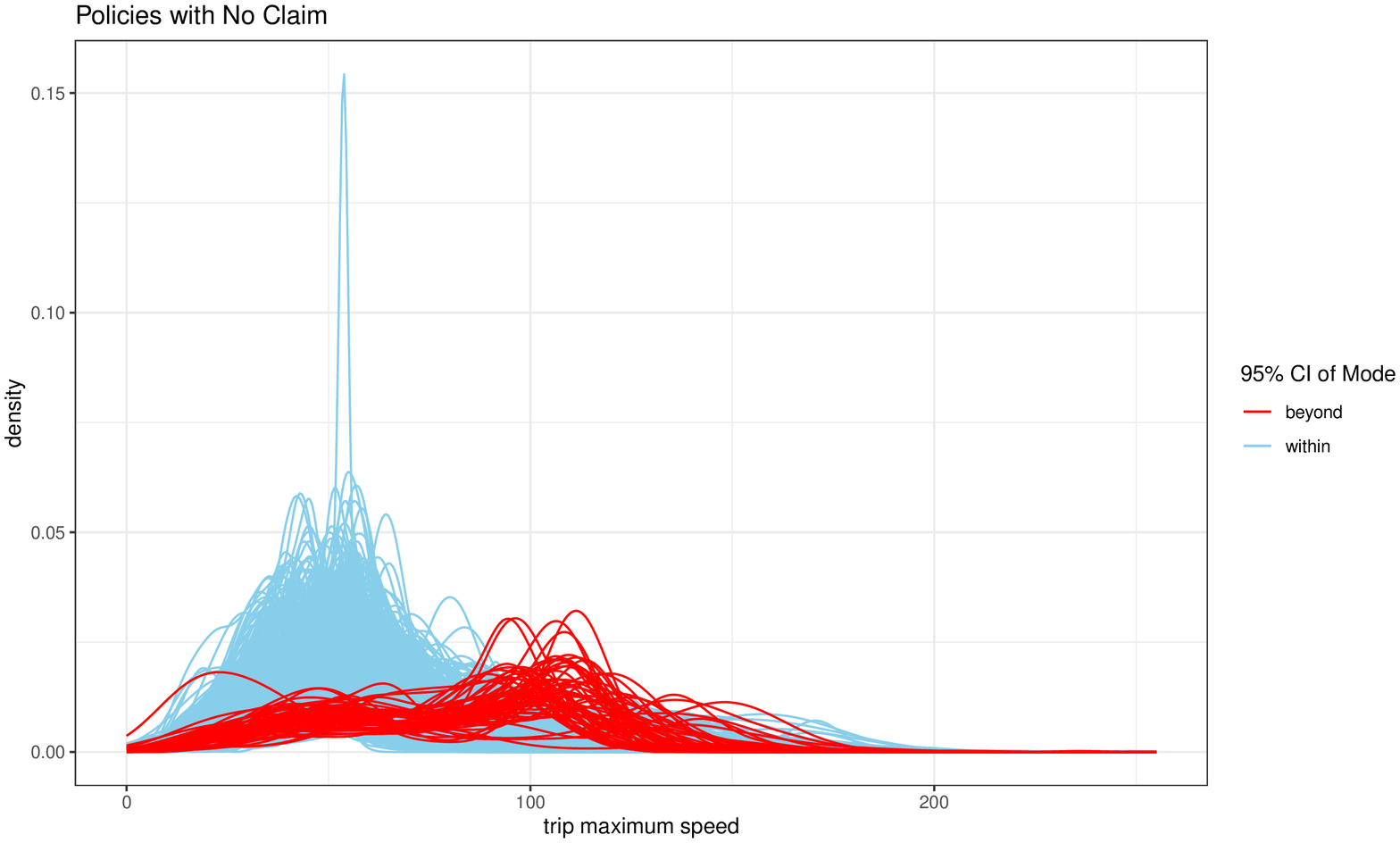}
        \caption{}
    \end{subfigure}
    \label{fig: trip-max-speed}
\end{figure*}

Figures \ref{fig: trip-mean-speed} and \ref{fig: trip-max-speed} show the trip-based mean and maximum speeds, respectively, for policies with and without claims.  Each line represents a policy's empirical density of mean/maximum speed per trip.  We apply a simple clustering of whether the density's mode is within (blue) or beyond (red) the 95\% confidence interval of the group.  We observe quite a heterogeneity in driving behaviour, evidenced by first the multimodality in individual empirical density, and second the various density shapes even among the policies that have or have not filed claims.

\begin{figure*}[h]
\caption{Sample Driving Behaviour (Trip Mean and Maximum Speed) and the Corresponding Speed Transition Matrix.}
    \centering
    \begin{subfigure}[b]{0.48\textwidth}
        \centering
        \includegraphics[width=\textwidth]{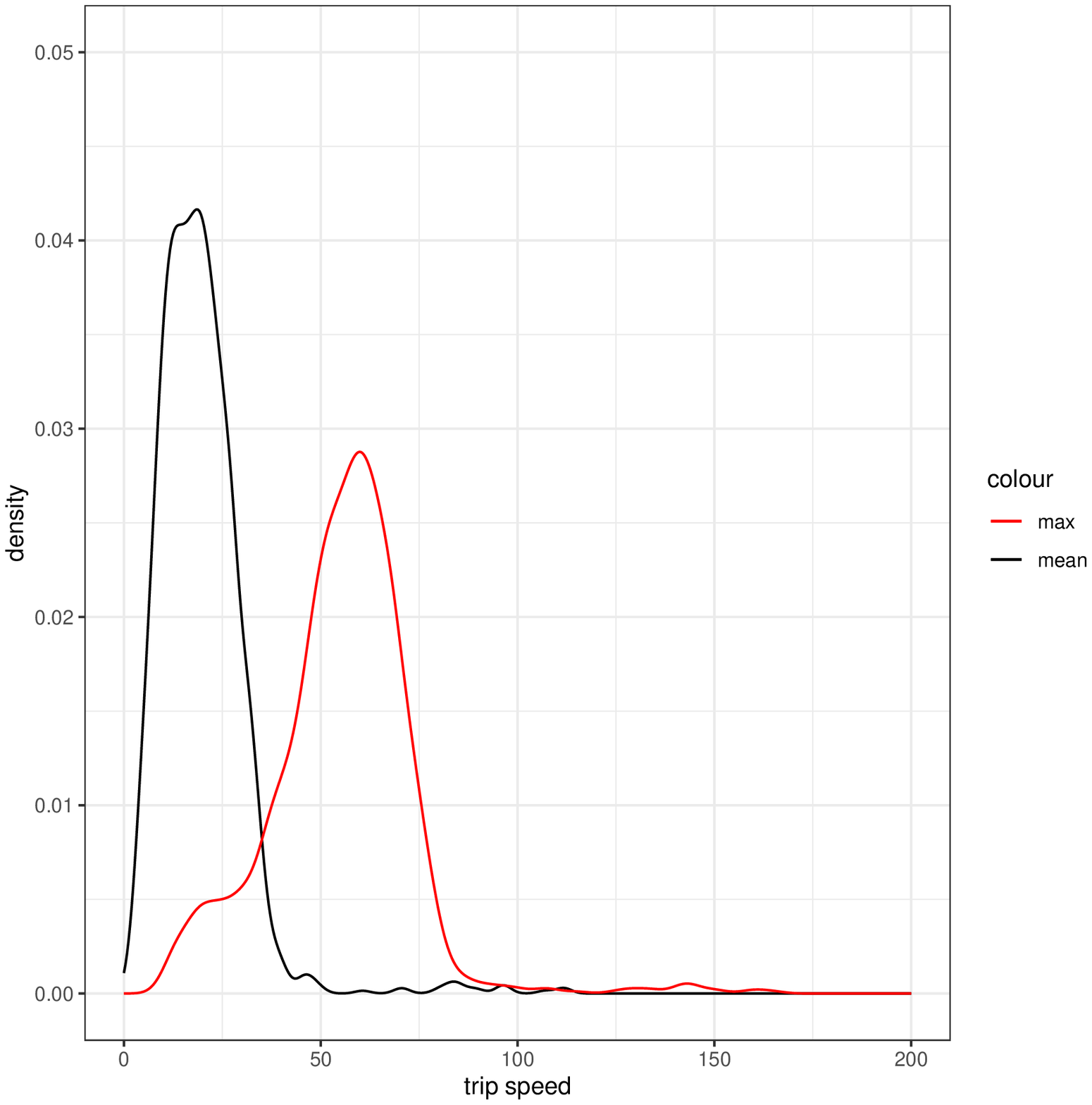}
        \caption*{}
    \end{subfigure}
    \hfill
    \begin{subfigure}[b]{0.48\textwidth}  
        \centering 
        \includegraphics[width=\textwidth]{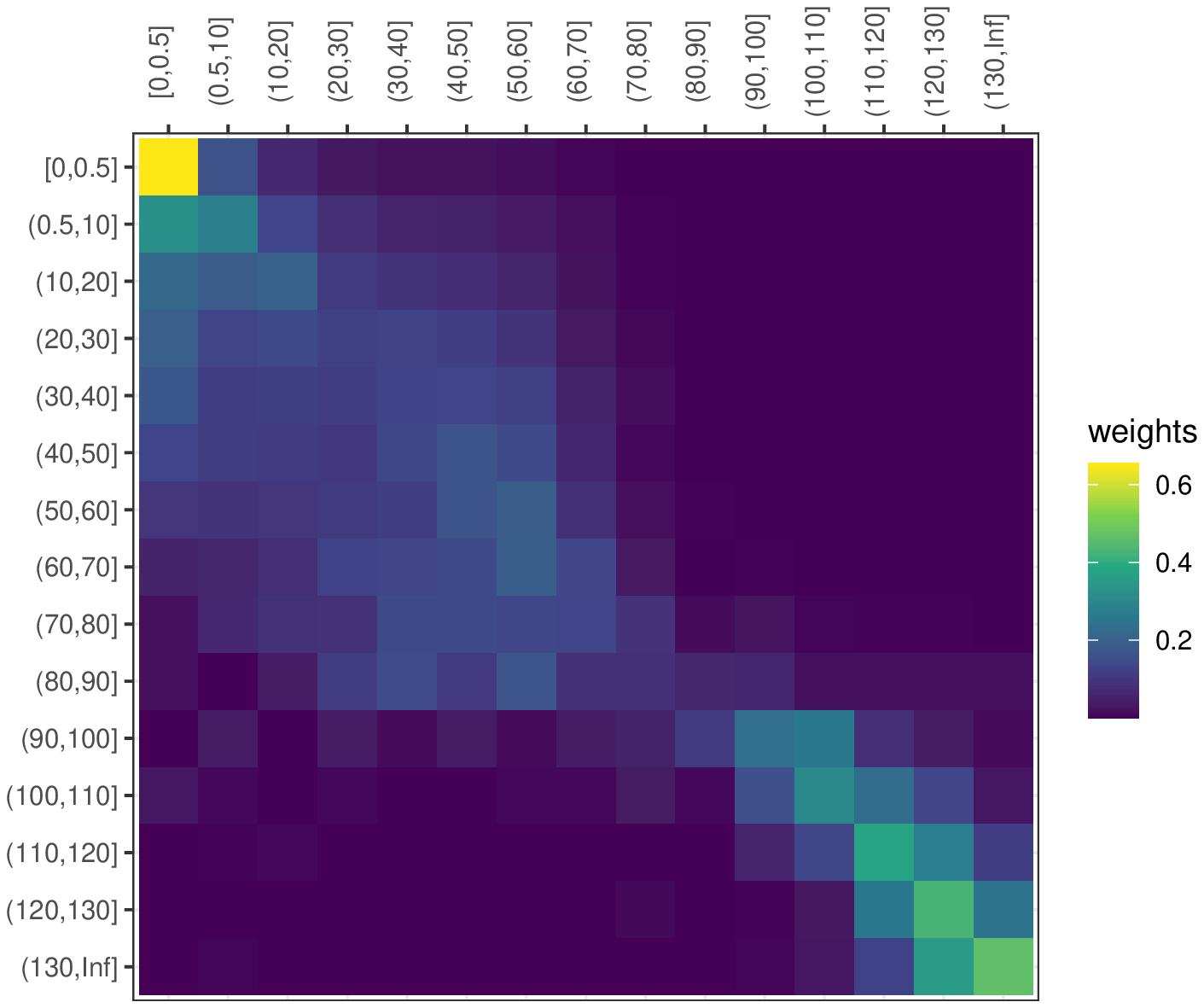}
        \caption*{}
    \end{subfigure}
    \caption*{Driver A}
    \hfill
    \vskip\baselineskip
        \begin{subfigure}[b]{0.48\textwidth}
        \centering
        \includegraphics[width=\textwidth]{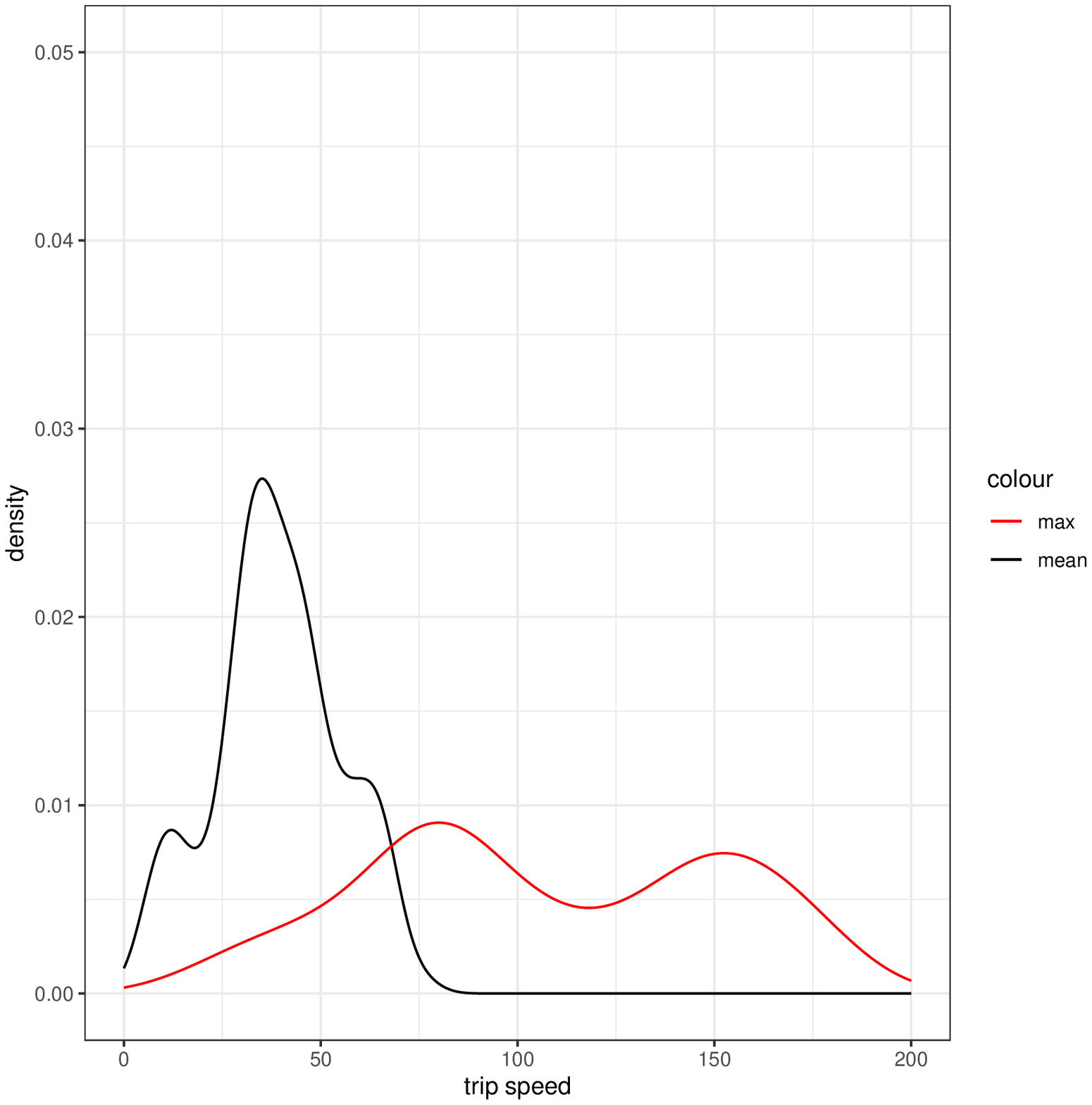}
        \caption*{}
    \end{subfigure}
    \hfill
    \begin{subfigure}[b]{0.48\textwidth}  
        \centering 
        \includegraphics[width=\textwidth]{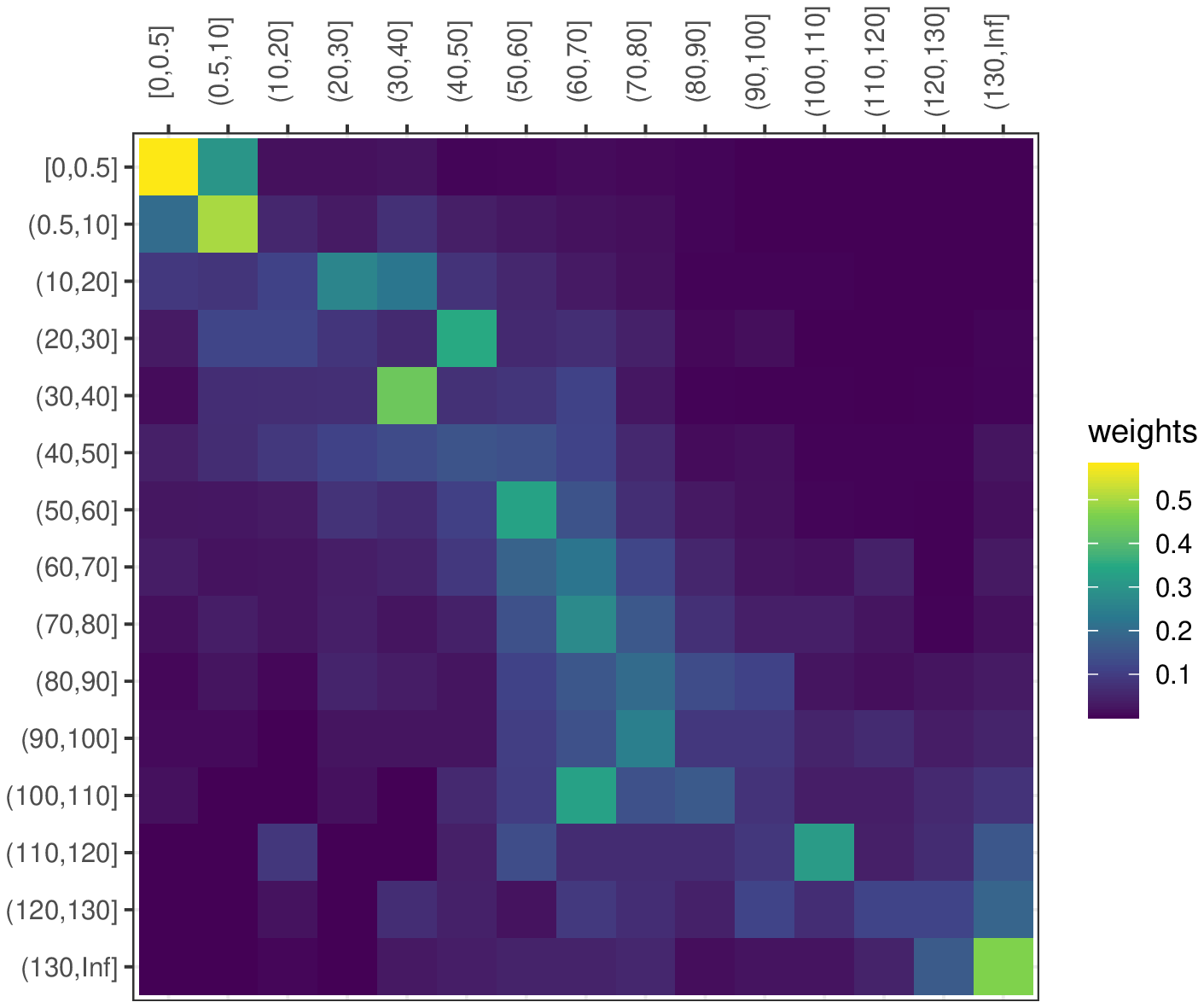}
        \caption*{}
    \end{subfigure}
    \caption*{Driver B}
    \label{fig: sample-transition-matrix-heatmap}
\end{figure*}

While the heterogeneity within each group may average out and lead to some difference between the two groups on an aggregate level as in Figure \ref{fig: grouped-histograms-telematics}, the predictive power (for claim occurrence) of some of the telematics covariates may be undermined.  As will be shown in Section \ref{NBGLM}, common summary statistics of speed, such as average and maximum, have reduced predictive power especially when summarized across the entire policy period (i.e., over multiple trips).  This suggests the need of a more flexible feature engineering and extraction for telematics data if we would like to make full use of them for auto-insurance ratemaking.  For example, the proposed speed transition matrix takes into account all trips, all speed records, and the connections between them (in terms of transitions) over the whole policy period.  As the only restriction is each row (i.e., the starting speed bin) sums to unity, it allows for multiple concentrated spots (i.e., with the same weights/probabilities) and hence can capture a variety of patterns in individual driving behaviour.  For illustrative purposes, we show in Figure \ref{fig: sample-transition-matrix-heatmap} two considerably different driving patterns.  While both drivers have not reported claims in their policy periods, the trip-based mean and maximum speed empirical densities of Driver B have more pronounced multimodality than those of Driver A.  This situation is captured and reflected by the proposed speed transition matrix, as illustrated by the heatmaps on the right: Driver A's matrix is very concentrated on the diagonal starting at 90 km/h, while Driver B has quite a diversified transition pattern at 120 km/h.  Moreover, the green grid from (100, 110] to (60, 70] indicates a relatively frequent deceleration at such a higher speed.

\begin{figure*}[h]
\caption{Sample Driving Behaviour I - Over Time: Left - Mean Speed, Mid - Maximum Speed, Right - Trip Starting Hour}
    \centering
    \begin{subfigure}[b]{0.31\textwidth}
        \centering
        \includegraphics[width=\textwidth]{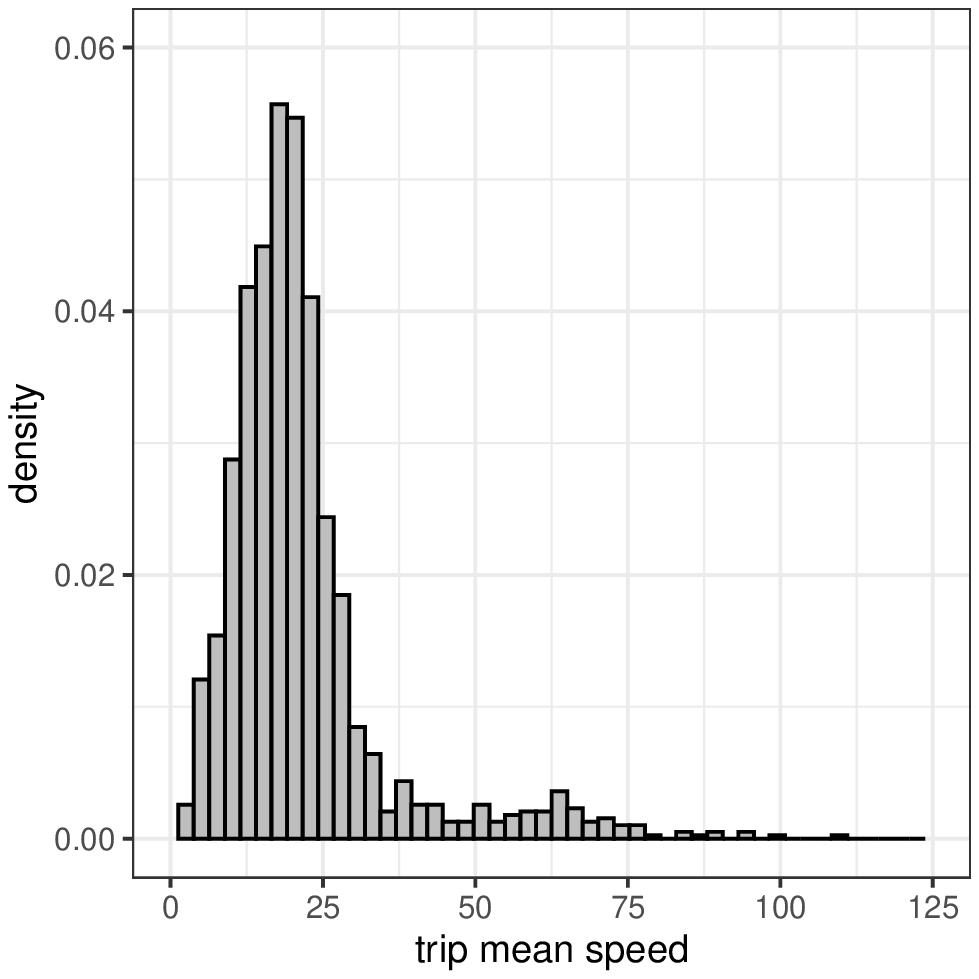}
        \caption*{}
    \end{subfigure}
    \hfill
    \begin{subfigure}[b]{0.31\textwidth}  
        \centering 
        \includegraphics[width=\textwidth]{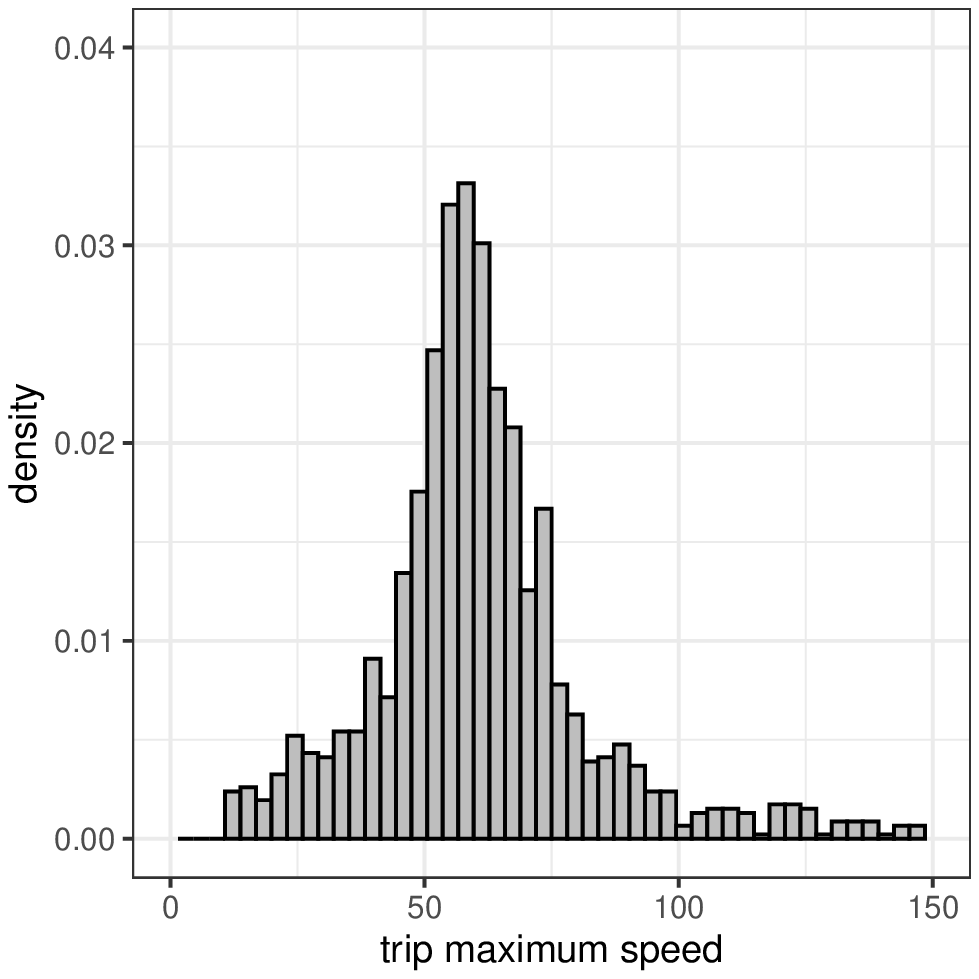}
        \caption{Driver C in 2018 - 0 Claims}
    \end{subfigure}
    \hfill
    \begin{subfigure}[b]{0.31\textwidth}  
        \centering 
        \includegraphics[width=\textwidth]{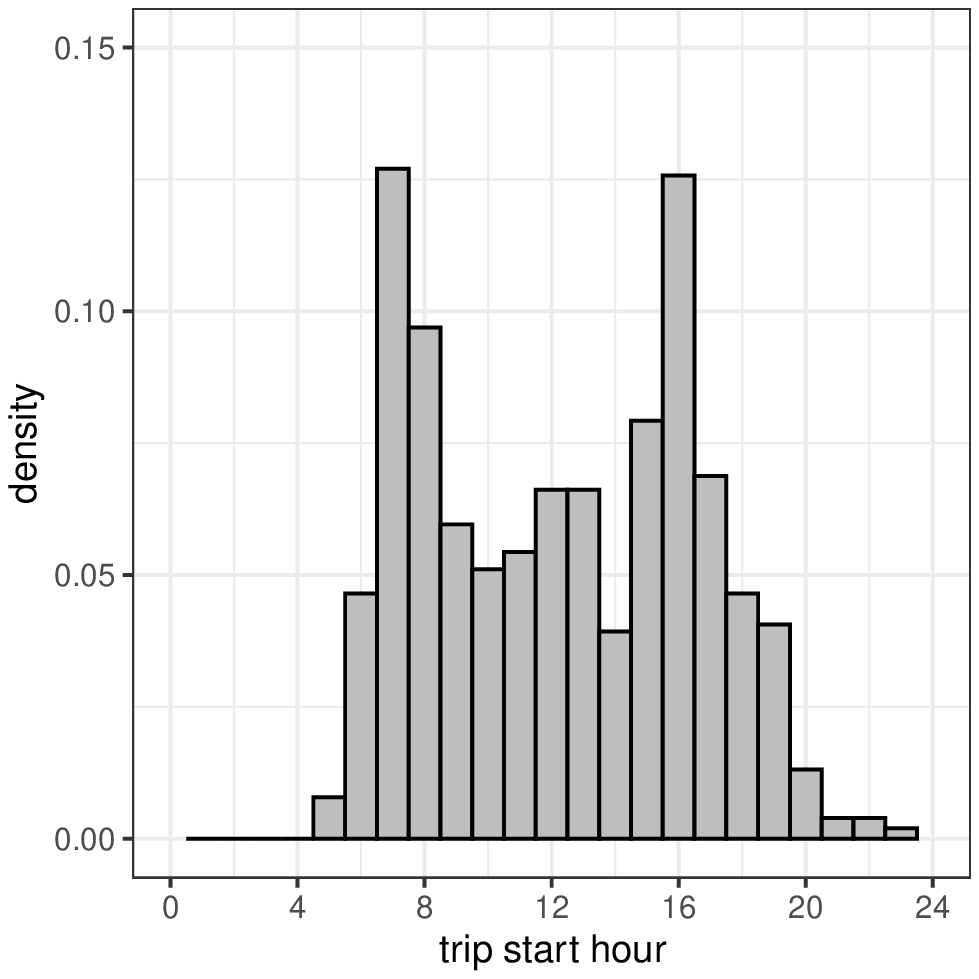}
        \caption*{}
    \end{subfigure}
    \vskip\baselineskip
        \begin{subfigure}[b]{0.31\textwidth}
        \centering
        \includegraphics[width=\textwidth]{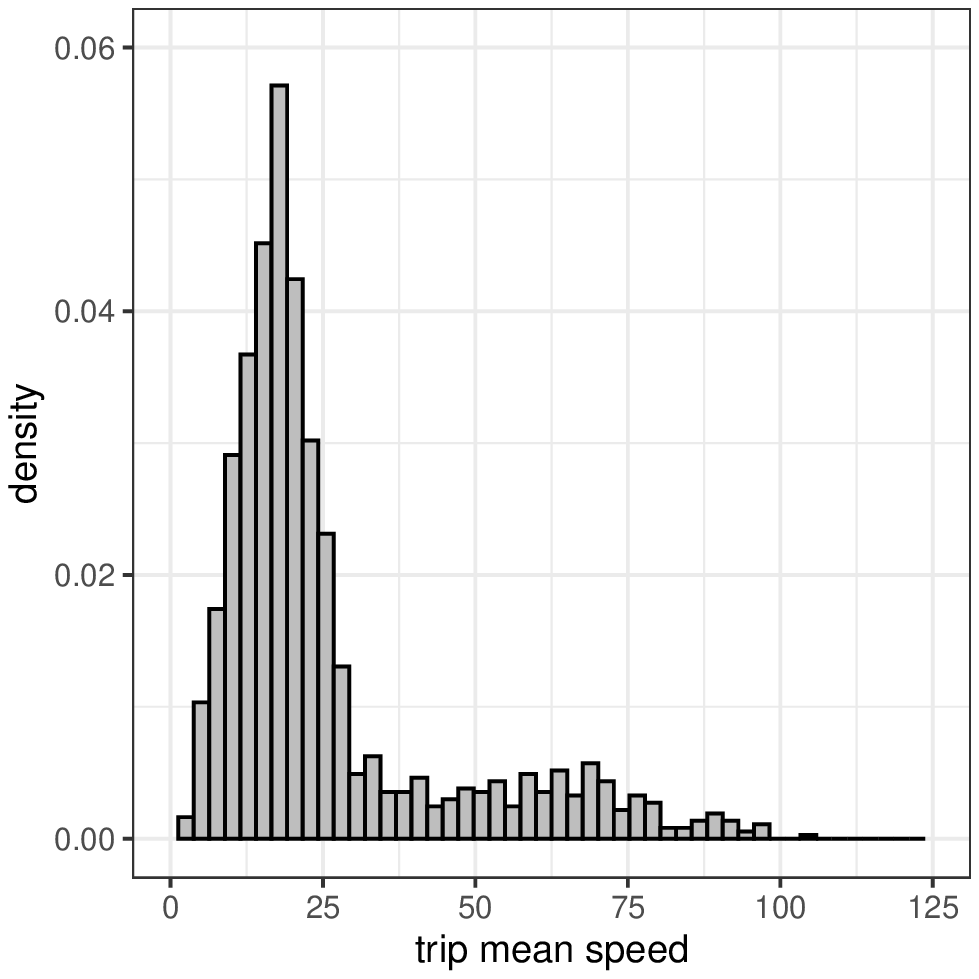}
        \caption*{}
    \end{subfigure}
    \hfill
    \begin{subfigure}[b]{0.31\textwidth}  
        \centering 
        \includegraphics[width=\textwidth]{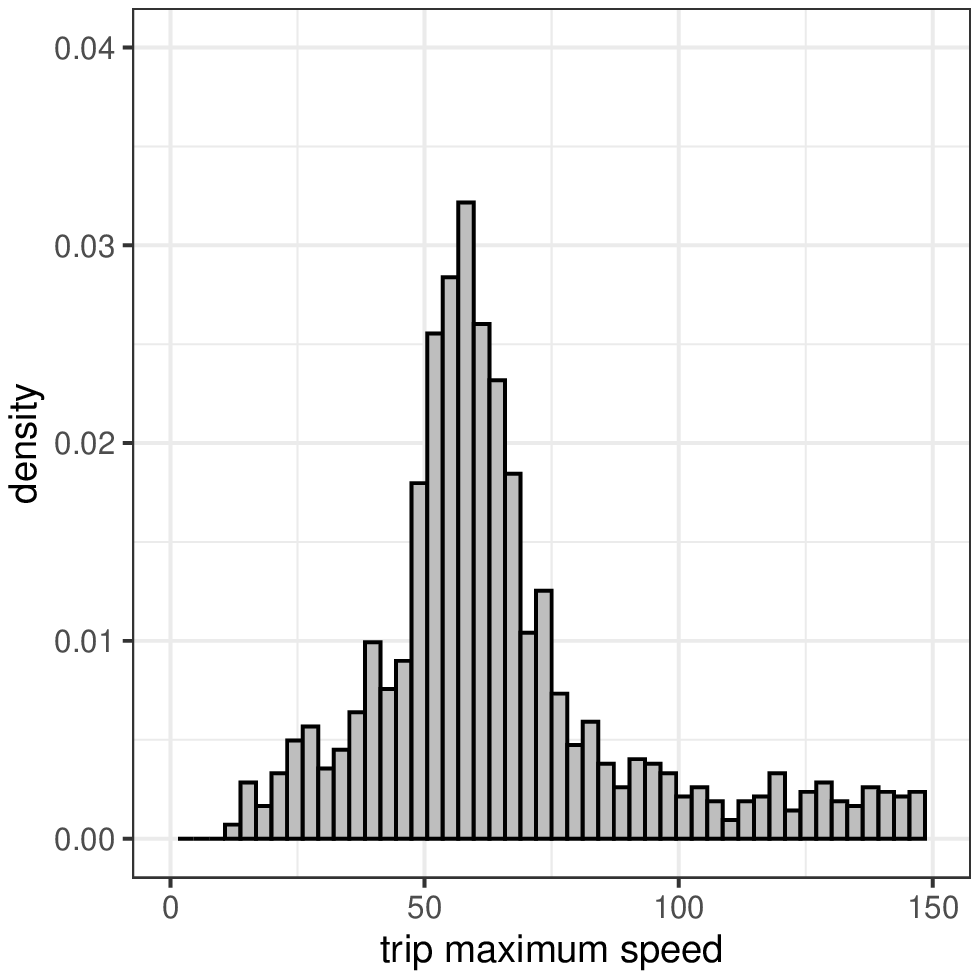}
        \caption{Driver C in 2019 - 2 Claims}
    \end{subfigure}
    \hfill
    \begin{subfigure}[b]{0.31\textwidth}  
        \centering 
        \includegraphics[width=\textwidth]{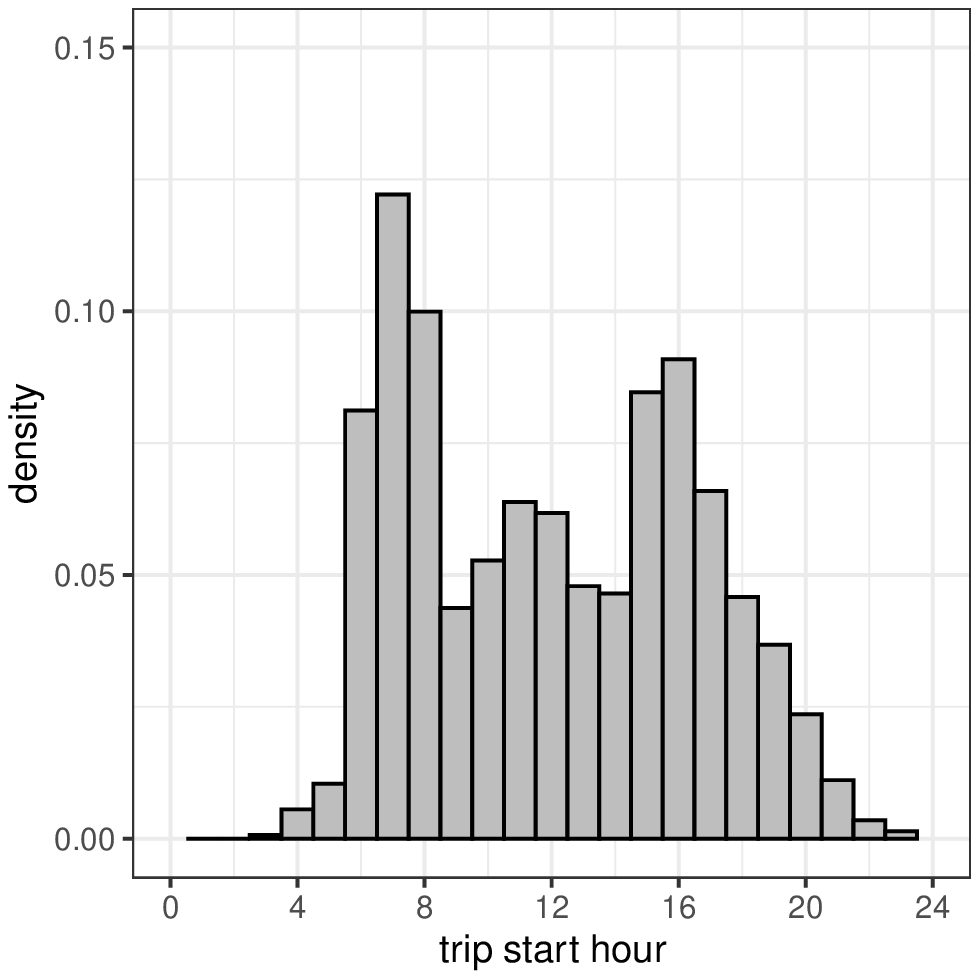}
        \caption*{}
    \end{subfigure}
    \vskip\baselineskip
        \begin{subfigure}[b]{0.31\textwidth}
        \centering
        \includegraphics[width=\textwidth]{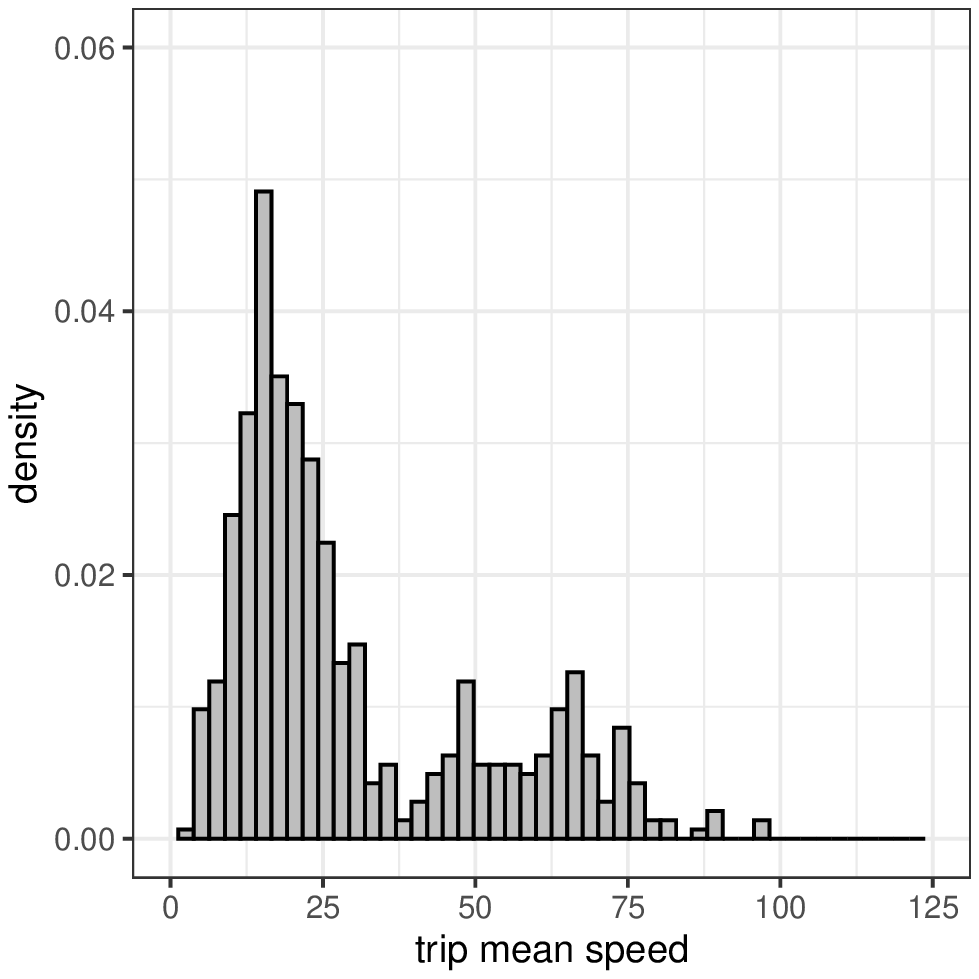}
        \caption*{}
    \end{subfigure}
    \hfill
    \begin{subfigure}[b]{0.31\textwidth}  
        \centering 
    \includegraphics[width=\textwidth]{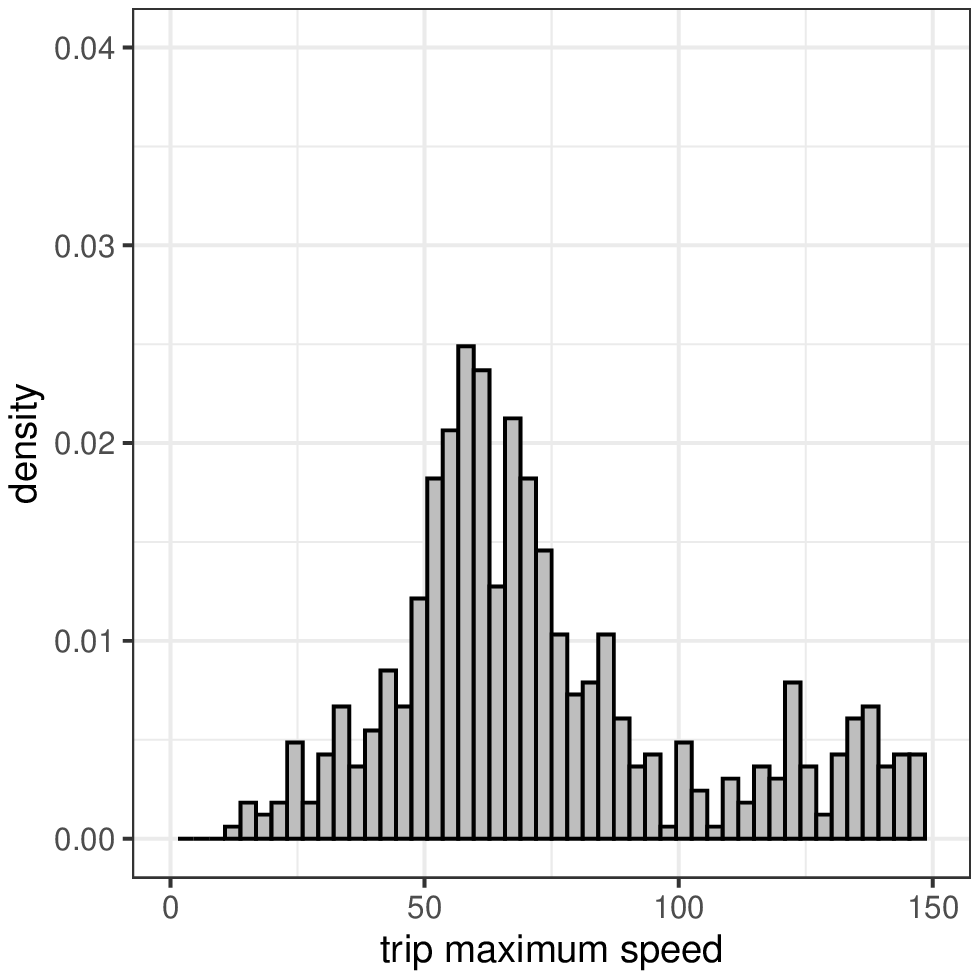}
        \caption{Driver C in 2020 - 0 Claims}
    \end{subfigure}
    \hfill
    \begin{subfigure}[b]{0.31\textwidth}  
        \centering 
    \includegraphics[width=\textwidth]{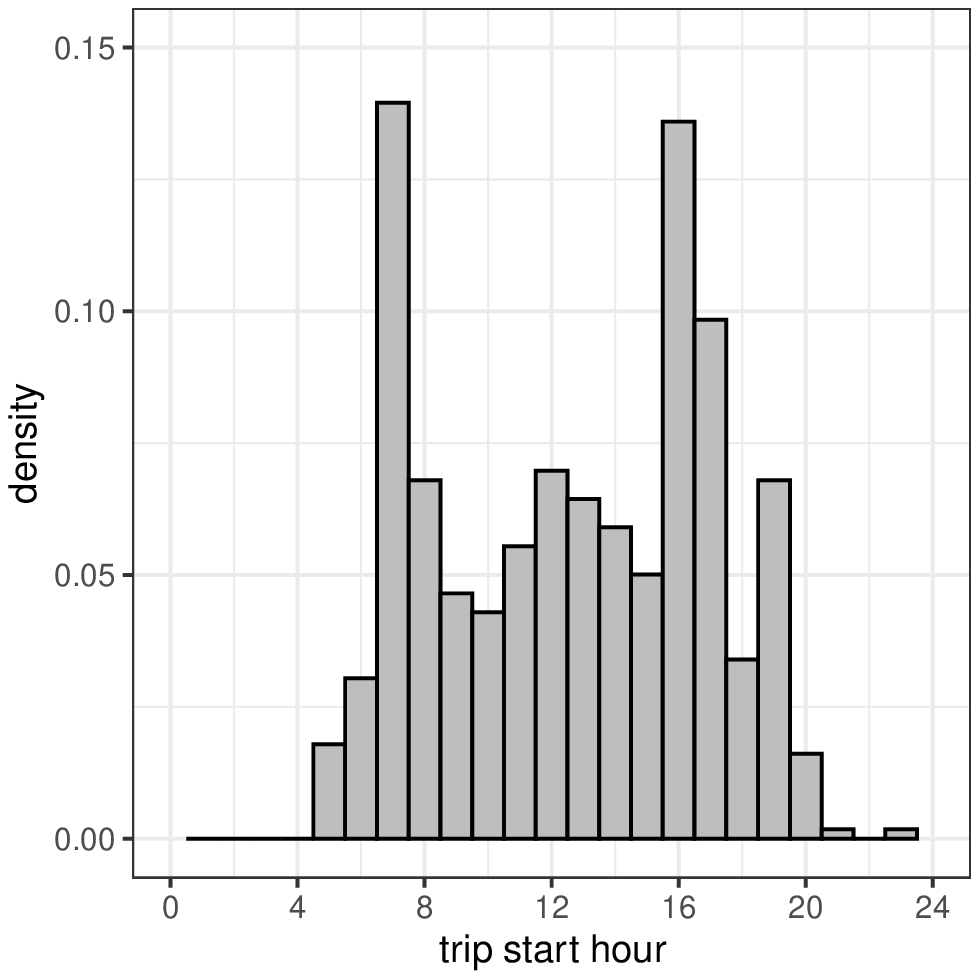}
        \caption*{}
    \end{subfigure}
    \label{fig: sample-individual-3}
\end{figure*}

\begin{figure*}[h]
\caption{Sample Driving Behaviour II - Over Time: Left - Mean Speed, Mid - Maximum Speed, Right - Trip Starting Hour}
    \centering
    \begin{subfigure}[b]{0.31\textwidth}
        \centering
        \includegraphics[width=\textwidth]{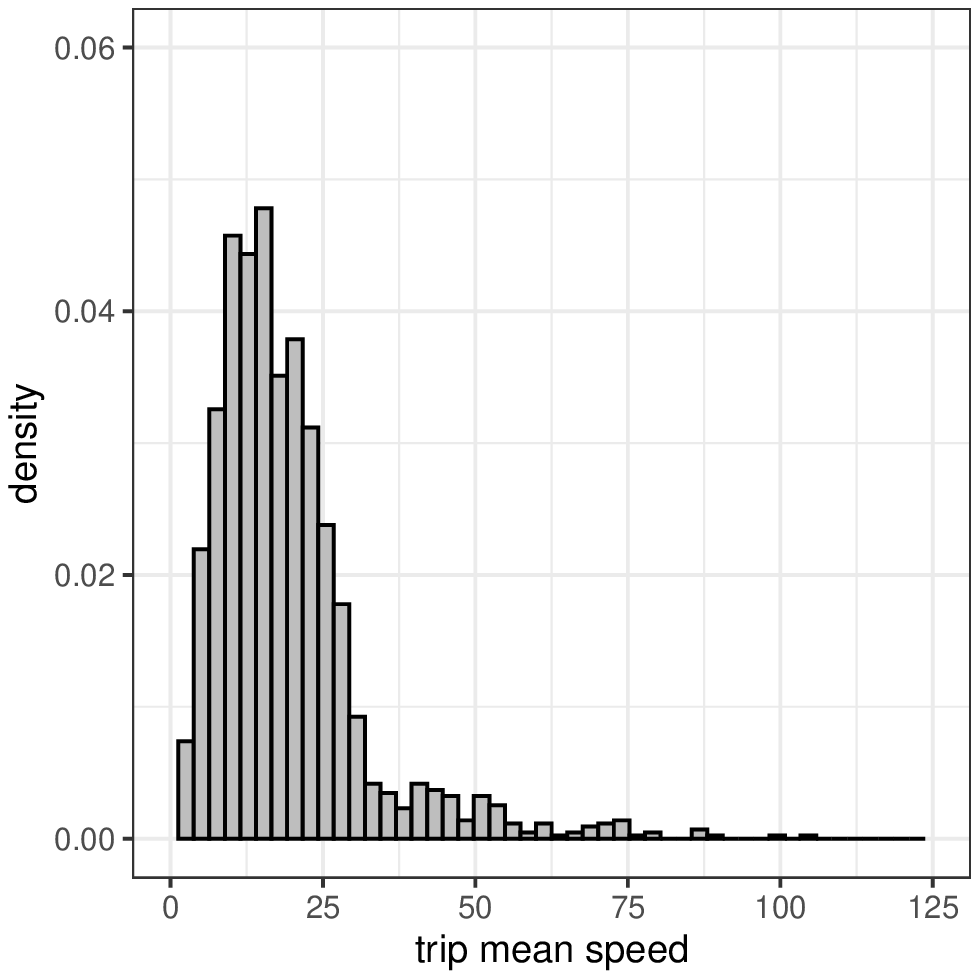}
        \caption*{}
    \end{subfigure}
    \hfill
    \begin{subfigure}[b]{0.31\textwidth}  
        \centering 
        \includegraphics[width=\textwidth]{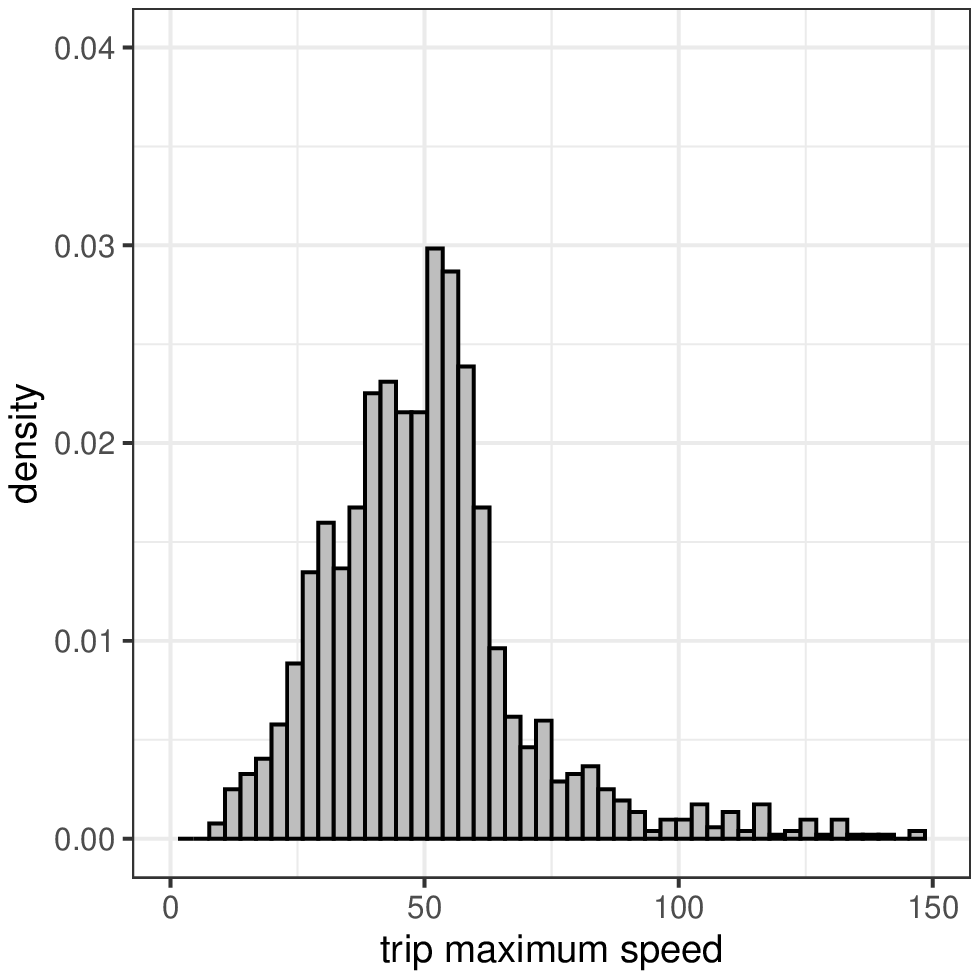}
        \caption{Driver D in 2018 - 0 Claim}
    \end{subfigure}
    \hfill
    \begin{subfigure}[b]{0.31\textwidth}  
        \centering 
        \includegraphics[width=\textwidth]{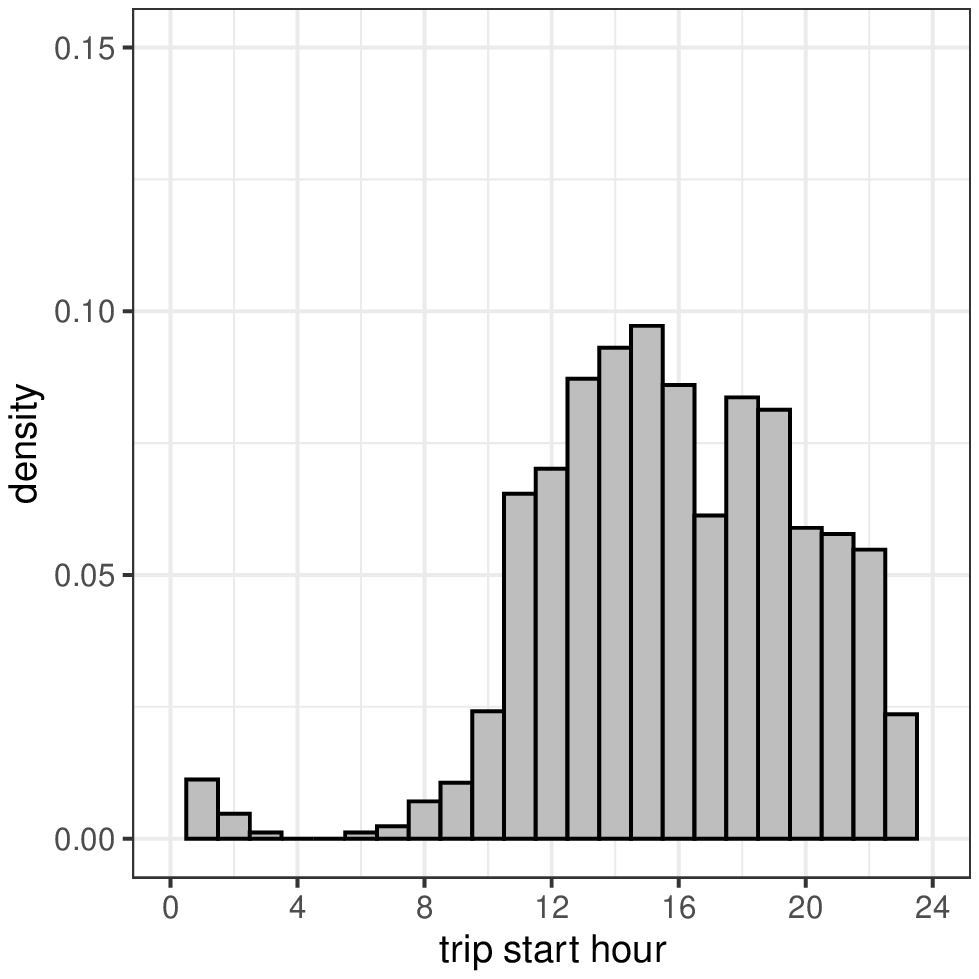}
        \caption*{}
    \end{subfigure}
    \vskip\baselineskip
        \begin{subfigure}[b]{0.31\textwidth}
        \centering
        \includegraphics[width=\textwidth]{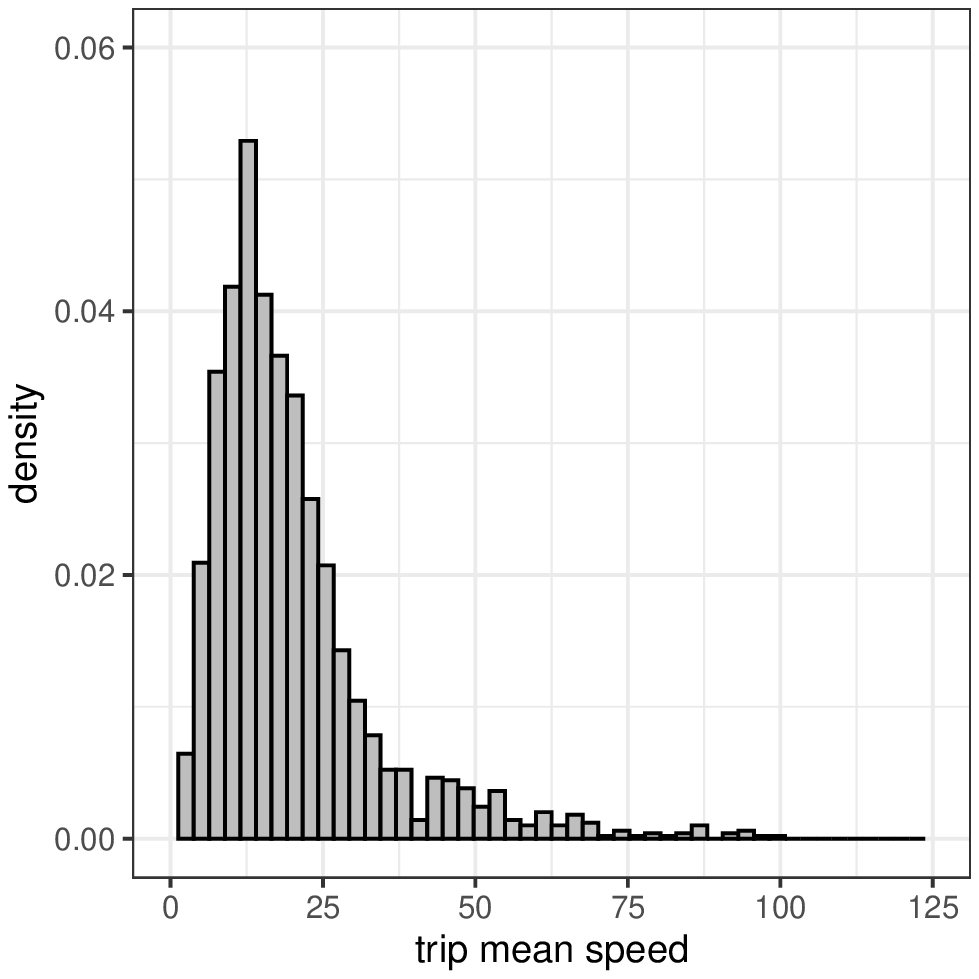}
        \caption*{}
    \end{subfigure}
    \hfill
    \begin{subfigure}[b]{0.31\textwidth}  
        \centering 
        \includegraphics[width=\textwidth]{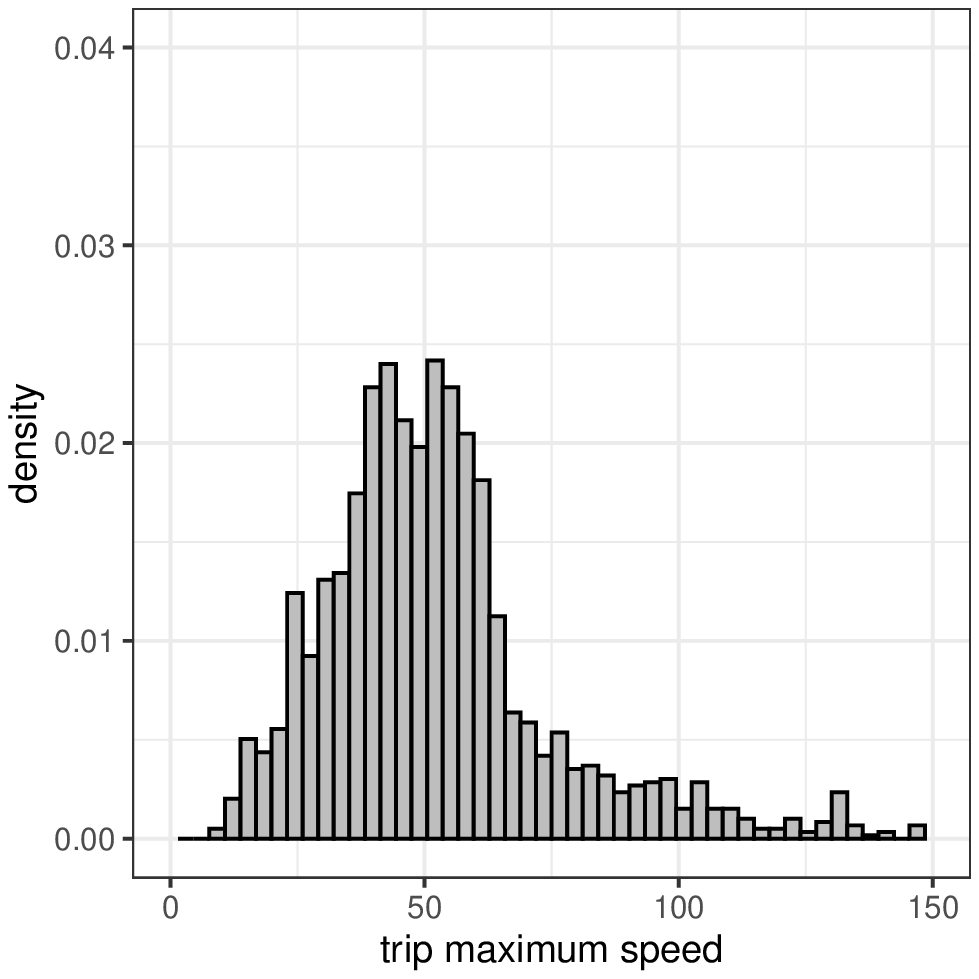}
        \caption{Driver D in 2019 - 0 Claim}
    \end{subfigure}
    \hfill
    \begin{subfigure}[b]{0.31\textwidth}  
        \centering 
        \includegraphics[width=\textwidth]{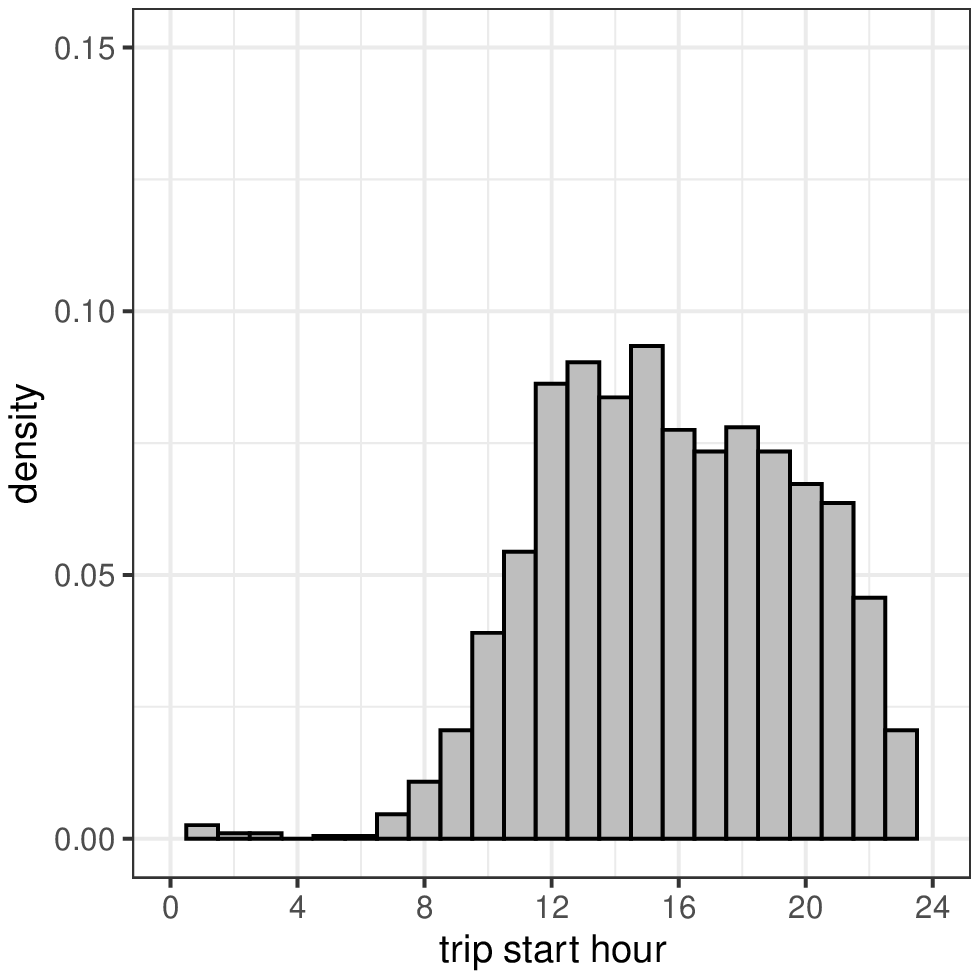}
        \caption*{}
    \end{subfigure}
    \vskip\baselineskip
        \begin{subfigure}[b]{0.31\textwidth}
        \centering
        \includegraphics[width=\textwidth]{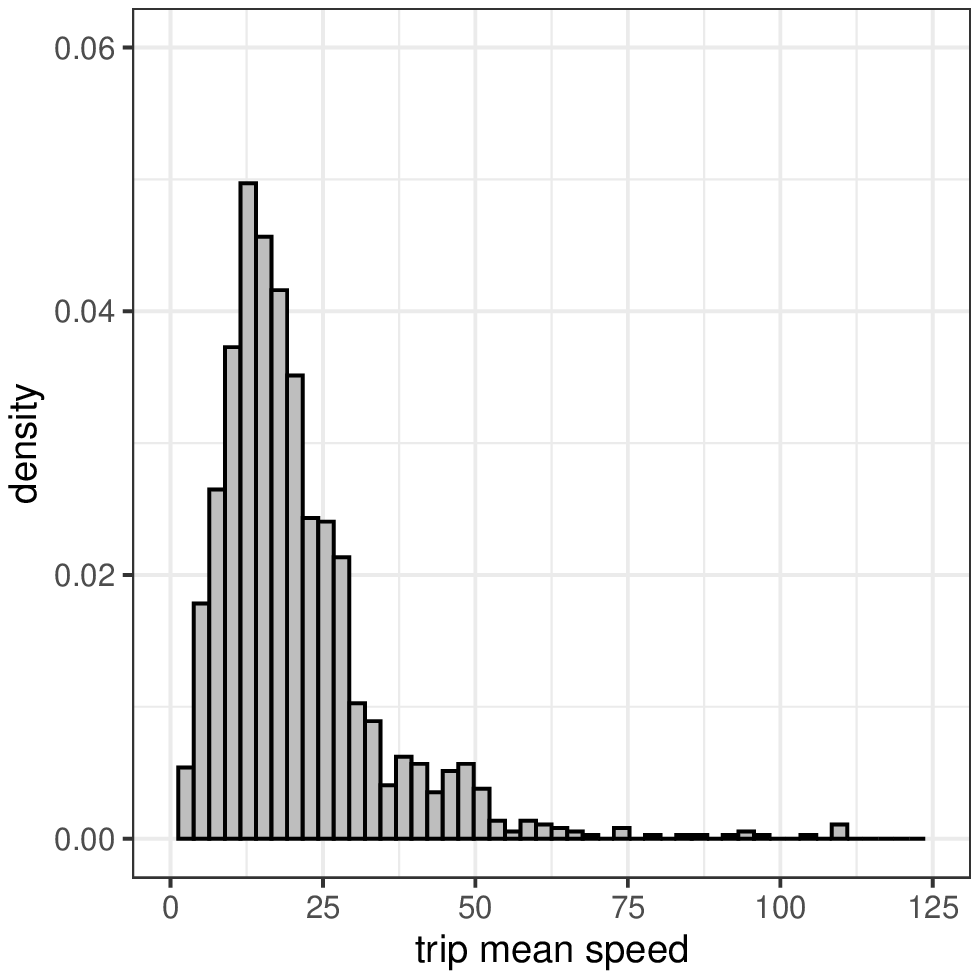}
        \caption*{}
    \end{subfigure}
    \hfill
    \begin{subfigure}[b]{0.31\textwidth}  
        \centering 
    \includegraphics[width=\textwidth]{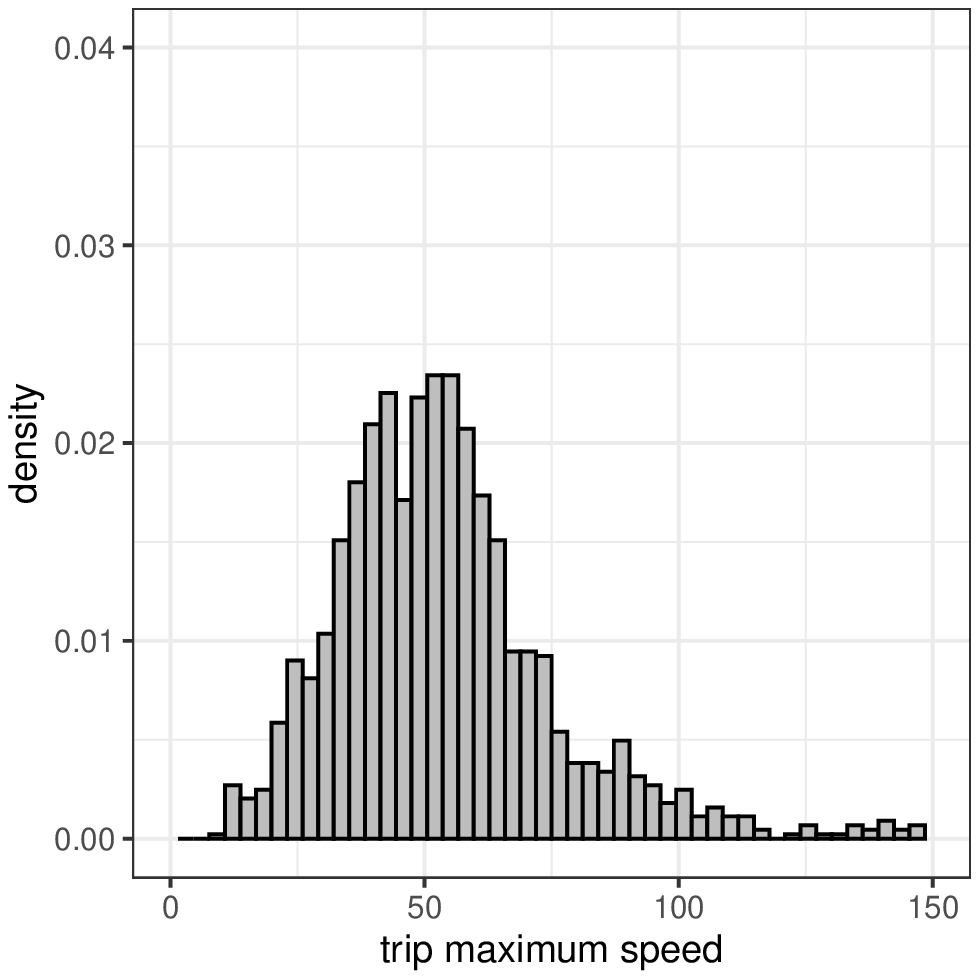}
        \caption{Driver D in 2020 - 0 Claim}
    \end{subfigure}
    \hfill
    \begin{subfigure}[b]{0.31\textwidth}  
        \centering 
    \includegraphics[width=\textwidth]{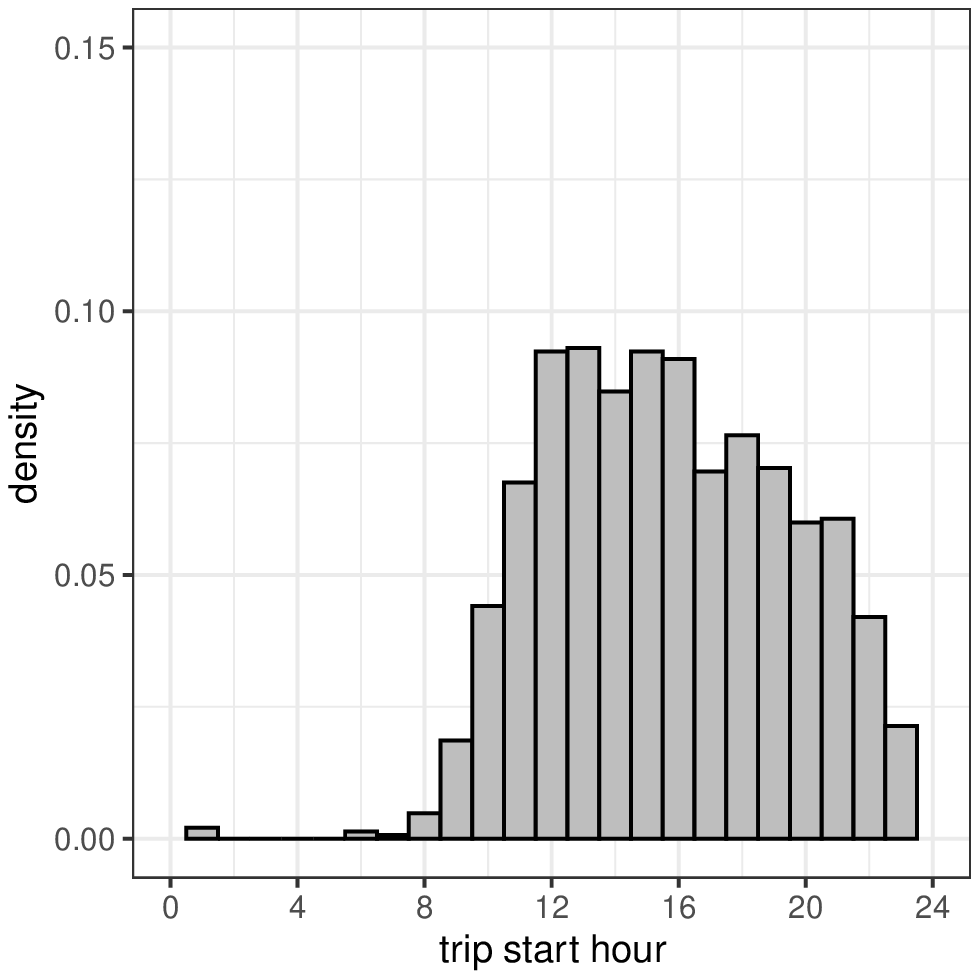}
        \caption*{}
    \end{subfigure}
    \label{fig: sample-individual-4}
\end{figure*}

Moreover, Figures \ref{fig: sample-individual-3} and \ref{fig: sample-individual-4} show the driving behaviour over three policy periods of two selected drivers, respectively.  For each driver, while the histograms have similar shapes throughout the years, representing driving habit and regular travel schedule, they are not exactly the same, demonstrating some variation in driving behaviour.  Hence, simply using historical statistics such as averages of telematics covariates to predict future claims can lead to biased results (\citet{fang_mocha_2021}).  This suggests the use of a time-series framework.

\section{Negative Binomial Regression and Drivers' Learning Effects}
\label{NBGLM}
 
This section analyzes the relationship between claim counts (response) and various covariates, both the traditional and the telematics ones.  In particular we will consider the Negative Binomial regression and perform feature selection via a 5-fold cross-validation.

\subsection{Empirical Evidence of Drivers' Learning Effects from Telematics Data}
\label{Empirical-Evidence-of-Drivers'-Learning-Effects-from-Telematics-Data}

Before we model the claim counts, we first study the harsh event arrivals to understand how driving behaviour is changing.  Harsh driving events may be considered as `near-misses'.  A near-miss can be defined as `sudden braking and rapid steering operations by the driver without resulting an accident' (\citet{arai_accidents_2001}).  Aside from modelling claim probability and claim counts with near-misses, several literatures have also considered the modelling of near-misses, considering them as proxies for the rarely observed claim arrivals (\citet{guillen_can_2020}, \citet{sun_assessing_2020}, \citet{sun_driving_2021}).  A good definition of near-miss events is itself a research topic, and questions include what the detection threshold of harsh events should be and how closely related the near-misses and claims are.  In our data, harsh driving events and their detection thresholds are predetermined.  Our empirical results show that it takes increasingly longer for a severe harsh event to arrive as total driving time increases, regardless of age, while there is little to no change for other harsh events detected at a lower threshold.  This may be how learning effect - the idea that as drivers spend more time on the road, they become more experienced and can react to road conditions better - is reflected in driving behaviour.  While we have not found a formal definition of learning effect in the literature, in our analysis, we define it as the decreasing rate of severe harsh event arrivals as cumulative driving time (or distance) increases.

\textit{Remark: To the best of our knowledge, the idea of learning effect was first suggested by \citet{boucher_exposure_2017} in the actuarial context, but it has been later rejected by a longitudinal model proposed in \citet{boucher_longitudinal_2020}.  While our data is not a panel data and it does not contain the policyholders' prior driving experience, we could not provide a definite statement about the validity of those conclusions. However, as it will be discussed below, a cubic spline on (logged) total driving time (see Figure \ref{fig: cubic-spline}) confirms that a linear relationship between claims and logged driving time/distance should be sufficient.  Although we do not have information on drivers' prior experience, our dataset contains mainly drivers who are driving for the first time with the newly installed telematics tracking devices.  Hence it is more conservative to say that we are studying the learning effect since (telematics) device installation, and this should be less dependent on how long the drivers have driven before.  Being aware that they are being monitored and their behavior is recorded will definitely influence drivers' behavior, especially for those who are more sensitive to increases in insurance premiums.}

For each policy, we record the rank, type and cumulative driving time of each detected harsh event.  We have five types of harsh events: harsh acceleration, harsh braking, left cornering, right cornering and severe harsh events.  Recall that the first four types are detected at a lower threshold of 0.5G and the last type is detected at 1.5G.  For each event type $j$, we perform a linear regression on the log-log scale:

\begin{align*} 
\ln(\hat y_{ijk}) &= \hat \alpha_j + \hat \beta_j \ln(t_{ijk}) \\
\hat y_{ijk} &= (t_{ijk})^{\hat \beta_j}\times \exp(\hat \alpha_j)
\end{align*}

where for event type $j$, $\hat{y}_{ijk}$ is the expected rank of the $k$-th harsh event of policyholder $i$, which is also the expected number of harsh events up till time $t_{ijk}$, and $t_{ijk}$ is his/her cumulative driving time at the occurrence of the $k$-th event.  We also highlight the fact that $y_{ijk}$ equals $k$ if policyholder $i$ has at least $k$ harsh events of type $j$, and does not exist otherwise; whilst $\hat{y}_{ijk}$ does not necessarily equal to $k$ as it is the predicted value from the fitted regression.  While linear regression is not the best model for discrete event numbers, our aim here is merely to fit a trend between harsh event arrivals and cumulative driving time.  As the total number of harsh events can only increase with time, the coefficient $\hat \beta_j$ must be positive.  In particular, if $\hat \beta_j$ is close to 1, the relationship between harsh event arrivals and total driving time is directly proportional, whilst if $\hat \beta_j$ is significantly less than 1, it takes increasingly longer for an event to arrive as total driving time increases which demonstrates some learning effect.

The fitted coefficients for each type of harsh event are shown in Table \ref{table: harsh-event-learning-effect}.  As the telematics observation period for each policy varies, each policyholder's total number of harsh events also varies.  To ensure robustness of our analysis, we filter for the event numbers where there are at least five occurrences (e.g. $y_{ij,1000} = 1000$ is retained if at least five policies have reached 1000 harsh event type $j$).  We observe that $\hat \beta_j$ range between 0.78 and 1.03 for the first four types of harsh events, while they are significantly smaller for severe events, ranging between 0.26 and 0.49.  Moreover, we also investigate if there is a difference in the behaviour of drivers with and without claims.  While the general trends are similar, we observe that the coefficients fitted from policies with claims are all larger than those fitted from policies with no claims.

We can quantify the learning effect by means of the second derivative of the power function for the number of severe harsh events.  The second derivative takes the form:

\begin{align*} 
\hat{f}''(t) &= t^{\hat \beta - 2}\times (\exp(\hat \alpha_j) \hat \beta(\hat \beta - 1)) = \hat c \times t^{\hat p}
\end{align*}

where $t$ is cumulative driving time, and $\hat \alpha$ and $\hat \beta$ are the fitted intercept and coefficient, respectively, from the regression of severe events.  Since $\hat \beta$ is less than 1, the second derivative is a negative function and specifies the rate of decrease in severe event arrivals as cumulative driving time increases.  In Table \ref{table: quantify-learning-effect} we report the coefficient $\hat c = \exp(\hat \alpha_j) \hat \beta(\hat \beta - 1)$ and power $\hat p = \hat \beta - 2$, with cumulative driving time in units of hours.  It is noteworthy that the learning effect demonstrated by policies with claims is smaller than that without claims, as reflected by a smaller absolute value of both the coefficient and power.

To answer the question on whether more driving can continue to decrease severe harsh event rate for experienced drivers, we also compare the learning effects of different ages.  Naturally we assume that older drivers possess more driving experience.  We focus on policies with a valid driver's age (63\% of our data) and categorize the policies into three age groups: `young' (below 40), `mid' (40-59), and `old' (60 and above).  For each event type $j$, we perform a linear regression on the log-log scale with an interaction term of cumulative driving time and age group:

\begin{align*} 
\ln(\hat y_{ijk}) &= \hat \alpha_j + \ln(t_{ijk})\left(\sum_{a \in \{young, mid, old\}}\hat \beta_{ja} \times \mathbbm{1}_{\{age=a\}}\right)
\end{align*}

where $\mathbbm{1}_{\{age=a\}} = 1$ is an indicator variable equal to 1 if the age group is $a$ and 0 otherwise.  The fitted coefficients $\hat \beta_{ja}$ of the interaction term follow the same interpretation as $\hat \beta_{j}$, and the results are shown in Table \ref{table: harsh-event-learning-effect-age}.  Similar observations are made: first, $\hat \beta_{ja}$ range between larger values (0.69 to 1.13) for the first four types of harsh events, while they are much smaller for severe events (0.17 to 0.62); second, the coefficients fitted from policies with claims are all larger than those fitted from policies with no claims.  More importantly, for a given harsh event type $j$, $\hat \beta_{ja}$ have similar magnitudes across age groups.  This is particularly important for severe harsh events, indicating the presence and consistency of learning effect regardless of age.

In conclusion, the results above can be interpreted as learning effect.  As drivers spend more time on the road, they become more experienced and can react to road conditions better.  As a consequence, there are less severe harsh events occurring as cumulative driving time increases, and this effect holds true across all age groups.  However, no matter how experienced a driver is, not all harsh events can be avoided in reality, and some of them may actually be necessary, such as harsh braking to avoid a collision (\citet{guillen_near-miss_2021}).  Hence, we do not see a change as significant for the harsh events detected at a lower threshold.  Moreover, we observe that riskier drivers (those who filed at least one claim) demonstrate a smaller learning effect than safer drivers in general.  Their more aggressive and unimproved driving behaviour may be one of the many factors that contributes to claim occurrence.

\begin{table}[h]
\caption{Fitted Coefficient $\hat \beta_j$ for Each Type of Harsh Event for All Policies: a coefficient close to 1 shows little to no learning effect, while a coefficient significantly less than 1 demonstrates existence of learning effect.  Rows indicated as $\geq 5$ refer to the event numbers where there are at least five occurrences.}
\centering
\begin{threeparttable}
\begin{tabular}{ll|rrrrr}
 &  & \multicolumn{1}{l}{\textbf{Acceleration}} & \multicolumn{1}{l}{\textbf{Deceleration}} & \multicolumn{1}{l}{\textbf{Left Cornering}} & \multicolumn{1}{l}{\textbf{Right Cornering}} & \multicolumn{1}{l}{\textbf{Severe Event}} \\ \hline\hline
\multirow{2}{*}{\textbf{All Policies}} & \textbf{all} & 0.9455 & 0.9390 & 1.0272 & 0.9312 & 0.4430 \\
 & \textbf{$\geq 5$} & 0.7920 & 0.9101 & 0.8925 & 0.7789 & 0.2872 \\ \hline
\multirow{2}{*}{\textbf{No Claim}} & \textbf{all} & 0.9058 & 0.9142 & 0.9056 & 0.8308 & 0.4291 \\
 & \textbf{$\geq 5$} & 0.7340 & 0.8664 & 0.8144 & 0.7576 & 0.2632 \\ \hline
\multirow{2}{*}{\textbf{Claimed}} & \textbf{all} & 0.9831 & 0.9591 & 1.1361 & 1.0349 & 0.4878 \\
 & \textbf{$\geq 5$} & 0.8449 & 0.9526 & 0.9646 & 0.7976 & 0.3630 \\ \hline
\end{tabular}
\begin{tablenotes}
\small
\item Under the Wald test, all estimated coefficients are statistically different from 1 with significance level of 0.5\%.
\end{tablenotes}
\end{threeparttable}
\label{table: harsh-event-learning-effect}
\end{table}

\begin{table}[h]
\caption{Coefficient $\hat c$ and Power $\hat p$ for the Second Derivative of the Power Function of Severe Harsh Events.  Cumulative driving time is in unit of hours.  Rows indicated as $\geq 5$ refer to the event numbers where there are at least five occurrences.}
\centering
\begin{tabular}{ll|rr}
 &      & \multicolumn{1}{c}{$\hat c$} & \multicolumn{1}{c}{$\hat p$}\\ \hline\hline
\multirow{2}{*}{\textbf{All Policies}} & \textbf{all} & -0.1903&-1.5571\\
 & \textbf{$\geq 5$} & -0.2446&-1.7128\\ \hline
\multirow{2}{*}{\textbf{No Claim}} & \textbf{all} & -0.2037&-1.5709\\
 & \textbf{$\geq 5$} & -0.2601&-1.7368\\ \hline
\multirow{2}{*}{\textbf{Claimed}} & \textbf{all} & -0.1497&-1.5122\\
 & \textbf{$\geq 5$} & -0.1872&-1.6370\\ \hline
\end{tabular}
\label{table: quantify-learning-effect}
\end{table}

\begin{table}[h]
\caption{Fitted Coefficient $\hat \beta_ja$ for Each Type of Harsh Event and Each Age Group for Policies with a Valid Driver's Age: a coefficient close to 1 shows little to no learning effect, while a coefficient significantly less than 1 demonstrates existence of learning effect.  Rows indicated as $\geq 5$ refer to the event numbers where there are at least five occurrences.  Age groups are: `young' (below 40), `mid' (40-59), and `old' (60 and above).}
\centering

\begin{subtable}[t]{\textwidth}
\begin{tabular}{ll|rrrrr}
 &  & \multicolumn{1}{l}{\textbf{Acceleration}} & \multicolumn{1}{l}{\textbf{Deceleration}} & \multicolumn{1}{l}{\textbf{Left Cornering}} & \multicolumn{1}{l}{\textbf{Right Cornering}} & \multicolumn{1}{l}{\textbf{Severe Event}} \\ \hline\hline
\multirow{2}{*}{\textbf{All Policies}} & \textbf{all} & 1.0357& 0.8670& 0.8997& 0.7657& 0.3122\\
 & \textbf{$\geq 5$} & 0.8628& 0.8577& 0.8653& 0.7365& 0.2021\\ \hline
\multirow{2}{*}{\textbf{No Claim}} & \textbf{all} & 0.8890& 0.8473& 0.7964& 0.7168& 0.2029\\
 & \textbf{$\geq 5$} & 0.8261& 0.8371& 0.7812& 0.6985& 0.1813\\ \hline
\multirow{2}{*}{\textbf{Claimed}} & \textbf{all} & 1.1104& 0.8684& 0.9930& 0.8129& 0.5654\\
 & \textbf{$\geq 5$} & 0.8695& 0.8597& 0.9423& 0.7699& 0.2956\\ \hline
\end{tabular}
\end{subtable}

 \vspace*{5mm}

\begin{subtable}[t]{\textwidth}
\begin{tabular}{ll|rrr|rrr}    
  && \multicolumn{3}{c|}{\textbf{Acceleration}}& \multicolumn{3}{c}{\textbf{Deceleration}}\\\hline\hline
  && young & mid & old & young & mid & old \\\hline
 \multirow{2}{*}{\textbf{All Policies}}&\textbf{all} & 1.0290 & 1.0281 & 1.0532 & 0.8697 & 0.8599 & 0.8722 \\
 &\textbf{>=5} & 0.8551 & 0.8457 & 0.8855 & 0.8596 & 0.8511 & 0.8632 \\\hline
 \multirow{2}{*}{\textbf{No Claim}} &\textbf{all}& 0.8795 & 0.8893 & 0.8425 & 0.8469 & 0.8472 & 0.8392 \\
 &\textbf{>=5} & 0.8162 & 0.8283 & 0.7854 & 0.8365 & 0.8375 & 0.8301 \\\hline
 \multirow{2}{*}{\textbf{Claimed}} &\textbf{all}& 1.0992 & 1.0958 & 1.1266 & 0.8728 & 0.8496 & 0.8812 \\
 &\textbf{>=5} & 0.8378 & 0.8004 & 0.8763 & 0.8638 & 0.8428 & 0.8737 \\\hline
\end{tabular}
\end{subtable}

\vspace*{5mm}

\begin{subtable}[t]{\textwidth}
\begin{tabular}{ll|rrr|rrr|rrr} 
  &&\multicolumn{3}{c|}{\textbf{Left Cornering}}& \multicolumn{3}{c|}{\textbf{Right Cornering}}& \multicolumn{3}{c}{\textbf{Severe Event}}\\\hline\hline
  &&young & mid & old & young & mid & old & young & mid & old \\\hline
  \multirow{2}{*}{\textbf{All Policies}}&\textbf{all} 
&0.8933 & 0.8748 & 0.9288 & 0.7612 & 0.7521 & 0.7854 & 0.3297 & 0.2801 & 0.2914 \\
  &\textbf{>=5} &0.8638 & 0.8469 & 0.8962 & 0.7354 & 0.7284 & 0.7551 & 0.2026 & 0.2025 & 0.1966 \\\hline
  \multirow{2}{*}{\textbf{No Claim}} &\textbf{all}
&0.7975 & 0.8032 & 0.8367 & 0.7131 & 0.7230 & 0.7186 & 0.2068 & 0.1984 & 0.1887 \\
  &\textbf{>=5} &0.7833 & 0.7876 & 0.8152 & 0.6929 & 0.7058 & 0.7017 & 0.1816 & 0.1816 & 0.1722 \\\hline
  \multirow{2}{*}{\textbf{Claimed}} &\textbf{all}
&1.0052* & 0.9532 & 1.0262 & 0.7935 & 0.7524 & 0.8216 & 0.6181 & 0.5187 & 0.5498 \\
  &\textbf{>=5} &0.9631 & 0.9159 & 0.9848* & 0.7652 & 0.7258 & 0.7861 & 0.3003 & 0.2951 & 0.2980 \\\hline
\end{tabular}
\end{subtable}
\label{table: harsh-event-learning-effect-age}
\end{table}

\subsection{Motivation for Using the Negative Binomial}
\label{Motivation-for-Using-the-Negative-Binomial}

While various literatures have employed the Poisson regression in modelling claim counts (\citet{boucher_exposure_2017}, \citet{ayuso_improving_2019}, \citet{huang_automobile_2019}, \citet{guillen_near-miss_2021}, \citet{gao_boosting_2022}), we propose to use the Negative Binomial regression instead.  First, the choice of Negative Binomial distribution is motivated by the existence of zero-inflation in the claim data, which will lead to an overdispersion with respect to Poisson distribution.  Our claim count has a sample mean of 0.4465 and a sample variance of 0.7909.  Clearly the widely used Poisson regression is not a proper candidate model in this situation.  Second, the use of Generalized Linear Model (GLM) instead of more flexible models such as a Generalized Additive Model (GAM) is to prevent overfitting, especially when the number of observations is limited compared to the number of covariates.

\textit{Remark: The seemingly high claims frequency and its variance can be attributed to two reasons.  First, our dataset originates from a major European insurer with all policies written in Romania, where the claims experience, road conditions, and safety levels may differ from those in North America.  Second, the number of claims comprises both the motor third-party liability (MTPL) and motor damage coverages, with claims frequencies for the two reaching as high as 8\% and 30\%, respectively, in European countries (as of 2016). \footnote{Charts 37-38 in European Motor Insurance Markets prepared by Insurance Europe.  See https://www.insuranceeurope.eu/publications/465/european-motor-insurance-markets/} \footnote{News article by Agerpres, Romania's national news agency.  See https://www.agerpres.ro/english/2022/02/09/frequency-of-rca-damages-in-romania-among-highest-at-european-level--861872}}

Our first analysis compares the Poisson and the Negative Binomial regressions, both using a log link function and taking into account only the traditional covariates (with \texttt{policy\_period} as the offset and \texttt{region2} as the location factor covariate).  We assess both the goodness-of-fit and predictive power of the two models with a 80-20 train/test split.  The train and test sets have close sample means (0.4490 and 0.4364 respectively) to ensure that the latter is representative of the original data, and hence is a set on which out-of-sample testing is valid.  In Table \ref{table: GLM-Poisson-NB} we report the in-sample and out-of-sample Root-Mean-Squared-Error (RMSE) and Mean-Absolute-Error (MAE).  While the performances of both models seem close, RMSE and MAE are susceptible to outliers, hence we also reported the Chi-Square Statistics, given by

$$\chi^2 = \sum_{g=0}^m \frac{(O_g - E_g)^2}{E_g}$$

where for each bin $g$ (i.e., the unique claim counts $0, 1, 2, ..., m$), $O_g$ is the observed value and $E_g$ is the expected value from the fitted model.  As revealed by the in-sample and out-of-sample Chi-Square Statistics in Tables \ref{table: in-sample-chi2-Poisson-NB} and \ref{table: out-sample-chi2-Poisson-NB} respectively, in fact Poisson regression has a worse fit as it misses both the zero-inflation and the right-tail.

\begin{table}[h]
\caption{Model Performance of Poisson and Negative Binomial Regressions}
\centering
\begin{tabular}{|ll|r|r|}
\hline
 & \textbf{}  & \textbf{Poisson} & \textbf{Negative Binomial} \\ \hline
 \textbf{In-sample} & \textbf{RMSE}  & 0.8879  & 0.8939  \\  
 & \textbf{MAE}  & 0.6320  & 0.6320  \\
 & \textbf{Chi-square} & 414.2863  & 3.5415  \\ \hline
 \textbf{Out-of-sample} & \textbf{RMSE}  & 0.8478  & 0.8514  \\ 
 & \textbf{MAE}  & 0.6180  & 0.6168  \\
 & \textbf{Chi-square} & 155.5979  & 5.1729  \\ \hline
\end{tabular}
\label{table: GLM-Poisson-NB}
\end{table}

\begin{table}[h]
\caption{In-sample Chi-Square Statistic of Poisson and Negative Binomial Regressions}
\centering
\begin{tabular}{|rr|rr|}
\hline
\multicolumn{1}{|r|}{\textbf{Claim count}} & \textbf{Observed} & \multicolumn{2}{c|}{\textbf{Expected}} \\ \cline{3-4} 
\multicolumn{1}{|r|}{} &  & \multicolumn{1}{r|}{\textbf{Poisson}} & \textbf{Negative Binomial} \\ \hline
\multicolumn{1}{|r|}{0} & 844 & \multicolumn{1}{r|}{751.88} & 842.09 \\
\multicolumn{1}{|r|}{1} & 201 & \multicolumn{1}{r|}{323.91} & 207.91 \\
\multicolumn{1}{|r|}{2} & 72 & \multicolumn{1}{r|}{76.16} & 70.85 \\
\multicolumn{1}{|r|}{3} & 34 & \multicolumn{1}{r|}{12.86} & 26.84 \\
\multicolumn{1}{|r|}{4} & 7 & \multicolumn{1}{r|}{1.84} & 10.80 \\
\multicolumn{1}{|r|}{5} & 5 & \multicolumn{1}{r|}{0.28} & 4.54 \\
\multicolumn{1}{|r|}{6} & 4 & \multicolumn{1}{r|}{0.07} & 3.97 \\ \hline
\multicolumn{2}{|l|}{\textbf{Chi-square statistics}} & \multicolumn{1}{r|}{\textbf{414.2863}} & \textbf{3.5415} \\ \hline
\end{tabular}
\label{table: in-sample-chi2-Poisson-NB}
\end{table}

\begin{table}[h]
\caption{Out-of-sample Chi-Square Statistic of Poisson and Negative Binomial Regressions.  The binned Chi-square statistics combines the two bins of claim counts 4 and 6 to reduce fluctuation.}
\centering
\begin{tabular}{|rr|rr|}
\hline
\multicolumn{1}{|c|}{\textbf{Claim count}} & \multicolumn{1}{c|}{\textbf{Observed}} & \multicolumn{2}{c|}{\textbf{Expected}} \\ \cline{3-4} 
\multicolumn{1}{|l|}{} & \multicolumn{1}{l|}{} & \multicolumn{1}{c|}{\textbf{Poisson}} & \multicolumn{1}{c|}{\textbf{Negative Binomial}} \\ \hline
\multicolumn{1}{|r|}{0} & 209 & \multicolumn{1}{r|}{186.84} & 209.87 \\
\multicolumn{1}{|r|}{1} & 52 & \multicolumn{1}{r|}{81.18} & 51.92 \\
\multicolumn{1}{|r|}{2} & 22 & \multicolumn{1}{r|}{19.23} & 17.72 \\
\multicolumn{1}{|r|}{3} & 3 & \multicolumn{1}{r|}{3.24} & 6.72 \\
\multicolumn{1}{|r|}{4} & 4 & \multicolumn{1}{r|}{0.44} & 2.70 \\
\multicolumn{1}{|r|}{6} & 1 & \multicolumn{1}{r|}{0.06} & 2.07 \\ \hline
\multicolumn{2}{|l|}{\textbf{Chi-square statistics}} & \multicolumn{1}{r|}{\textbf{56.2159}} & \textbf{4.2768} \\ \hline
\multicolumn{2}{|l|}{\textbf{Chi-square statistics (binned)}} & \multicolumn{1}{r|}{\textbf{82.1170}} & \textbf{3.7304} \\ \hline
\end{tabular}
\label{table: out-sample-chi2-Poisson-NB}
\end{table}

\subsection{Traditional and Telematics Covariates}
\label{Traditional-and-Telematics-Covariates}

To identify the important claim predictors, we perform a 5-fold cross-validation (CV) for different Negative Binomial regression candidates, each using a (sub)set of the available covariates.  In a $k$-fold cross validation, the dataset is split into $k$ subsets.  In each fold, the model is trained on $k-1$ subsets, and the trained model is tested on the remaining subset.  The $k$-fold CV then reports the average of the $k$ results.

Our dataset is randomly partitioned into 5 subsets such that their mean claim counts are close (0.4364, 0.4192, 0.4573, 0.4384 and 0.4845 respectively), which again ensures that each subset is representative of the original dataset.  As we have restricted to the Negative Binomial distribution, we also report the Negative Binomial deviance, which is 

\begin{equation*}
    D = 2 \sum_{i=1}^n \left\{y_i \log\left(\frac{y_i}{\mu_i}\right) + (1 + y_i) \log\left(\frac{1+\mu_i}{1+y_i}\right)\right\}
\end{equation*}

where if $y_i = 0$, the term $y_i \log\left(\frac{y_i}{\mu_i}\right)$ is taken to be zero, with $y_i$ and $\mu_i$ being the $i$-th observation and its prediction respectively. Additionally, we evaluate the predictive performance of the regression candidates using proper scoring rules for count data summarized in Table \ref{table: scoring-rules} \citep{czado_predictive_2009}.  Scoring rules offer concise metrics for assessing the quality of probabilistic forecasts, by assigning a numerical score $s(P,x)$ based on the predictive distribution $P$ and the observation $x$.  In particular, a scoring rule is considered proper if $s(Q,Q) \leq s(P,Q)$ for all $P$ and $Q$ with $s(P,Q)$ the expected value of $s(P, \cdot)$ under $Q$ \citep{gneiting_strictly_2007}.  In practice, scores are usually reported as averages over appropriate sets of probabilistic forecasts.  Therefore, for each CV fold, we calculate the average scores, and the $k$-fold CV presents the average of the $k$ values.  For each of these seven metrics, a lower value indicates a better performance.

 \begin{table}
 \caption{Proper Scoring Rules for Count Data}
    \centering
    \begin{tabular}{|l|A|}\hline
         \textbf{Score}& \multicolumn{2}{l|}{\textbf{Formula}}\\\hline
         quadratic (qs)& \text{qs}(P,x) & = -2 p_x + ||p||^2\\
         spherical (sphs)& \text{sphs}(P,x) & = -\frac{p_x}{||p||}\\
         ranked probability (rps)& \text{rps}(P,x) & = \sum_{k=0}^\infty \{P_k - 1(x\leq k)\}^2\\
         Dawid-Sebastiani (dss)& \text{dss}(P,x) & = \left(\frac{x-\mu_P}{\sigma_P}\right)^2 + 2\log \sigma_P\\\hline
    \end{tabular}
    \label{table: scoring-rules}
\end{table}

In Table \ref{table: GLM-traditional} we begin with \texttt{mod0}, a null model with only exposure \texttt{policy\_period} as an offset, and \texttt{mod1}, which includes all traditional covariates (with \texttt{region1}) but no exposure.  In \texttt{mod2} and \texttt{mod2b} we compare the performance when the exposure \texttt{policy\_period} is included, either as an offset (i.e., regression coefficient being 1) or an ordinary covariate.  The inclusion of \texttt{policy\_period} is clearly essential and beneficial, but the extra flexibility in using it as a covariate does not bring about an improvement in model fit and prediction; hence our analysis proceeds with \texttt{policy\_period} as an offset.  In \texttt{mod3} we replace \texttt{region1} with \texttt{region2} which has fewer levels.  As expected, the new model generalizes better and has a better out-of-sample performance, and so we proceed with \texttt{region2}.

\begin{table}[h]
\caption{5-Fold Cross-Validation Performance of Models with Traditional Covariates Only.}
\centering
\begin{tabular}{ll|rrrrr}
\multicolumn{1}{l}{} & \textbf{Variables} &  \textbf{mod0}&\textbf{mod1} & \textbf{mod2} & \textbf{mod2b} & \textbf{mod3} \\ \hline\hline
\multicolumn{1}{l|}{Traditional} & \texttt{region1} &  &X & X & X &  \\
\multicolumn{1}{l|}{} & \texttt{region2} &   &&  &  & X \\
\multicolumn{1}{l|}{} & \texttt{max\_weight} &  &X & X & X & X \\
\multicolumn{1}{l|}{} & \texttt{car\_value} &  &X & X & X & X \\
\multicolumn{1}{l|}{} & \texttt{num\_seats} &  &X & X & X & X \\
\multicolumn{1}{l|}{} & \texttt{renewal} &  &X & X & X & X \\
\multicolumn{1}{l|}{} & \texttt{use} &  &X & X & X & X \\
\multicolumn{1}{l|}{} & \texttt{vehicle\_age} &  &X & X & X & X \\
\multicolumn{1}{l|}{} & \texttt{policy\_period} &   offset&& offset & X & offset \\ \hline\hline
 & \textbf{deviance} &  0.8520&0.8817& 0.8548& 0.8527& 0.8498\\
 & \textbf{RMSE} &  0.8821&0.8987& 0.8960& 0.8982& 0.8922\\
 & \textbf{MAE} &  0.6360&0.6439& 0.6330& 0.6323& 0.6321\\
 & \textbf{qs}&  -0.5597&-0.5548& -0.5595& -0.5598&-0.5600\\
 & \textbf{sphs}&  -0.7468&-0.7453& -0.7467& -0.7467&-0.7468\\
 & \textbf{rps}&  0.3535&0.3593& 0.3555& 0.3556&0.3546\\
 & \textbf{dss}&  0.7144&0.8338& 0.7240& 0.7004&0.6939\\
 \end{tabular}
\label{table: GLM-traditional}
\end{table}

\begin{table}[h]
\caption{5-Fold Cross-Validation Performance of Models with Traditional Covariates and Telematics Exposure.}
\centering
\begin{tabular}{ll|rrrrrr}
 & \textbf{Variables} & \textbf{t\_mod1} & \textbf{t\_mod1b} & \textbf{t\_mod1c} & \textbf{t\_mod2} & \textbf{t\_mod2b} & \textbf{t\_mod2c} \\ \hline\hline
\multicolumn{1}{l|}{Traditional} & \texttt{region1} &  &  &  &  &  &  \\
\multicolumn{1}{l|}{} & \texttt{region2} & X & X & X & X & X & X \\
\multicolumn{1}{l|}{} & \texttt{max\_weight} & X & X & X & X & X & X \\
\multicolumn{1}{l|}{} & \texttt{car\_value} & X & X & X & X & X & X \\
\multicolumn{1}{l|}{} & \texttt{num\_seats} & X & X & X & X & X & X \\
\multicolumn{1}{l|}{} & \texttt{renewal} & X & X & X & X & X & X \\
\multicolumn{1}{l|}{} & \texttt{use} & X & X & X & X & X & X \\
\multicolumn{1}{l|}{} & \texttt{vehicle\_age} & X & X & X & X & X & X \\
\multicolumn{1}{l|}{} & \texttt{policy\_period} &  &  &  &  &  &  \\ \hline
\multicolumn{1}{l|}{Telematics} & \texttt{total\_distance} & offset & X & logged &  &  &  \\
\multicolumn{1}{l|}{} & \texttt{total\_time} &  &  &  & offset & X & logged \\ \hline\hline
 & \textbf{deviance} & 0.8737& 0.8606& 0.8460& 0.8463&0.8483&0.8353\\
 & \textbf{RMSE} & 1.0367& 0.9158& 0.8997& 0.9203&0.8947&0.8891\\
 & \textbf{MAE} & 0.6416& 0.6399& 0.6234& 0.6193&0.6281&0.6178\\
 & \textbf{qs}& -0.5628& -0.5582& -0.5632&-0.5647&-0.5607&-0.5647\\
 & \textbf{sphs}& -0.7480& -0.7462& -0.7480&-0.7485&-0.7471&-0.7484\\
 & \textbf{rps}& 0.3618& 0.3574& 0.3520&0.3528&0.3531&0.3496\\
 & \textbf{dss}& 1.0042& 0.7501& 0.7632&0.8342&0.7251&0.7231\\
\end{tabular}
\label{table: GLM-telematics-exposure}
\end{table}

\begin{table}[h]
\caption{5-Fold Cross-Validation Performance of Models with Both Traditional and Telematics Covariates. \texttt{tm} indicates \textbf{t}elematics and speed transition \textbf{m}atrix, while \texttt{tt} indicates the former two and driving \textbf{t}ime.}
\centering
\begin{tabular}{ll|rrrrrr}
 & \textbf{Variables} & \textbf{t\_mod3} & \textbf{t\_mod4} & \textbf{tm\_mod1} & \textbf{tm\_mod2} & \textbf{tt\_mod1} & \textbf{tt\_mod2} \\ \hline\hline
\multicolumn{1}{l|}{Traditional} & \texttt{region1} &  &  &  &  &  &  \\
\multicolumn{1}{l|}{} & \texttt{region2} & X &  & X &  & X &  \\
\multicolumn{1}{l|}{} & \texttt{max\_weight} & X &  & X &  & X &  \\
\multicolumn{1}{l|}{} & \texttt{car\_value} & X &  & X &  & X &  \\
\multicolumn{1}{l|}{} & \texttt{num\_seats} & X &  & X &  & X &  \\
\multicolumn{1}{l|}{} & \texttt{renewal} & X &  & X &  & X &  \\
\multicolumn{1}{l|}{} & \texttt{use} & X &  & X &  & X &  \\
\multicolumn{1}{l|}{} & \texttt{vehicle\_age} & X &  & X &  & X &  \\
\multicolumn{1}{l|}{} & \texttt{policy\_period} &  &  &  &  &  &  \\ \hline
\multicolumn{1}{l|}{Telematics} & \texttt{prop\_roadtype} & X & X & X & X & X & X \\
\multicolumn{1}{l|}{} & \texttt{avg\_speed} & X & X & X & X & X & X \\
\multicolumn{1}{l|}{} & \texttt{max\_speed} & X & X & X & X & X & X \\
\multicolumn{1}{l|}{} & \texttt{pc1} &  &  & X & X & X & X \\
\multicolumn{1}{l|}{} & \texttt{pc2} &  &  & X & X & X & X \\
\multicolumn{1}{l|}{} & \texttt{prop\_time} &  &  &  &  & X & X \\
\multicolumn{1}{l|}{} & \texttt{num\_acc} & X & X & X & X & X & X \\
\multicolumn{1}{l|}{} & \texttt{num\_brake} & X & X & X & X & X & X \\
\multicolumn{1}{l|}{} & \texttt{num\_left} & X & X & X & X & X & X \\
\multicolumn{1}{l|}{} & \texttt{num\_right} & X & X & X & X & X & X \\
\multicolumn{1}{l|}{} & \texttt{num\_severe} & X & X & X & X & X & X \\
\multicolumn{1}{l|}{} & \texttt{total\_distance} &  &  &  &  &  &  \\
\multicolumn{1}{l|}{} & \texttt{total\_time} & logged & logged & logged & logged & logged & logged \\ \hline\hline
 & \textbf{deviance} & 0.8359& 0.8297& 0.8331& 0.8258&0.8357&0.8285\\
 & \textbf{RMSE} & 0.8956& 0.9024& 0.8922& 0.8923&0.8890&0.8863\\
 & \textbf{MAE} & 0.6162& 0.6200& 0.6128& 0.6157&0.6087&0.6120\\
 & \textbf{qs}& -0.5663& -0.5671& -0.5669&-0.5678&-0.5672&-0.5677\\
 & \textbf{sphs}& -0.7490& -0.7493& -0.7493&-0.7496&-0.7495&-0.7497\\
 & \textbf{rps}& 0.3501& 0.3500& 0.3489&0.3479&0.3488&0.3480\\
 & \textbf{dss}& 0.7458& 0.7002& 0.7498&0.7034&0.7676&0.7098\\
\end{tabular}
\label{table: GLM-both}
\end{table}

The inclusion of traditional covariates does not bring much improvement over the null model, hence we begin to include telematics covariates.  In Table \ref{table: GLM-telematics-exposure} we replace \texttt{policy\_period} with either \texttt{total\_distance} travelled (\texttt{t\_mod1}, \texttt{t\_mod1b} and \texttt{t\_mod1c}) or \texttt{total\_time} driven (\texttt{t\_mod2}, \texttt{t\_mod2b} and \texttt{t\_mod2c}) as the exposure.  In each case, exposure is included either as an offset or an ordinary covariate (with or without taking logarithm).  Several observations can be made: first, logged expected claim counts seems to have a linear relationship with the logged exposure rather than with the original feature, since the models with the original performed worse; second, both total driving time and distance are better exposure features compared to policy period as they provide a better picture on the extent of vehicle usage.  As the models using total time perform better than those using total distance, our analysis proceeds with logged \texttt{total\_time} as a covariate; more discussion on this decision (as opposed to including it as an offset) and result is given in Section \ref{Discussions-on-the-Treatment-of-Total-Time-or-Distance-Travelled}.

In Table \ref{table: GLM-both}, \texttt{t\_mod3} includes both traditional and telematics covariates (except transition matrix and proportions of driving in different timeslots), whilst \texttt{t\_mod4} includes only telematics covariates.   \texttt{tm\_mod1} and \texttt{tm\_mod2} build on \texttt{t\_mod3} and \texttt{t\_mod4} respectively, and include the first two PCs of the speed transition matrix \texttt{pc1} and \texttt{pc2}.  \texttt{tt\_mod1} and \texttt{tt\_mod2} further include the proportions of driving in different timeslots \texttt{prop\_time}.  While the inclusion of telematics covariates has improved model performance, it is yet unclear what the best model is due to the excessive number of covariates.  We proceed to perform feature selection.

One of the most commonly used variable selection method for GLMs is the stepwise algorithm.  However, since we evaluate each model via a 5-fold cross-validation, how a stepwise selection should be performed is not immediately obvious.  We employ a majority voting scheme: first, stepwise both-side (forward \textit{and} backward) selection based on Akaike Information Criteria (AIC) is performed on the training set in each fold, and the selected features are recorded; then, the total number of times a feature has been selected (ranging from 0 - not chosen at all, to 5 - chosen in every fold) is summarized, and the features which have been selected more than three times are included in the final model.  This can eliminate covariates that are important to only a small portion of the data and reduce overfitting.  Based on the finalized subset of covariates, performance metrics are calculated on the testing set in each fold.  Note that this finalized subset is not necessarily the same subset selected on each fold, e.g. if covariate A is selected four times in total, then there is at least one fold B where its training set does not select this covariate, yet its testing set will be evaluated with this covariate.  This should provide a fairer assessment of model performance.

We perform the majority voting procedure on various sets of covariates and the voting results are shown in Table \ref{table: GLM-majority-votes}.  We state explicitly the zero votes to identify the covariates included for stepwise selection.  The set of selected features may differ depending on what covariates are included as they compete to explain the most variability in the response.  The important traditional covariates are vehicle usage, maximum weight and (monetary) value of the vehicle, whilst to our surprise, vehicle age is relatively unimportant.  Maximum speed and number of harsh braking are the more important telematics covariates.  However, in the presence of speed transition matrices and driving in different timeslots, which have become the most important telematics covariates, the importance of maximum speed drops slightly, while average speed is eliminated from the selection.  A potential explanation is that not only information about average and maximum speed have been captured by the matrix, the matrix is able to provide even more information on the driving speed.  It is noteworthy that in \texttt{tt\_mod1s}, number of harsh braking falls to exactly three votes (the majority criteria) while number of severe events receives one vote.  Due to the dependence between the different timeslots and harsh event rates, the former may have captured some predictive power of the harsh events.  Although the first PC, by definition, explains the most variability in the transition matrix, it is the second PC that provides the most predictive power for the response.  Driving in different roadtypes and numbers of other harsh events are deemed unimportant in all models.

\textit{Remark: Motivated by the discussions in Section \ref{Portfolio-Level-Distribution}, we examine the effect on the regressions models when substituting the self-reported counties/regions of residence with the most-driven ones.  Although we observe a marginal improvement in performance metrics, it's worth noting that the most-driven regions (both in granularity as \texttt{region1} and \texttt{region2}) are still not chosen based on the majority votes criteria.  As a consequence, the results have been omitted from the paper.}

We observe that \texttt{tt\_mod1s}, which includes both PCs and the driving timeslots, achieves the best result for five out of the seven performance metrics considered (and ranking second and third best for deviance and sphs, respectively).  By comparing \texttt{mod3s} to \texttt{t\_mod4s}, \texttt{tm\_mod2s} and \texttt{tt\_mod2s}, we observe that the models with only telematics covariates can already outperform the models with only traditional covariates, but the best performance is attained by a combination of both after all.  This verifies the benefits of adding telematics data to claim prediction, but the value of traditional covariates, such as the characteristics of the vehicle, still cannot be ignored, especially due to the limitation of our data.

\begin{table}[h]
\caption{5-Fold Cross-Validation Performance of Models Resulted from Majority Votes.  Covariates with a number are included for stepwise selection, whilst covariates with \textbf{three votes and more} are included in the final model.  \texttt{policy\_period} is included as an offset in the first model, and logged \texttt{total\_time} is chosen in all the other models.}
\centering
\makebox[\textwidth]{
\begin{tabular}{ll|rrrrrrr}
 & \textbf{Variables} & \textbf{mod3s} & \textbf{t\_mod3s} & \textbf{t\_mod4s} & \textbf{tm\_mod1s} & \textbf{tm\_mod2s} & \textbf{tt\_mod1s} & \textbf{tt\_mod2s} \\ \hline\hline
\multicolumn{1}{l|}{Traditional} & \texttt{region1} &  &  &  &  &  &  &  \\
\multicolumn{1}{l|}{} & \texttt{region2} & 0 & 0 &  & 0 &  & 0 &  \\
\multicolumn{1}{l|}{} & \texttt{max\_weight} & 4 & 5 &  & 4 &  & 4&  \\
\multicolumn{1}{l|}{} & \texttt{car\_value} & 5 & 5 &  & 4 &  & 4 &  \\
\multicolumn{1}{l|}{} & \texttt{num\_seats} & 2 & 0 &  & 0 &  & 0 &  \\
\multicolumn{1}{l|}{} & \texttt{renewal} & 2 & 0 &  & 0 &  & 0 &  \\
\multicolumn{1}{l|}{} & \texttt{use} & 1 & 5 &  & 3 &  & 4 &  \\
\multicolumn{1}{l|}{} & \texttt{vehicle\_age} & 0 & 0 &  & 0 &  & 0 &  \\
\multicolumn{1}{l|}{} & \texttt{policy\_period} & offset &  &  &  &  &  &  \\ \hline
\multicolumn{1}{l|}{Telematics} & \texttt{prop\_roadtype} &  & 1 & 1 & 0 & 0 & 0 & 0 \\
\multicolumn{1}{l|}{} & \texttt{avg\_speed} &  & 2 & 4 & 0 & 0 & 0 & 0 \\
\multicolumn{1}{l|}{} & \texttt{max\_speed} &  & 5 & 5 & 4 & 5 & 4& 4 \\
\multicolumn{1}{l|}{} & \texttt{pc1} &  &  &  & 0 & 1 & 0 & 0 \\
\multicolumn{1}{l|}{} & \texttt{pc2} &  &  &  & 5 & 5 & 5 & 5 \\
\multicolumn{1}{l|}{} & \texttt{prop\_time} &  &  &  &  &  & 5 & 5 \\
\multicolumn{1}{l|}{} & \texttt{num\_acc} &  & 0 & 0 & 0 & 0 & 0 & 0 \\
\multicolumn{1}{l|}{} & \texttt{num\_brake} &  & 4 & 4 & 4& 4 & 3& 4\\
\multicolumn{1}{l|}{} & \texttt{num\_left} &  & 0 & 0 & 0 & 0 & 0 & 0 \\
\multicolumn{1}{l|}{} & \texttt{num\_right} &  & 0 & 0 & 0 & 0 & 0 & 0 \\
\multicolumn{1}{l|}{} & \texttt{num\_severe} &  & 0 & 0 & 0 & 0 & 1 & 1\\
\multicolumn{1}{l|}{} & \texttt{total\_distance} &  &  &  &  &  &  &  \\
\multicolumn{1}{l|}{} & \texttt{total\_time} &  & logged & logged & logged & logged & logged & logged \\ \hline\hline
 & \textbf{deviance} & 0.8403& 0.8174& 0.8241& \textbf{0.8135}&0.8158&0.8144&0.8177\\
 & \textbf{RMSE} & 0.8914& 0.8797& 0.8892& 0.8777&0.8863&\textbf{0.8752}&0.8789\\
 & \textbf{MAE} & 0.6325& 0.6102& 0.6152& 0.6087&0.6125&\textbf{0.6049}&0.6084\\
 & \textbf{qs}& -0.5619& -0.5686& -0.5681&-0.5689&-0.5691&\textbf{-0.5691}&-0.5690\\
 & \textbf{sphs}& -0.7474& -0.7498& -0.7497&-0.7498&-0.7500&-0.7499&\textbf{-0.7501}\\
 & \textbf{rps}& 0.3527& 0.3457& 0.3478&0.3449&0.3457&\textbf{0.3449}&0.3457\\
 & \textbf{dss}& 0.6657& 0.6631& 0.6798&0.6438&0.6561&\textbf{0.6435}&0.6588\\
\end{tabular}}
\label{table: GLM-majority-votes}
\end{table}

In Table \ref{table: GLM-final-coefficients} we report the estimated regression coefficients of the final model \texttt{tt\_mod1s} fitted to the entire dataset.  \texttt{car\_value} is in dollars of domestic currency, every 10,000 dollars increase in price leads to $\exp(0.01074) -1 = 1.08\%$ increase in expected claim count.  \texttt{max\_weight} is in kilograms, and every 10 kilograms increase leads to a small, 0.28\% decrease of expected claim count.  The reference class of vehicle usage \texttt{use} is `others', hence personal use demonstrates a higher risk.  The coefficient estimate of (logged) \texttt{total\_time} is 0.55 instead of 1, where the latter will correspond to an offset.  \texttt{max\_speed}, \texttt{pc2} and \texttt{num\_brake} all have a positive effect on expected claim counts.  As explained in Section \ref{Portfolio-Level-Distribution}, \texttt{pc2} has small, negative values along the diagonal of the matrix, which are the moderate speed bins and closer transitions, hence an increase in \texttt{pc2} actually represents reduced stability in speed transitions.  Finally, as we have excluded \texttt{prop\_20\_24}, it acts as the reference group and all other proportions should be interpreted with respect to it.  As all the other coefficients are negative, driving in all other timeslots is expected to be less risky than driving between 8 p.m. to midnight.

\begin{table}[h]
\centering
\caption{Estimated Regression Coefficients of Selected Model}
\begin{threeparttable}
\begin{tabular}{ll|rrrrl}
 & \textbf{Variables} & \textbf{Estimate} & \textbf{Std. Error} & \textbf{z-value} & \textbf{p-value} &  \\ \hline
\multicolumn{1}{l|}{Traditional} & \texttt{intercept} & -3.6660& 1.5050& -2.4350& 0.0149& *\\
\multicolumn{1}{l|}{} & \texttt{car\_value} & 0.0000& 0.0000& 1.9180& 0.0552& .\\
\multicolumn{1}{l|}{} & \texttt{max\_weight} & -0.0003& 0.0002& -1.8340& 0.0667& .\\
\multicolumn{1}{l|}{} & \texttt{use\_personal} & 0.2305& 0.1211& 1.9040& 0.0570& .\\ \hline
\multicolumn{1}{l|}{Telematics} & \texttt{log(total\_time)} & 0.5495& 0.1016& 5.4100& 0.0000& ***\\
\multicolumn{1}{l|}{} & \texttt{max\_speed} & 0.0040& 0.0020& 1.9480& 0.0514& .\\
\multicolumn{1}{l|}{} & \texttt{pc2} & 0.0345& 0.0137& 2.5230& 0.0116& *\\
\multicolumn{1}{l|}{} & \texttt{num\_brake}& 0.0011& 0.0007& 1.5280& 0.1264&\\
\multicolumn{1}{l|}{} & \texttt{prop\_0\_4} & -0.0921& 0.0353& -2.6100& 0.0091& **\\
\multicolumn{1}{l|}{} & \texttt{prop\_4\_8} & -0.0309& 0.0116& -2.6600& 0.0078& **\\
\multicolumn{1}{l|}{} & \texttt{prop\_8\_12} & -0.0290& 0.0119& -2.4360& 0.0148& *\\
\multicolumn{1}{l|}{} & \texttt{prop\_12\_16}& -0.0315& 0.0130& -2.4120& 0.0159& *\\
\multicolumn{1}{l|}{} & \texttt{prop\_16\_20}& -0.0372& 0.0151& -2.4630& 0.0138& *\\ \hline\hline
\end{tabular}
\begin{tablenotes}
\small
\item * Coefficient for \texttt{car\_value} is 1.074$\times$ 10$^{-6}$
\item Significance codes: 0 `***', 0.001 `**', 0.01 `*', 0.05 `.', 0.1 ` ', 1
\end{tablenotes}
\end{threeparttable}
\label{table: GLM-final-coefficients}
\end{table}

\subsection{Robustness of Speed Transition Matrix to Different Bin Widths}
\label{Robustness-of-Speed-Transition-Matrix-wrt-h}

We have used a speed transition matrix with a constant bin width of 10 km/h throughout our analysis.  In this section we study the change in the PC's predictive power (hence model performance) with respect to a change in the bin width, provide suggestions on choosing a good bin width, and illustrate that the predictive power is fairly robust to different widths.  While one can definitely choose a different bin width for each of the speed bins, we only study constant bin widths for convenience.

\begin{figure*}[h]
\caption{Out-of-sample Model Performance with Respect to Different Speed Bin Widths. Red line indicates the performance of the traditional covariates only model \texttt{t\_mod3s}.}
    \centering
    \begin{subfigure}[b]{\textwidth}
        \centering
        \includegraphics[width=0.6\textwidth]{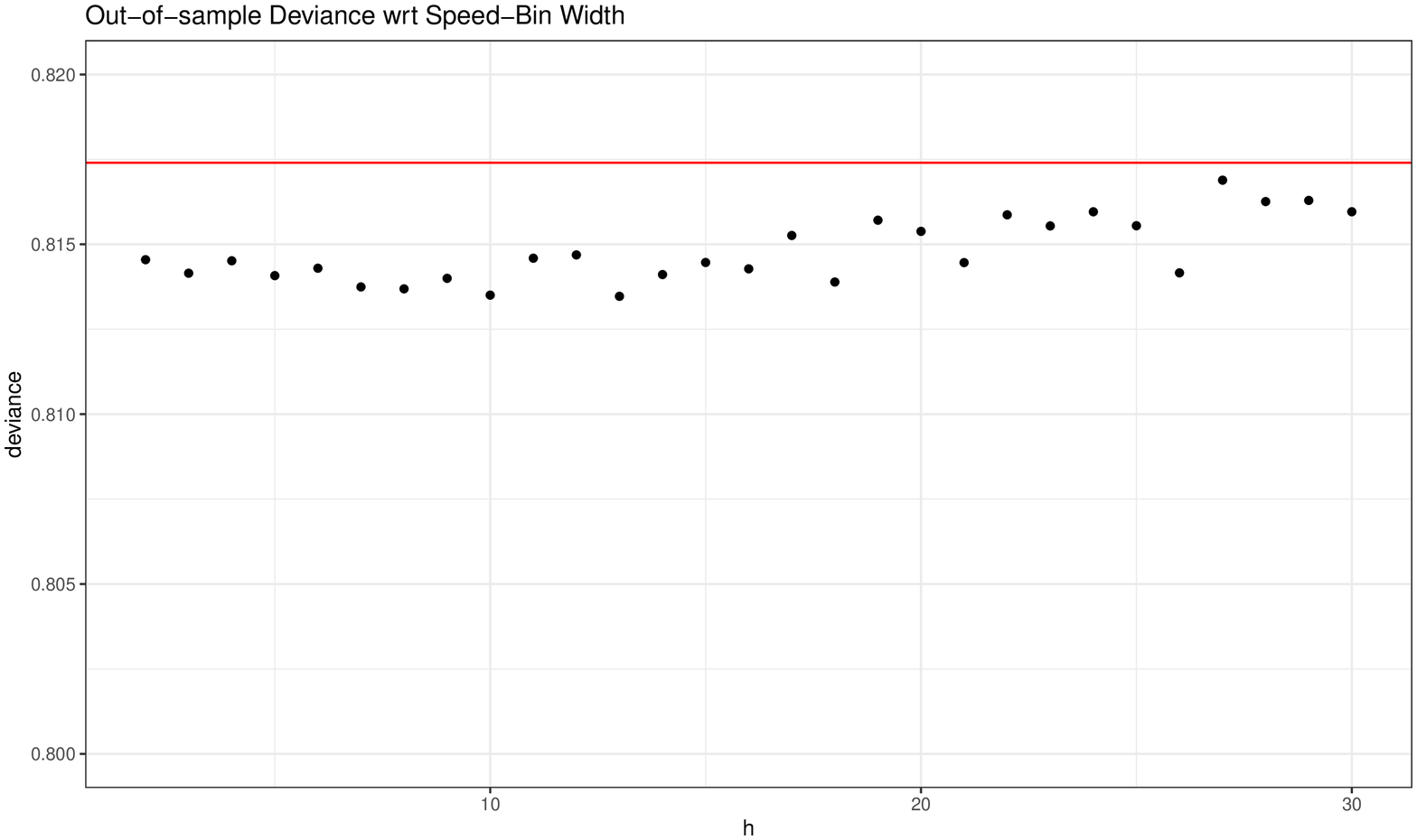}
        \caption*{}
    \end{subfigure}
    \vskip\baselineskip
    \begin{subfigure}[b]{\textwidth}   
        \centering 
        \includegraphics[width=0.6\textwidth]{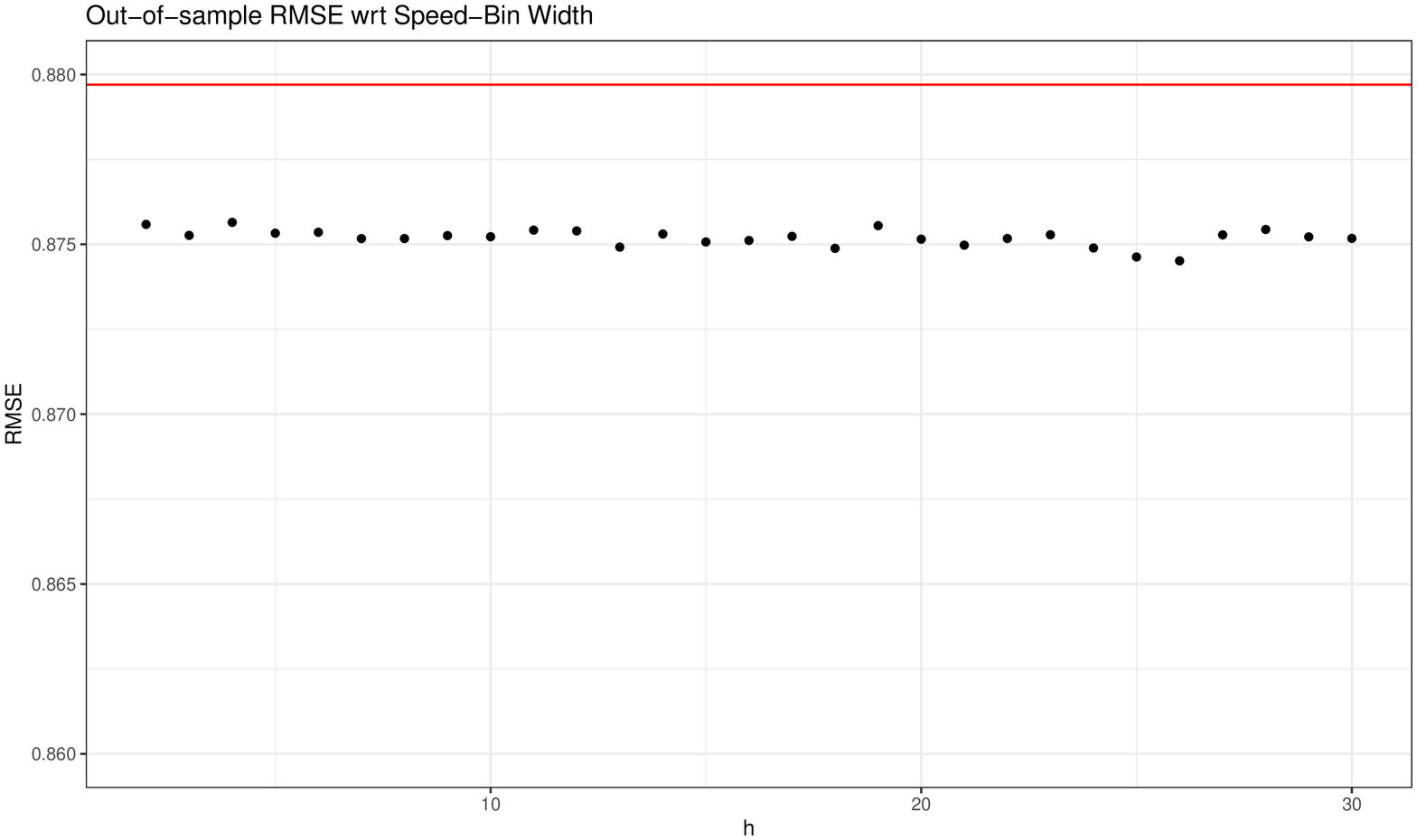}
        \caption*{}
    \end{subfigure}
    \vskip\baselineskip
    \begin{subfigure}[b]{\textwidth}   
        \centering 
        \includegraphics[width=0.6\textwidth]{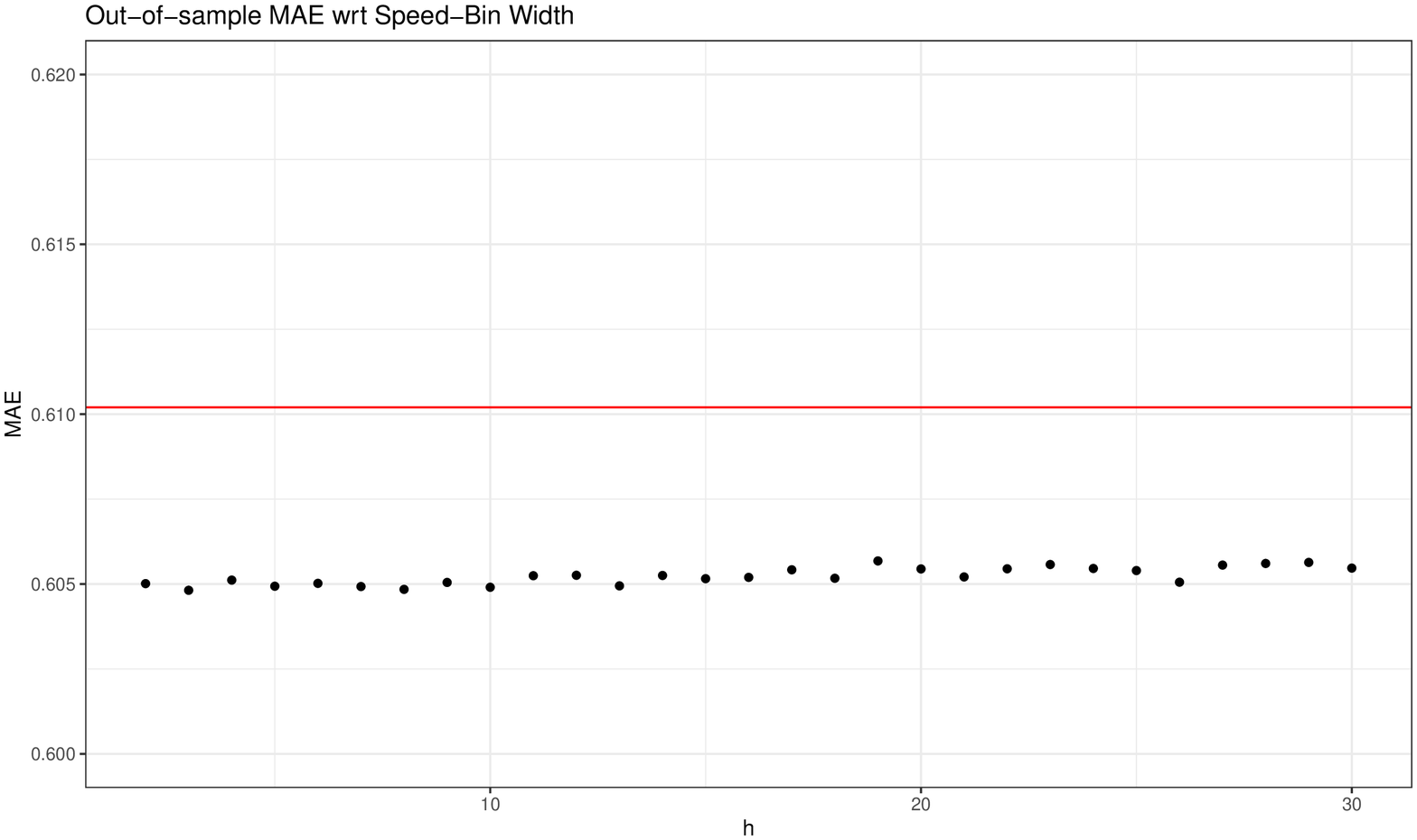}
        \caption*{}
    \end{subfigure}
    \label{fig: robustness-plots}
\end{figure*}

For a bin width of (positive integer) $h$ km/h, we create a $m$-by-$m$ speed transition matrix as: [0, 0.5), [0.5, $h$), [$h$, $2h$), ..., [$(m-1)h$, Inf), where $(m-1)h$ is the largest integer less than or equal to 130.  While the first bin again captures the stationary state, the last bin may not align with the speed limit.  Without loss of generality, we consider the same feature sets in Table \ref{table: GLM-majority-votes}.  For all integer $h$ between 2 to 30, the best out-of-sample deviance is attained by \texttt{tm\_mod1s}, and the best out-of-sample RMSE and MAE are attained by \texttt{tt\_mod1s}.  In Figure \ref{fig: robustness-plots} we compare the model performances against those of the benchmark model \texttt{t\_mod3s} as indicated by the red line, which is the model with only traditional covariates.  On the one hand, we observe that the values are all close and well below the benchmark, indicating that the model performance (or predictive power) is fairly robust to different bin widths.  On the other hand, the values provide suggestions on choosing a good bin width.  First, there should be a balance between information granularity and dimensionality: as we include only the first two PC's and only the second one has strong predictive power, \texttt{pc2} cannot capture sufficient variance of the matrix if its size is too large (i.e., $h$ being too small); whilst the matrix cannot capture the transition patterns and becomes uninformative if the bin widths $h$ is too large (e.g., in the extreme case, only [0, 0.5) and [0.5, Inf)).  Second, the bin widths also determine how closely we can align with the speed limit 130 km/h, and the model performances are better when $h$ is a factor of 130.  For example, while the patterns captured by the PC's of transition matrices of $h$ being 26 km/h and 27 km/h are consistent, the former produces a better model performance as its last bin aligns with the speed limit.  An alignment with the speed limit is preferred as it generally captures speeding events; moreover, as the bin is larger (i.e., when lower bound is much less than 130), the transitions within high speeds are neglected.

\begin{figure*}[h]
\caption{First Principal Component (PC) of Speed Transition Matrices with Different Bin Widths.  Patterns are consistent but the (last) bins can be very different.}
    \centering
    \begin{subfigure}[b]{0.47\textwidth}
        \centering
        \includegraphics[width=\textwidth]{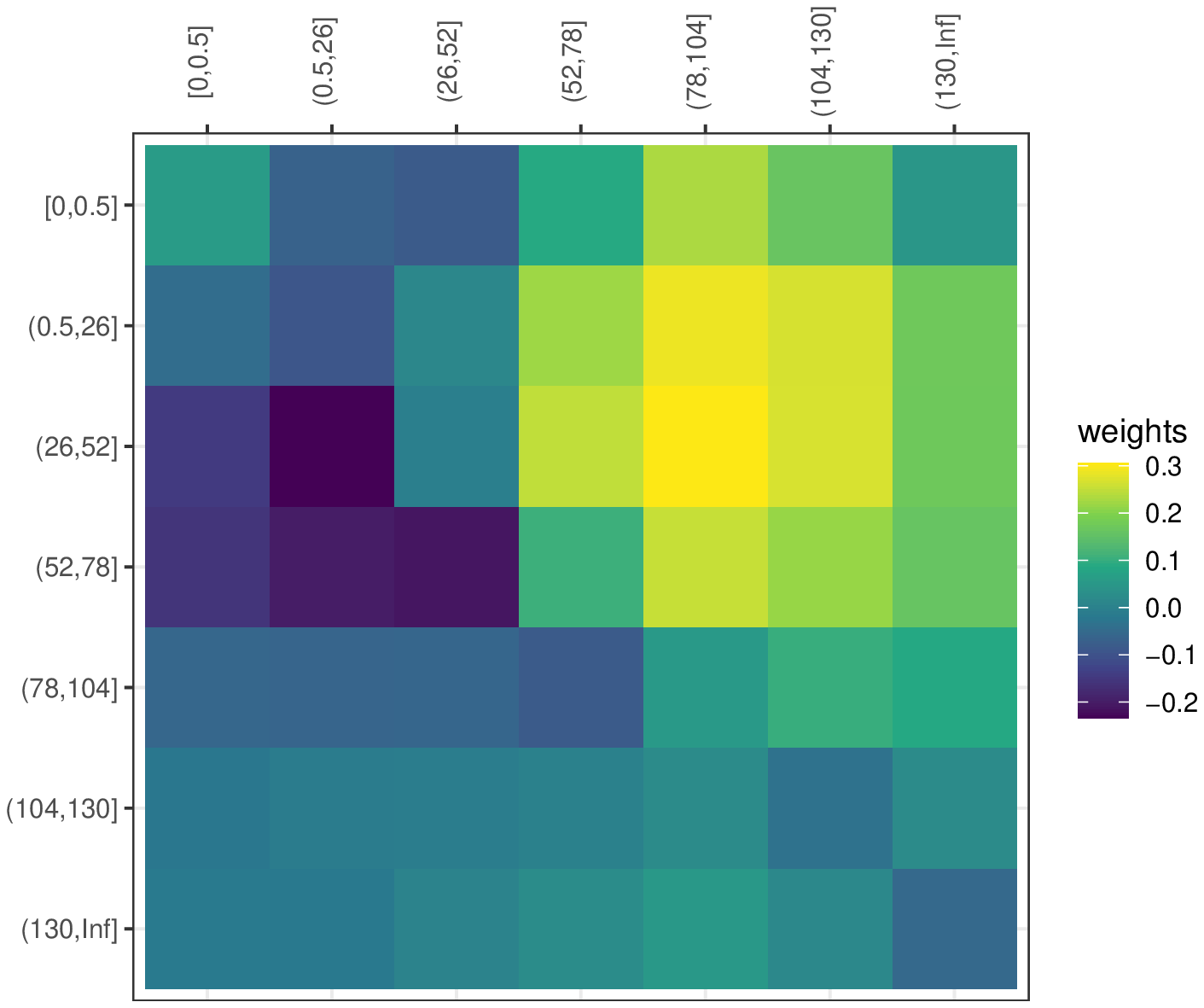}
        \caption{Bin Width of 26 km/h}
    \end{subfigure}
    \hfill
    \begin{subfigure}[b]{0.47\textwidth}  
        \centering 
        \includegraphics[width=\textwidth]{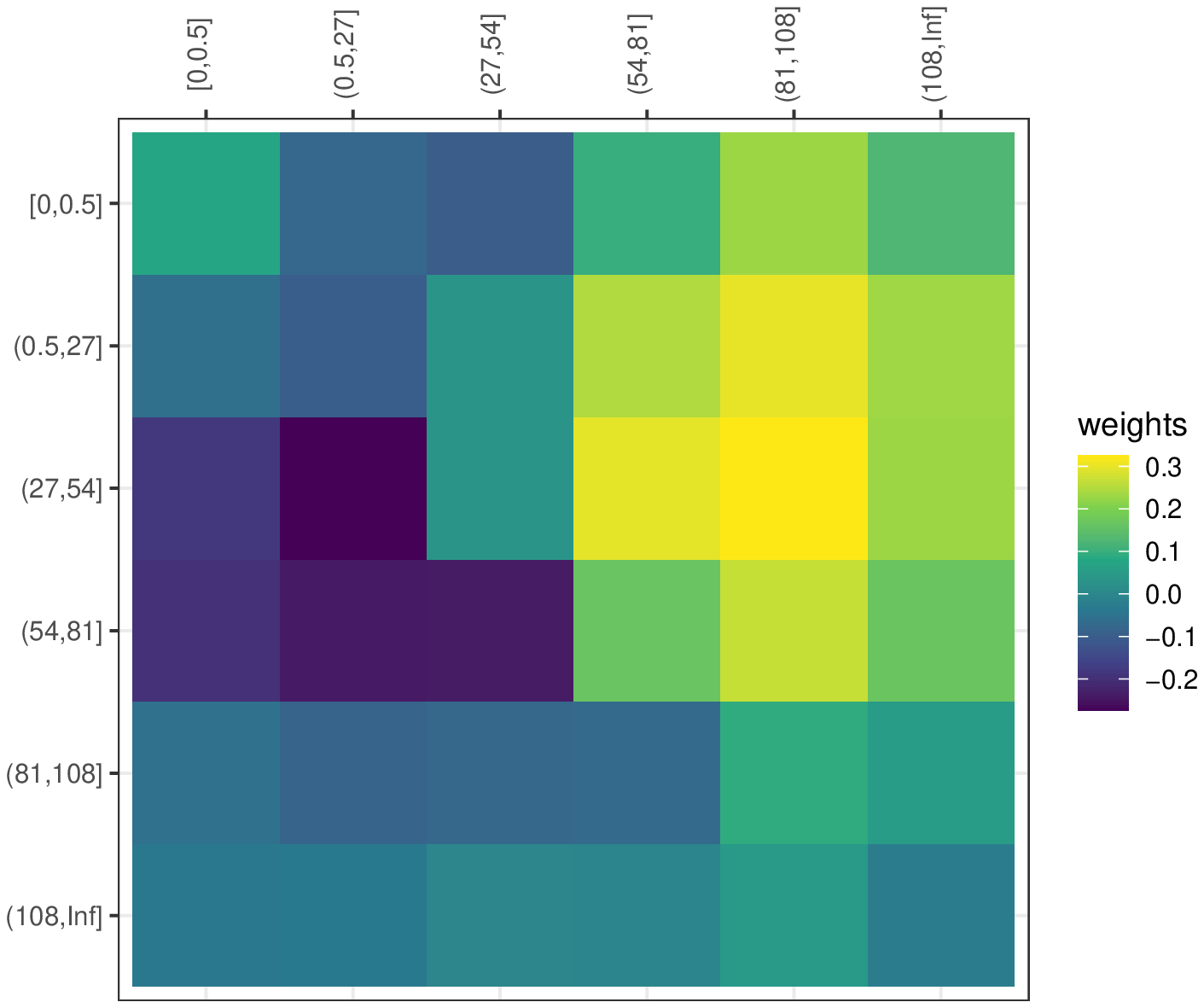}
        \caption{Bin Width of 27 km/h}
    \end{subfigure}
    \label{fig: h26-vs-h27}
\end{figure*}

\subsection{Discussions on the Treatment of Total Time or Distance Travelled}
\label{Discussions-on-the-Treatment-of-Total-Time-or-Distance-Travelled}

In classical ratemaking, policy coverage period is usually treated as an offset in modelling claim frequency, e.g. with coefficient 1 in a GLM.  When telematics data becomes available to insurers, this term is usually replaced by total distance travelled.  This gives rise to Pay-As-You-Drive (PAYD) Insurance (\citet{boucher_pay-as-you-drive_2013} and \citet{paefgen_multivariate_2014}), in which ratemaking takes into account how much the policyholders drive, in addition to the policyholders' traditional covariates.  While one may continue to use total driving time or distance as an offset (\citet{ayuso_improving_2019}, \citet{denuit_multivariate_2019}, \citet{guillen_near-miss_2021}), some literatures have treated them as a covariate with coefficients estimated from the data (\citet{boucher_pay-as-you-drive_2013}, \citet{paefgen_multivariate_2014}, \citet{verbelen_unravelling_2018}).  This additional flexibility usually leads to a better model fit.  However, discussion on the interpretation behind such model has halted since the rise of Pay-How-You-Drive (PHYD) Insurance - an insurance product which builds on top of PAYD Insurance and considers policyholders' driving behaviour as well.  In this work, we would like to once again raise the attention on the fact that expected claim counts are not directly proportional with total driving time or distance.

In the previous section, we have evaluated models treating total driving time or total distance travelled as an offset, as an ordinary covariate, or as a covariate after taking natural logarithm.  Consistent with existing works, model performance is better with the covariate approach.  For the best model \texttt{tt\_mod1s}, the coefficients of logged total driving time is 0.5495, regardless of the unit of time.  Under the Wald test, this estimated coefficient is statistically different from 1 at significance level of 0.0009\%.  For illustrative purpose, we state the log-link Negative Binomial regression with coefficient 0.5,

\begin{align*} 
\ln(\hat N_{i}) &= 0.5 \log(t_i) + \hat \beta_0 + \hat \beta_1 x_{1i} + \hat \beta_2 x_{2i} + ... \\
\hat N_{i} &= \sqrt{t_i} \times \exp(\hat \beta_0 + \hat \beta_1 x_{1i} + \hat \beta_2 x_{2i} + ...)
\end{align*}

where $\hat{N_i}$ is the expected claim counts for policyholder $i$, $t_i$ is his/her total driving time, and $x_{pi}$ are the other covariates of this policyholder.  Under this model, claim counts are expected to increase as a function of the square root of total driving time instead of directly proportional.  Hence with all other covariates constant, the driver's riskiness is still increasing with total driving time, but the rate of increase is reducing.

\textit{Remark: Since PAYD Insurance is based on mileage instead of driving time, we also report on the relationship between claim counts and mileage for completeness of our analysis.  The best model \textbf{tt\_mod1s} is refitted by replacing logged total driving time with logged total distance travelled, and the coefficient is found to be 0.4518.  Hence we again conclude that claim counts are expected to increase with mileage, but less than directly proportional.}

\begin{figure*}[h]
\caption{Best Model \texttt{tt\_mod1s} with Cubic Splines on Total Driving Time, Without (Left) and With Logarithm (Right) Applied.}
    \centering
    \begin{subfigure}[b]{0.47\textwidth}
        \centering
        \includegraphics[width=\textwidth]{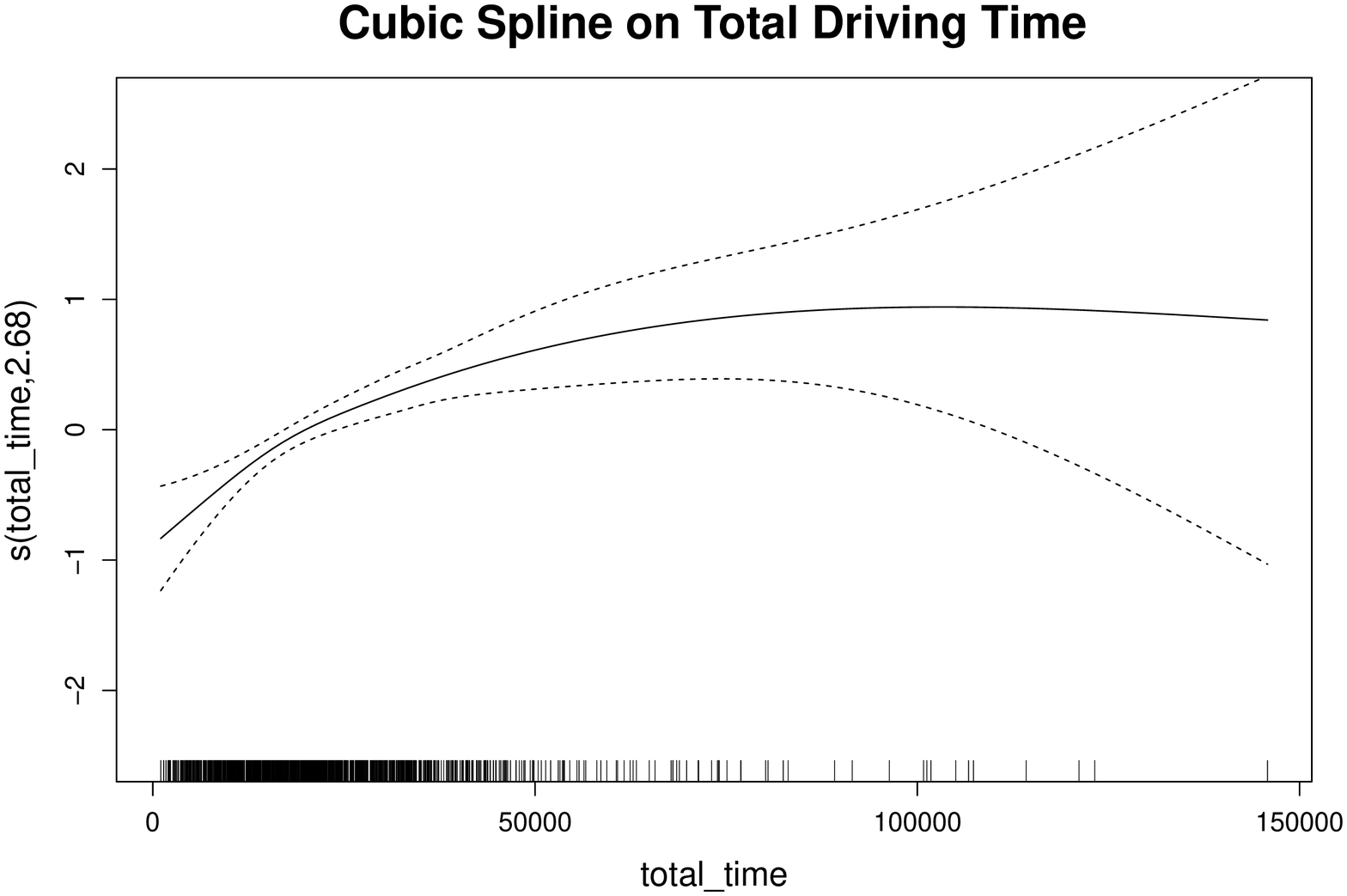}
        \caption*{}
    \end{subfigure}
    \hfill
    \begin{subfigure}[b]{0.47\textwidth}  
        \centering 
        \includegraphics[width=\textwidth]{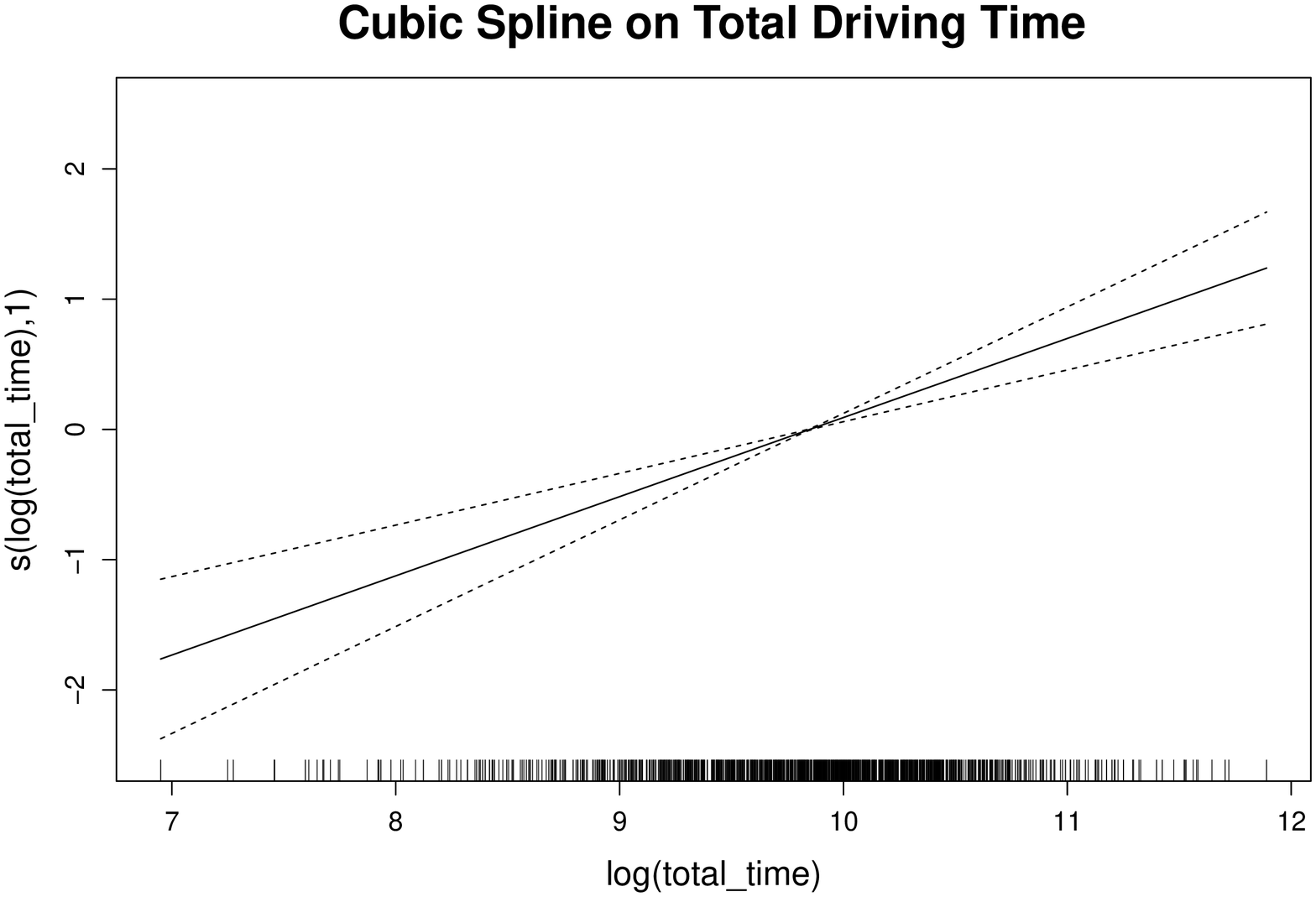}
        \caption*{}
    \end{subfigure}
    \label{fig: cubic-spline}
\end{figure*}

The same logged covariate structure is employed in \citet{boucher_pay-as-you-drive_2013}, where the authors consider a Poisson regression to model claim counts (from different lines of business) but with traditional covariates only.  Our results are in line with theirs, where the estimated coefficients of logged total distance travelled are close to 0.5 in all cases.  There are two reasons why we do not consider a more flexible function of the covariate, such as the cubic smoothing spline in \citet{boucher_exposure_2017}.  First it does not seem necessary, as shown in Figure \ref{fig: cubic-spline}, where we apply cubic spline on either total time or logged total time alongside the selected covariates from \texttt{tt\_mod1s} (in linear form).  We observe that cubic spline on total time shows a square-root-like shape, whilst that on logged total time is linear.  Hence, a linear relationship between claim counts and logged total time should be sufficient.  The second reason is to ensure a monotonically increasing relationship between (expected) claim counts and the total driving time or distance with minimal effort.  This is an important condition as the expected claim counts should never decrease with increasing driving.  The risk resulted from an earlier part of the trip, say the first 100 km or the first hour driven, should not be affected by the later part of the trip.  As a consequence and also given the fact that risk cannot be negative, the overall risk should be at least as big as that resulted from any interval of a trip.  Moreover, this condition is relevant from a ratemaking perspective, otherwise policyholders can simply drive infinitely long to minimize their premium.

Using a log-link in GLM with a logged covariate is equivalent to fitting a power function of the covariate, hence the relationship is monotonically increasing as long as the fitted coefficient is positive.  This observation of average claim counts increasing less than directly proportional to total driving time or total distance travelled is first made in \citet{lemaire_automobile_1985}, \citet{litman_distance_2011} and \citet{ferreira_measuring_2012}, where they conclude that a doubling, or even tripling of mileage increases average claim counts by less than a double.  One explanation is that the areas travelled can have different riskiness, e.g. familiar neighbourhood versus new destinations (assuming all other factors such as road and weather conditions equal).  Another explanation is the drivers' learning effect introduced in Section \ref{Empirical-Evidence-of-Drivers'-Learning-Effects-from-Telematics-Data}.  In that section, we have demonstrated a similar, (approximately) square-root trend in severe harsh event arrivals: it also takes increasingly longer for a severe harsh event (detected at 1.5G) to occur as total driving time increases.  The gain of experience and improvement in driving behaviour are reflected first in severe harsh event arrivals, and ultimately in claim arrivals.

From a practical perspective, the use of total time or distance travelled as an offset leads to poorer claims modelling and eventually poorer ratemaking.  In particular, the use of total driving time as an offset leads to a miss in both tails of the claims distribution when compared to the use of logged covariate.  We compare the goodness-of-fit and predictive power of the two models with the same 80-20 train/test split in Section \ref{Motivation-for-Using-the-Negative-Binomial}.  By comparing a linear function to a power function with power between 0 and 1, it is clear that the linear function (which is the offset model) underestimates the risk from short-term driving while overestimates the risk from long-term driving.  As a consequence, its prediction distribution overestimates both claim 0 and 4+, as shown in Tables \ref{table: in-sample-dist} and \ref{table: out-sample-dist}, whilst the covariate model gives closer fit and prediction both in- and out-of-sample.

\begin{table}[h]
\caption{In-sample Prediction Distributions of \texttt{tt\_mod1s} with Total Driving Time as Offset or (Logged) Covariate.}
\centering
\begin{tabular}{|rr|rr|}
\hline
\multicolumn{1}{|r|}{\textbf{Claim count}} & \textbf{Observed} & \multicolumn{2}{c|}{\textbf{Expected}} \\ \cline{3-4}
\multicolumn{1}{|r|}{} &  & \multicolumn{1}{c|}{\textbf{Offset}} & \textbf{Covariate} \\ \hline
\multicolumn{1}{|r|}{0} & 844 & \multicolumn{1}{r|}{848.50} & 841.90 \\
\multicolumn{1}{|r|}{1} & 201 & \multicolumn{1}{r|}{197.84} & 210.12 \\
\multicolumn{1}{|r|}{2} & 72 & \multicolumn{1}{r|}{67.47} & 68.79 \\
\multicolumn{1}{|r|}{3} & 34 & \multicolumn{1}{r|}{27.21} & 25.74 \\
\multicolumn{1}{|r|}{4} & 7 & \multicolumn{1}{r|}{12.22} & 10.60 \\
\multicolumn{1}{|r|}{5} & 5 & \multicolumn{1}{r|}{5.95} & 4.73 \\
\multicolumn{1}{|r|}{6} & 4 & \multicolumn{1}{r|}{7.83} & 5.13 \\ \hline
\multicolumn{1}{|l|}{\textbf{Total Claims}} & 524 & \multicolumn{1}{r|}{\begin{tabular}[c]{@{}r@{}}539.94\\ (+3.04\%)\end{tabular}} & {\begin{tabular}[c]{@{}r@{}}521.73\\ (-0.43\%)\end{tabular}}\\ \hline
\end{tabular}
\label{table: in-sample-dist}
\end{table}

\begin{table}[h]
\caption{Out-of-sample Prediction Distributions of \texttt{tt\_mod1s} with Total Driving Time as Offset or (Logged) Covariate.}
\centering
\begin{tabular}{|rr|rr|}
\hline
\multicolumn{1}{|c|}{\textbf{Claim count}} & \multicolumn{1}{c|}{\textbf{Observed}} & \multicolumn{2}{c|}{\textbf{Expected}} \\ \cline{3-4} 
\multicolumn{1}{|l|}{} & \multicolumn{1}{l|}{} & \multicolumn{1}{c|}{\textbf{Offset}} & \multicolumn{1}{c|}{\textbf{Covariate}} \\ \hline
\multicolumn{1}{|r|}{0} & 209 & \multicolumn{1}{r|}{213.29} & 210.54 \\
\multicolumn{1}{|r|}{1} & 52 & \multicolumn{1}{r|}{49.07} & 52.32 \\
\multicolumn{1}{|r|}{2} & 22 & \multicolumn{1}{r|}{16.39} & 16.99 \\
\multicolumn{1}{|r|}{3} & 3 & \multicolumn{1}{r|}{6.47} & 6.29 \\
\multicolumn{1}{|r|}{4} & 4 & \multicolumn{1}{r|}{2.85} & 2.57 \\
\multicolumn{1}{|r|}{6} & 1 & \multicolumn{1}{r|}{2.94} & 2.29 \\ \hline
\multicolumn{1}{|l|}{\textbf{Total Claims}} & 127 & \multicolumn{1}{r|}{\begin{tabular}[c]{@{}r@{}}130.26\\ (+2.56\%)\end{tabular}} & {\begin{tabular}[c]{@{}r@{}}129.18\\ (+1.71\%)\end{tabular}}\\ \hline
\end{tabular}
\label{table: out-sample-dist}
\end{table}

Moreover, the intrinsic difference between policy coverage period and total time or distance travelled suggests the use of different treatments.  First, policy period is usually known beforehand, with slight modifications to account for policy cancellation, whilst total time or distance travelled are unknown until policy coverage is over.  Second and more importantly, we usually use 1 to denote the most common coverage period and match the unit of time of the model (e.g. one year for annually renewable policies), with values smaller and larger to represent shorter and longer periods.  However, it is hard to decide in advance what the most common driving time or distance travelled should be, and this can easily change from year to year.  Hence, while policy coverage period can be included as an offset, it is more appropriate to treat total time or distance travelled as a covariate.

\section{Conclusion}
\label{Conclusion}

In this paper, we analyze an auto-insurance dataset with telematics data collected from a major European insurer.  On the one hand, through a discussion of the telematics data structure and related data quality issues, we have explained some practical challenges in processing and incorporating telematics information in loss modelling and ratemaking.  We hope that this can serve as a preview and an alert for interested researchers.  On the other hand, through an exploratory data analysis on both the portfolio and individual levels, we have demonstrated the existence of heterogeneity in individual driving behaviour even within the groups of policyholders with and without claims.  Through a 5-fold cross-validation, our regression analysis has reiterated the importance of telematics data in claims modelling: the model with only telematics covariates can already outperform the model with only traditional covariates, but the best performance is achieved by a combination of both.  In particular, we have proposed a speed transition matrix to describe speed time series, and concluded that large speed transitions, together with higher maximum speed attained, nighttime driving and increased harsh braking, are associated with increased claim counts.  Moreover, we have discovered that expected claim counts (or risk) are not directly proportional with driving time or distance, but increase in a decreasing rate instead; a similar pattern is observed in harsh events detected at a higher threshold, which we define and quantify as learning effect.  While the current work improves the understanding of telematics data, it also suggests directions for future work:
\begin{itemize}
    \item There are limitations in our dataset.  For example, some traditional covariates are missing or have unreliable values, while GPS signals are recorded per minute, which hinders the use of some telematics information such as the precise acceleration magnitudes.  Moreover, the detection thresholds of harsh events are predetermined, and what a reasonable threshold should be is itself an interesting question to explore.  While there is little that can be done on our side, it will rely on both a better data-collecting process from the insurers and an increased collaboration with the industry.
    \item We have modelled claim counts on an aggregate, policy period level.  It will be interesting to explore driving riskiness on a more granular level, such as on a seasonal, daily, or even trip basis.  This will allow for more frequent updates of risk classification and insurance premium, as well as improved safety through post-trip driver feedback.
    \item We have considered telematics information observed over the same period as policy coverage period, hence claim occurrence.  While this approach, as opposed to using historical average, can better illustrate the relationship between claims and telematics information and reduce bias, it requires the prediction of covariates and demands a flexible modelling of the telematics data.  This will be an important problem to consider in future research.
\end{itemize}

\section*{Acknowledgement}

The authors thank the editor and anonymous referees for their valuable comments and suggestions. The authors acknowledge the financial support provided by the Natural Sciences and Engineering Research Council of Canada [RGPIN 284246, RGPIN-2017-06684].


\bibliographystyle{apalike}
\bibliography{references}

\end{document}